\PassOptionsToPackage{usenames,dvipsnames,svgnames,table}{xcolor}
\documentclass[twocolumn,11pt, letter, natbib, twocolappendix]{aastex62}
\pdfoutput=1

\usepackage{graphicx}
\usepackage{times}
 \usepackage{xspace}
\usepackage{amsmath}
\usepackage{mathtools}
\usepackage{stix}

\renewcommand{\baselinestretch}{0.935} 

\newcommand{\unit}[1]{\mathrm{\,#1}}
\newcommand{\Msun}{\ensuremath{M_{\odot}}\xspace}
\renewcommand{\AA}{\Angstrom}

\newcommand{\HST}{\textit{HST}}
\newcommand{\KMOSTD}{KMOS$^{\hbox{\scriptsize{3D}}}$\xspace}

\newcommand{\deltPA}{\ensuremath{\Delta \mathrm{PA}}\xspace}

\newcommand{\Halpha}{\ensuremath{\textrm{H}\alpha}\xspace}
\newcommand{\Hbeta}{\ensuremath{\textrm{H}\beta}\xspace}
\newcommand{\OIII}{\ensuremath{\textsc{[Oiii]}}\xspace}

\newcommand{\Mstar}{\ensuremath{M_*}\xspace}
\newcommand{\Mgas}{\ensuremath{M_{\mathrm{gas}}}\xspace}
\newcommand{\Mbar}{\ensuremath{M_{\mathrm{baryon}}}\xspace}
\newcommand{\Mdyn}{\ensuremath{M_{\mathrm{dyn}}}\xspace}
\newcommand{\keff}{\ensuremath{k_{\mathrm{eff}}}\xspace}

\newcommand{\ktot}{\ensuremath{k_{\mathrm{tot}}}\xspace}

\newcommand{\Av}{\ensuremath{A_{\mathrm{V}}}\xspace}

\newcommand{\re}{\ensuremath{R_E}\xspace}  

\newcommand{\Vrms}{\ensuremath{V_{\mathrm{rms}}}\xspace}
\newcommand{\rms}{\ensuremath{{\mathrm{rms}}}\xspace}

\newcommand{\sigmavint}{\ensuremath{\sigma_{V, 0}}\xspace}
\newcommand{\sigmavobs}{\ensuremath{\sigma_{V, \, \mathrm{obs}}}\xspace}
\newcommand{\sigmavmodel}{\ensuremath{\sigma_{V, \, \mathrm{model}}}\xspace}

\newcommand{\sigmaonedobs}{\ensuremath{\sigma_{V,\mathrm{1D,obs}}}\xspace}

\newcommand{\vtosig}{\ensuremath{V/\sigmavint}\xspace}
\newcommand{\vtosigre}{\ensuremath{(V/\sigmavint)_{\re}}\xspace}
\newcommand{\vtosigtt}{\ensuremath{(V/\sigmavint)_{2.2}}\xspace}

\newcommand{\Vcirc}{\ensuremath{V_{\mathrm{circ}}}\xspace}
\newcommand{\Vtt}{\ensuremath{V_{2.2}}\xspace}

\newcommand{\fbar}{\ensuremath{f_{\mathrm{bar}}}\xspace}
\newcommand{\fdm}{\ensuremath{f_{\mathrm{DM}}}\xspace}

\newcommand{\Sigmabar}{\ensuremath{\Sigma_{\mathrm{bar},e}}\xspace}
\newcommand{\Sigmagas}{\ensuremath{\Sigma_{\mathrm{gas},e}}\xspace}

\newcommand{\deltlogM}{\ensuremath{\Delta\log_{10}M}\xspace}

\newcommand{\shalf}{\ensuremath{S_{0.5}}\xspace}

\newcommand{\MISFIT}{\texttt{MISFIT}\xspace}

\newcommand*{\MPE}{Max-Planck-Institut f\"{u}r extraterrestrische Physik (MPE), Giessenbachstr. 1, D-85748 Garching, Germany}
\newcommand*{\UCB}{Department of Astronomy, University of California,
  Berkeley, CA 94720, USA}
\newcommand*{\UPacific}{Department of Physics, University of the Pacific, 3601 Pacific Avenue, Stockton, CA 95211, USA}
\newcommand*{\UCLA}{Department of Physics \& Astronomy, University of
  California, Los Angeles, CA 90095, USA}
\newcommand*{\UCR}{Department of Physics \& Astronomy, University of California, Riverside, CA
  92521, USA}
\newcommand*{\UCSD}{Center for Astrophysics and Space Sciences, University of California, San Diego, La Jolla, CA 92093, USA}
\newcommand*{\UAriz}{Department of Astronomy / Steward Observatory, 933 North Cherry Avenue, Tucson, AZ 85721, USA}
\newcommand*{\CfA}{Harvard-Smithsonian Center for Astrophysics, 60 Garden Street, Cambridge, MA 02138, USA}
\newcommand*{\Wooster}{Department of Physics, The College of Wooster, 1189 Beall Avenue, Wooster, OH 44691, USA}
\newcommand*{\UCD}{Department of Physics, University of California, Davis, 1 Shields Avenue, Davis, CA 95616, USA}

\shorttitle{Kinematic and Structural Evolution of Star-Forming Galaxies at \lowercase{$1.4 \leq z \leq 3.8$}}
\shortauthors{Price et al.}

\begin{document}

\title{The MOSDEF Survey: Kinematic and Structural Evolution of Star-Forming Galaxies at \lowercase{$1.4 \leq z \leq 3.8$}}

\author{Sedona H. Price}\altaffiliation{Email: sedona@mpe.mpg.de}\affiliation{\MPE}\affiliation{\UCB}
\author{Mariska Kriek}\affiliation{\UCB}
\author{Guillermo Barro}\affiliation{\UPacific}
\author{Alice E. Shapley}\affiliation{\UCLA}
\author{Naveen A. Reddy}\altaffiliation{Alfred P. Sloan Fellow}\affiliation{\UCR}
\author{William R. Freeman}\affiliation{\UCR}
\author{Alison L. Coil}\affiliation{\UCSD}
\author{Irene Shivaei}\altaffiliation{Hubble Fellow}\affiliation{\UAriz}
 \author{Mojegan Azadi}\affiliation{\CfA}
\author{Laura de Groot}\affiliation{\Wooster}
\author{Brian Siana}\affiliation{\UCR}
\author{Bahram Mobasher}\affiliation{\UCR}
\author{Ryan L. Sanders}\affiliation{\UCD}
  \author{Gene C. K. Leung}\affiliation{\UCSD}
\author{Tara Fetherolf}\affiliation{\UCR}
  \author{Tom O. Zick}\affiliation{\UCB}
\author{Hannah \"Ubler}\affiliation{\MPE}
\author{Natascha M. F\"orster Schreiber}\affiliation{\MPE}


\begin{abstract}
\noindent
We present ionized gas kinematics for 681 galaxies at $z\sim1.4-3.8$ from the MOSFIRE 
Deep Evolution Field survey, measured using models that account for random galaxy-slit misalignments 
together with structural parameters derived from CANDELS \textit{Hubble Space Telescope} (\HST) imaging. 
Kinematics and sizes are used to derive dynamical masses. Baryonic masses are estimated 
from stellar masses and inferred gas masses from dust-corrected star formation rates (SFRs) 
and the Kennicutt-Schmidt relation. We measure resolved rotation for 105 galaxies. 
For the remaining 576 galaxies we use models based on \HST{} imaging structural parameters together 
with integrated velocity dispersions and baryonic masses to 
statistically constrain the median ratio of intrinsic ordered to disordered motion, \vtosig. 
We find that \vtosig increases with increasing stellar mass and decreasing specific SFR (sSFR). 
These trends may reflect marginal disk stability, where systems with higher gas fractions have thicker disks.
For galaxies with detected rotation we assess trends between their kinematics and mass, sSFR, 
and baryon surface density (\Sigmabar). 
Intrinsic dispersion correlates most with \Sigmabar and velocity correlates most with mass. 
By comparing dynamical and baryonic masses, we find that galaxies at $z\sim1.4-3.8$ are 
baryon dominated within their effective radii (\re), with \Mdyn/\Mbar increasing over time. 
The inferred baryon fractions within \re, \fbar, decrease over time, even at fixed mass, size, or surface density. 
At fixed redshift, \fbar does not appear to vary with stellar mass but increases 
with decreasing \re and increasing \Sigmabar. 
For galaxies at $z\gtrsim2$, the median inferred baryon fractions generally exceed $100\%$. 
We discuss possible explanations and future avenues to resolve this tension.
\end{abstract}



\section{Introduction}

\setcounter{footnote}{0}

A key open question in galaxy formation and evolution is how galaxy structures arise and evolve over time. 
Today's massive star-forming galaxies are assumed to form by 
the collapse of baryons within dark matter halos (e.g., \citealt{White78}, \citealt{Fall80}, \citealt{Blumenthal84}), 
resulting in thin, smooth stellar disks. 
However, the exact details of how baryons and dark matter interact throughout 
the galaxy disk formation process are not well understood. 
Constraining these physical processes and testing different formation models 
require direct studies of galaxies at earlier times. 

Recent work shows that massive star-forming galaxies in the early universe look very different 
from their local counterparts. In particular, at $z\sim1-3$, the peak of cosmic star formation rate 
(SFR) density in the universe (e.g., \citealt{Madau14}), massive star-forming galaxies 
are generally smaller (e.g., \citealt{Williams10}, \citealt{vanderWel14a}), have large clumps 
(e.g., \citealt{Elmegreen07}, \citealt{Law07}, \citealt{Genzel11}, \citealt{ForsterSchreiber11b}, 
\citealt{Guo15,Guo18}), have high gas fractions (e.g., \citealt{Daddi08}, \citealt{Tacconi08,Tacconi18}), 
and are baryon dominated within their disk extent (e.g., \citealt{vanDokkum15}, \citealt{Wuyts16}, 
\citealt{Genzel17}, \citealt{Lang17}, \citealt{Ubler17}, \citealt{Ubler18}). 
Despite these differences, high-redshift massive galaxies do appear to have rotating gas disks, 
though they are thicker (e.g., \citealt{Elmegreen06}) 
and have higher intrinsic velocity dispersions 
and increased turbulence compared to local massive star-forming galaxies 
(e.g., \citealt{ForsterSchreiber06, ForsterSchreiber09}, \citealt{Epinat08}, 
\citealt{Newman13}, \citealt{Green14}, \citealt{Wisnioski15}, \citealt{Simons16}).

Current theoretical models suggest that the thick, gas-rich disks of $z\sim1-3$ 
massive star-forming galaxies are assembled through smooth, 
cold-mode accretion or minor mergers (e.g., \citealt{Keres05, Keres09}, \citealt{Dekel06}, 
\citealt{Dave08}, \citealt{Dekel09}, \citealt{Oser10}, \citealt{Cacciato12}, \citealt{Ceverino12}, 
\citealt{Danovich15}, \citealt{Rodriguez-Gomez15}, \citealt{Correa18}). 
In this framework, the high turbulence and clumpy morphologies of these 
galaxies at $z\sim1-3$ could reflect higher average gas fractions compared to local galaxies 
(e.g., \citealt{Dekel09}, \citealt{Bournaud11}, \citealt{Genzel11}, \citealt{Genel12}).

Recent instrumentation advances, including multiplexing near-infrared (NIR) spectrographs such as 
MOSFIRE (\citealt{McLean10, McLean12}) and KMOS (\citealt{Sharples04, Sharples13}), 
have enabled large kinematic studies at high redshifts to test these theoretical models. 
Surveys with MOSFIRE and KMOS 
(e.g., MOSDEF, \citealt{Kriek15}; SIGMA, \citealt{Simons16}; ZFIRE, \citealt{Nanayakkara16}; 
\KMOSTD, \citealt{Wisnioski15}; KROSS, \citealt{Stott16}; and KDS, \citealt{Turner17}) 
now provide the kinematics for thousands of galaxies at $z\sim1-3$,
augmenting more detailed studies of smaller samples with higher, adaptive-optics-assisted spatial resolution 
(e.g., SINS/zC-SINF, \citealt{ForsterSchreiber09, ForsterSchreiber18}; 
MASSIV, e.g., \citealt{Contini12}; AMAZE/LSD, e.g., \citealt{Gnerucci11}). 
While these surveys have greatly expanded our understanding of high-redshift galaxies, 
many challenges to interpreting these results remain.

First, the majority of previous high-redshift kinematic studies are 
conducted using ground-based, seeing-limited instruments (e.g., MOSFIRE and KMOS). 
The low spatial resolution of these observations can mask 
rotation signatures in small and lower-mass galaxies \citep{Newman13}, 
impacting most galaxies with $\Mstar\lesssim10^{10}\Msun$ at $z\sim1-3$ 
as these galaxies are unresolved under seeing-limited conditions (\citealt{vanderWel14a}). 
Second, multiobject slit spectrographs (e.g., MOSFIRE) 
often have a constant position angle for all slits in a mask, resulting in 
random galaxy orientations within the slits that can further mask 
rotation signals. 
Third, fully constraining star-forming galaxy formation models requires observations of 
lower-mass galaxies at high redshifts. The progenitors of today's massive star-forming disk galaxies 
have masses $\Mstar\sim10^9-10^{10}\Msun$ at $z\sim2$ (e.g., \citealt{Leja13}, \citealt{vanDokkum13}), 
but currently kinematic observations of these early, low-mass galaxies 
are limited to very small sample sizes.

Fortunately, high-resolution space-based imaging allows us to 
constrain the structures of high-redshift galaxies by combining the information from the 
detailed imaging with seeing-limited spectra (e.g., \citealt{Price16}). 
By leveraging morphology from space-based imaging, 
it is possible to account for how much of the galaxy falls outside of the slit 
due to misalignment and for the effects of seeing blurring, 
which enables us to better understand the galaxy components and kinematics 
that are captured in the seeing-limited spectra. 
This approach can be applied to galaxies both with and without spatially resolved kinematics 
and can thus be used to study the internal dynamics for galaxies over a wide 
range of masses and sizes at high redshifts, as in \citealt{Price16}.

In this paper, we use observations from the MOSFIRE Deep Evolution Field (MOSDEF) 
survey \citep{Kriek15} to study the dynamical and baryonic masses and internal kinematic structures of 
a sample of 681 star-forming galaxies at $z\sim1.4-3.8$. 
Rotation is robustly detected in 105 galaxies, 
after we initially restrict our analysis to galaxies
that are spatially resolved along the slit 
and that have rough alignment between the galaxy major axis and the slit. 
We use detailed structural information from \textit{Hubble Space Telescope} (\HST) imaging 
from the CANDELS survey (\citealt{Grogin11}, \citealt{Koekemoer11}) and 
other ancillary information to statistically constrain the kinematics of 
the 95 spatially resolved, slit-aligned galaxies without detected rotation and 
of the remaining 481 spatially unresolved or slit-misaligned galaxies.
We investigate how the derived intrinsic rotation velocities ($V$), intrinsic velocity dispersions (\sigmavint), 
and ratio of intrinsic ordered to unordered motion (\vtosig) correlate with other galaxy properties, 
and how these kinematic properties change over cosmic time. 
We also compare the derived dynamical masses with the galaxies' baryonic masses 
to infer the evolution of the baryon and dark matter fraction in galaxies 
as a function of galaxy properties and redshift.

We adopt a $\Lambda$CDM cosmology with $\Omega_m=0.3$, $\Omega_{\Lambda}=0.7$, 
and $H_0=70\unit{km}\unit{s^{-1}}\unit{Mpc^{-1}}$ throughout this work.

\begin{figure*}
\vglue -9pt
\centering
\includegraphics[width=0.95\textwidth]{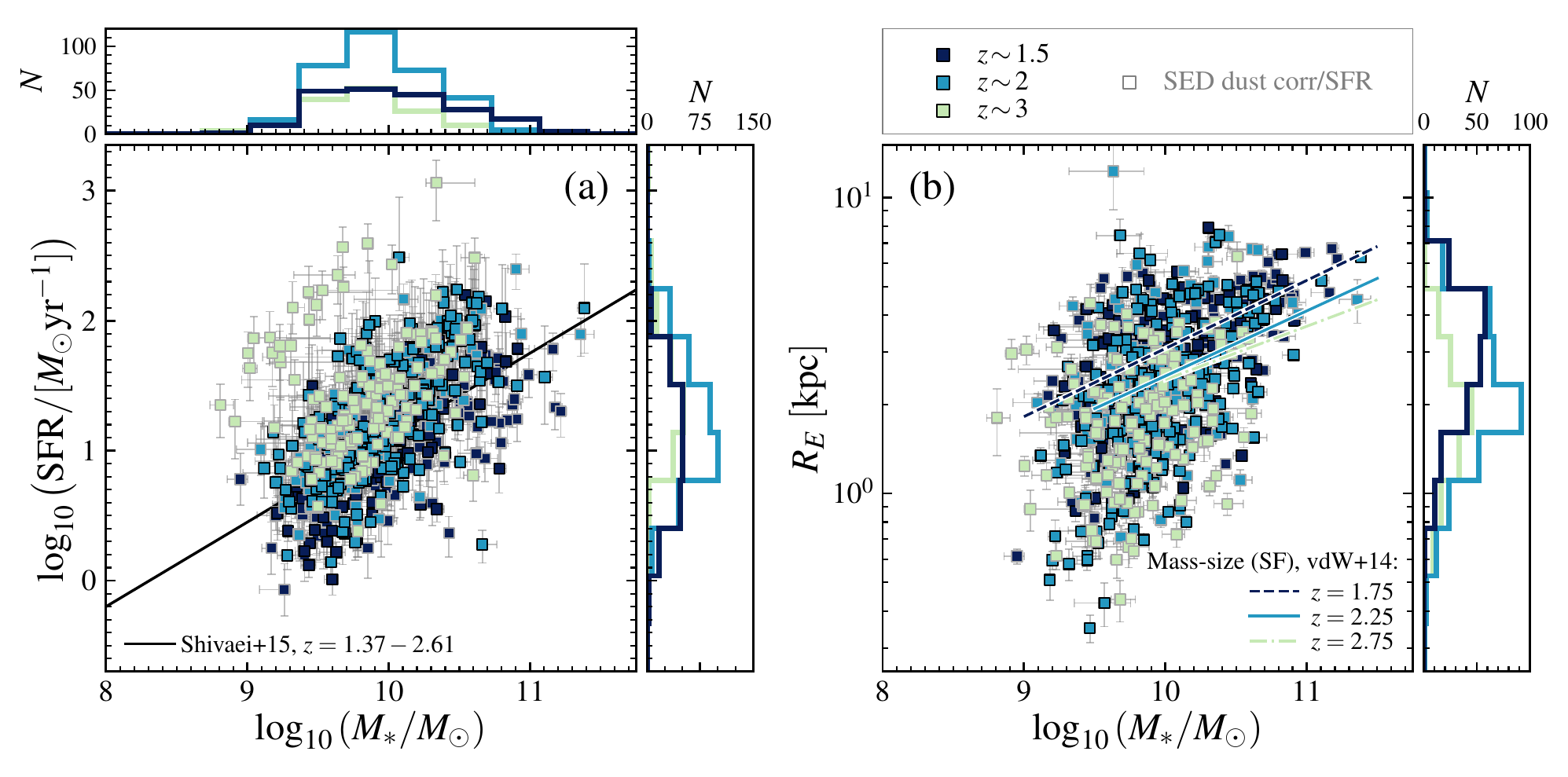}
\vglue -10pt
\caption{
Stellar mass, $\log_{10}(\Mstar/\Msun)$, versus \textit{(a)} $\log_{10}(\mathrm{SFR}/\Msun \mathrm{yr}^{-1})$ 
and \textit{(b)} effective radius, \re, for the galaxies in our sample.  
Galaxies at $z\sim1.5$, $z\sim2$, and $z\sim3$ are colored navy, teal, and light green, respectively. 
Histograms of  $\log_{10}(\Mstar/\Msun)$, $\log_{10}(\mathrm{SFR}/\Msun \mathrm{yr}^{-1})$, 
and \re show the distribution of galaxies in each redshift range. 
Galaxies without Balmer-decrement dust-corrected \Halpha SFRs are marked with gray outlines. 
The black solid line in the left panel shows the best-fit star-forming main sequence at 
$z\sim1.4-2.6$ from \citet{Shivaei15}. 
In the right panel, we show the best-fit mass-size relations for star-forming galaxies from 
\citet{vanderWel14a} at $z=1.75$ (navy dashed line), $z=2.25$ (teal solid line), and $z=2.75$ (light-green 
dashed-dotted line). Our sample spans a wide range of properties and generally follows the  
best-fit relations at these redshifts. 
}
\label{fig:sample}
\end{figure*}

\section{Data}
\label{sec:data}

\subsection{The MOSDEF Survey}
\label{sec:mosdef}

This work is based on the complete data set from the MOSDEF survey 
\citep{Kriek15}, which was carried out from 2012 December to 2016 May  
using the MOSFIRE spectrograph \citep{McLean12} on the 10 m Keck I telescope. 
In total, the survey obtained rest-frame optical, moderate-resolution ($R=3000-3600$) 
spectra for $\,\sim\!1500$ $H$-band-selected galaxies in the 
CANDELS (\citealt{Grogin11}, \citealt{Koekemoer11}) fields. 
The full survey details, including targeting, observational strategy, data reduction, success rate, sensitivities, 
redshift measurements, and other sample properties, are given in \citet{Kriek15}.

Structural parameters for the MOSDEF galaxies are measured by \citet{vanderWel14a} 
from \HST/F160W images from the CANDELS survey using \textsc{Galfit} \citep{Peng10a}. 
These parameters include the effective radius, \re (using the semi-major axis length as \re), 
the S\'ersic index, $n$ \citep{Sersic68}, the axis ratio, $b/a$, and the major-axis position angle.

The stellar masses and other stellar population parameters, including spectral energy distribution (SED) SFRs, 
are determined by fitting the $0.3-8.0 \mu\mathrm{m}$ multiwavelength photometry from the 3D-HST survey 
(\citealt{Brammer12}, \citealt{Skelton14}, \citealt{Momcheva16}) with stellar population models, in 
the same way as described in \citet{Price16} (see also \citealt{Kriek15}). Briefly, FAST \citep{Kriek09} is used to fit 
the flexible stellar population models (\citealt{Conroy09}, \citealt{Conroy10a}) to the multiwavelength 
photometry, using the MOSFIRE redshifts and assuming a \citet{Chabrier03} stellar initial mass function (IMF), 
a \citet{Calzetti00} dust attenuation curve, delayed exponentially declining star formation histories, and solar metallicity. 
We additionally correct the stellar masses for differences between the \textsc{Galfit} and total 
photometric magnitudes following \citet{Taylor10a} (as in Equation~(1) of \citealt{Price16}), 
ensuring self-consistency between 
the radii and stellar mass measurements.

Emission line fluxes (e.g., \Halpha, \OIII, \Hbeta) are calculated from the optimally extracted MOSFIRE 
1D spectra by fitting the lines and underlying continuum simultaneously with Gaussian profiles 
and a linear component. The \Halpha and \Hbeta fluxes are additionally corrected for the 
underlying Balmer absorption, using the best-fit stellar population models. 
More details on emission line measurements are given in \citet{Kriek15} and \citet{Reddy15}.

We use the following ranked approach to measure SFRs and estimate gas masses (\Mgas) for our sample. 
First, if both \Halpha and \Hbeta are detected ($\mathrm{S/N}\geq3$), 
the spectrum transmission at the lines is at least 50\% of the maximum transmission, 
and both lines have reliable fluxes (i.e., removing lines at the edge of the spectrum with unreliable fluxes), 
then Balmer-absorption-corrected \Halpha fluxes are corrected for dust attenuation using the Balmer 
decrement (320 galaxies, median $f_{\Halpha}/f_{\Hbeta} \sim 3.43$), 
assuming a \citet{Cardelli89} extinction curve (see \citealt{Reddy15}). 
\Halpha luminosities are then calculated from the dust-corrected \Halpha fluxes, and  
\Halpha SFRs are calculated from the luminosities using the relation of \citet{Hao11} 
for a \citet{Chabrier03} IMF \citep{Shivaei15}.

Next, if only one Balmer line (\Halpha or \Hbeta) is detected with signal-to-noise ratio 
$\mathrm{S/N}\geq3$ and has a reliable flux (283 galaxies), 
then we assume that the dust attenuation of the nebular regions is related to the continuum attenuation 
using the relation between $A_{V,\mathrm{neb, Calzetti}}$ and  $A_{V,\mathrm{cont,Calzetti}}$ by \citet{Price14}. 
As we have chosen to adopt the \citet{Cardelli89} curve for the nebular attenuation, we 
convert this relation to account for a \citet{Cardelli89} nebular and a \citet{Calzetti00} continuum attenuation 
curve.\footnote{As we determine the color excess from the observed Balmer decrement using
$E(\mathrm{B-V})_{\mathrm{neb}} = 
\log_{10}\left[(\Halpha/\Hbeta)_{\mathrm{obs}}/2.86\right]/
\left[0.4\,(k(\lambda_{\Hbeta})-k(\lambda_{\Halpha}))\right]$ for some reddening curve $k(\lambda)$, 
for the same observed Balmer decrement we have 
$E(\mathrm{B-V})_{\mathrm{neb,Cardelli}}=1.18E(\mathrm{B-V})_{\mathrm{neb,Calzetti}}$. 
\citet{Price14} Eq. 2 is equivalent to 
$E(\mathrm{B-V})_{\mathrm{neb}}=1.86\, E(\mathrm{B-V})_{\mathrm{cont}}$ as the \citet{Calzetti00} curve 
was adopted for both the nebular emission and stellar continuum, so this converts to 
$E(\mathrm{B-V})_{\mathrm{neb,Cardelli}}=2.2\, E(\mathrm{B-V})_{\mathrm{cont,Calzetti}}$. 
Since $A_{\mathrm{V}}=R_\mathrm{V} \times E(\mathrm{B-V})$, 
as $R_{\mathrm{V}}=3.1$ for the \citet{Cardelli89} curve
and $R_{\mathrm{V}}=4.05$ for the \citet{Calzetti00} curve, the final converted relation is 
 $A_{V,\mathrm{neb, Cardelli}} = 1.68 \,A_{V,\mathrm{cont,Calzetti}}$. }
We then use the resulting  
attenuation $A_{V,\mathrm{neb, Cardelli}}$ to correct the absorption-corrected Balmer line flux.  
The Balmer luminosities are then calculated from the dust-corrected line fluxes. 
If only \Hbeta is detected, we convert the dust-corrected \Hbeta luminosity to 
an \Halpha luminosity assuming $L_{\Halpha}/L_{\Hbeta} = 2.86$ \citep{Osterbrock06}. 
As with the above, \Halpha (or \Hbeta) SFRs are then determined from the dust-corrected Balmer luminosities 
using the \citet{Hao11}  relation.

Finally, if neither Balmer line is detected (78 galaxies), 
SFRs from SED fitting are adopted. 
In the absence of detected Balmer lines, these SED SFRs should be fairly reasonable to use, 
as \cite{Shivaei16} show that there is general consistency between Balmer decrement-corrected \Halpha 
and SED SFRs for a subset of the MOSDEF sample. 
However, we do note that \cite{Reddy15} find \Halpha SFRs are systematically higher than 
SED SFRs for galaxies with high SFRs or specific SFRs (sSFRs) for a separate MOSDEF sample subset.
To understand the impact of assuming different SFR indicators in this `ladder' method, 
we examine the agreement between the SFR measurements for the galaxies at $z\sim2$ 
that have detections of both \Halpha and \Hbeta in Appendix~\ref{sec:appendixB}.
We find that the indicators are generally in good agreement, though the SED SFRs are $\sim-0.23\unit{dex}$ lower 
than the Balmer-decrement \Halpha SFRs.  (We discuss further implications of these SFR offsets in 
Section~\ref{sec:caveats} and Appendix~\ref{sec:appendixB}.) 
While the SFR indicators are not perfectly matched, 
this staggered method nonetheless allows us to expand the sample selection 
from \citet{Price16} to include galaxies over a wider range of redshifts, 
as \Halpha measurements at $z\sim3$ are not accessible from the ground.

We then estimate gas masses for every galaxy using 
the relation $\Sigma_{\mathrm{SFR}} \propto \Sigma_{\mathrm{gas}}^N$, 
with $N=1.4$ following \citet{Kennicutt98}, 
as this slope is intermediate between previously measured Schmidt-law slopes at 
$z\sim1-3$ (see discussion in Section~\ref{sec:reconcile_masses}). 
Alternatively, gas masses could be estimated using the scaling relations by \citet{Tacconi18}. 
However, we choose to use the Kennicutt relation, as we primarily adopt \Halpha SFRs here and 
not UV+IR SFRs (though we discuss in Section~\ref{sec:scaling}  
how adopting gas masses from these scaling relations would impact our analysis).
Here we use $\Sigma_{\mathrm{SFR}}=\mathrm{SFR}/(2 \pi \re^2)$ 
and $\Sigma_{\mathrm{gas}}=\Mgas/(2 \pi \re^2)$,\footnote{We erroneously used 
$\Sigma_{\mathrm{SFR}}=\mathrm{SFR}/ (\pi \re^2)$
and $\Sigma_{\mathrm{gas}}=\Mgas/(\pi \re^2)$ in \citet{Price16}. 
Using the correct surface density definitions and the \citet{Kennicutt98} relation for a 
Chabrier IMF, the median $\log_{10}(\Mdyn/\Mbar)$ for our 
previous work should be $\deltlogM=-0.02$ (lower by $-0.06\unit{dex}$), 
implying no dark matter (instead of $\fdm=8\%$). 
Similarly, the intercept of the \Mbar-\shalf relation should be lower by 0.01 dex in $\log_{10}(\shalf)$.
Nonetheless, our primary result 
that galaxies at $z\sim1.5-2$ are very baryon dominated within their effective radii remains unchanged, 
as does the conclusion that a 
Chabrier IMF is more consistent with the dynamical masses than a Salpeter IMF.}$\ $where we adopt the best-available SFR and \re as the best-fit \textsc{Galfit} semi-major axis, 
assuming that the emission-line region coincides with the stellar continuum.

\begin{figure*}
\vglue -9pt
\centering
\includegraphics[width=0.9\textwidth]{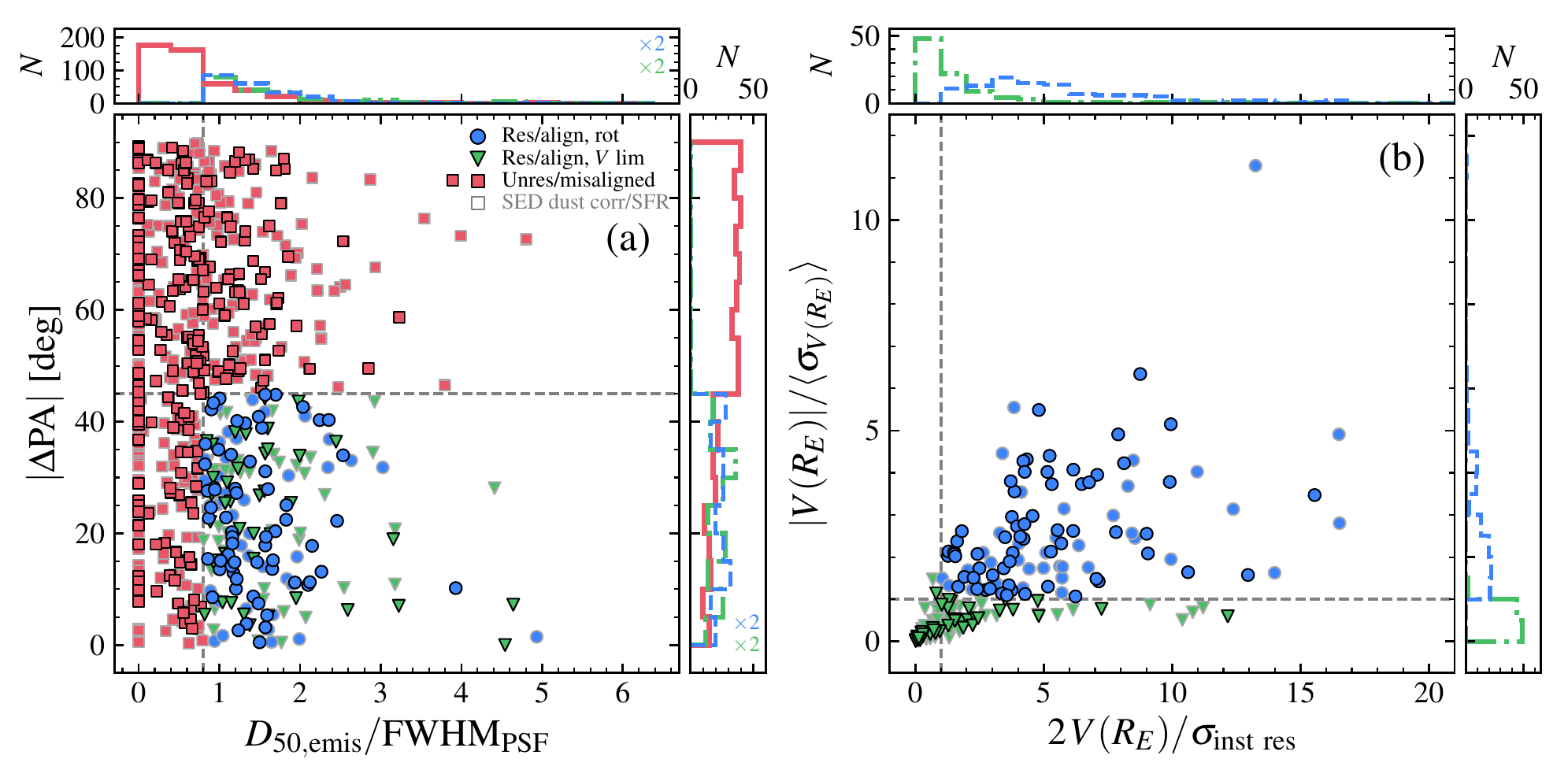}
\vglue -10pt
\caption{
Sample classification criteria: 
\textit{(a)} spatial resolution/alignment and \textit{(b)} velocity resolution. 
For the 2D kinematic fitting sample, we require the emission lines to be resolved 
($D_{50,\mathrm{emis}} \geq 0.8\times\mathrm{FWHM_{PSF}}$) 
and the galaxies to be aligned with the slit ($|\Delta\mathrm{PA}|<45^{\circ}$). 
Galaxies that fail the resolution/alignment cuts (red squares) are analyzed only using 1D kinematics.  
The resolved/aligned sample is then split on whether rotation is detected (blue circles) or not (green triangles), 
using the instrument resolution ($2V(\re)\geq\sigma_{\mathrm{inst\, res}}$) and 
uncertainty ($|V(\re)|/\langle err_{V(\re)}\rangle\geq1$). 
Histograms show the property distributions of the resolved/aligned, rotation-detected (blue dashed), 
rotation-limit (green dashed-dotted) and unresolved/misaligned (solid red) galaxies. 
In the left panel histograms, the number of aligned/resolved rotation and velocity-limit galaxies is 
multiplied by 2 for clarity. 
Galaxies without Balmer-decrement dust-corrected SFRs are 
marked with gray outlines, as in Figure~\ref{fig:sample}. 
}
\label{fig:sampleressplit}
\end{figure*}

\vfill \eject 

\subsection{Sample Selection}
\label{sec:sample}

We select a kinematics sample from the full MOSDEF survey using the following criteria. 
Galaxies are selected in the redshift ranges 
$1.34 \leq z \leq 1.75$, $2.075 \leq z \leq 2.6$, and $2.9 \leq z \leq 3.8$, 
to include galaxies in all three redshift ranges covered by the survey, 
with the best possible \Halpha coverage for the lower two ranges. 
We next require that each galaxy has at least one emission line of 
\Halpha, $\OIII 5007\AA$, or \Hbeta detected with 
$\mathrm{S/N} \geq 3$ and without severe skyline contamination. 
The sample is further restricted to only 
galaxies with \HST/F160W coverage, as we require structural parameter measurements. 
Where possible, we use \Halpha to measure kinematics to provide a fair comparison with previous studies  
using \Halpha kinematics (e.g., \citealt{ForsterSchreiber09}, \citealt{Wisnioski15}, \citealt{Stott16}), 
but if \Halpha fails the S/N or skyline contamination cuts, we include galaxies with clean $\OIII 5007\AA$ or \Hbeta 
detections (in order of preference). 
We prioritize $\OIII 5007\AA$ over \Hbeta to provide the most direct comparison with previous $z\sim3$ 
studies that use \OIII as the kinematic tracer (e.g., \citealt{Gnerucci11}, \citealt{Turner17}).

Additional selection criteria are applied to restrict the sample to galaxies with 
high-quality spectra and structural and stellar population parameters. 
First, we only include primary MOSDEF targets and exclude any serendipitously detected galaxies. 
Second, we only include galaxies with secure redshifts ($\mathtt{ZQUALFLAG} \geq 3$ in the final 
MOSDEF redshift catalog).
Third, we impose stellar population and structural parameter quality cuts to ensure that 
the models reasonably match the data. For the SED fitting, we exclude objects 
with best-fit reduced chi-square $\chi^2_{\mathrm{red}}>10$. We then 
exclude galaxies with structural parameter fits where 
(a) \textsc{Galfit} failed or reached a parameter limit or 
(b) the difference between the \textsc{Galfit} and \HST/F160W total magnitudes 
is greater than 0.5 mag. 
Fourth, we exclude active galactic nuclei (AGNs) from our sample based on X-ray luminosity, IRAC color, and 
rest-frame optical emission line ratios (\citealt{Coil15}, \citealt{Azadi17}), 
as kinematics of AGNs may include contributions from nonvirial motions in the nuclear region. 
Fifth, we exclude objects within the quiescent region in the 
UVJ diagram (\citealt{Wuyts07}, \citealt{Williams09}), 
to avoid objects where line emission comes from otherwise unidentified nuclear activity. 
Sixth, we remove any galaxy that appears to have an interacting counterpart at a similar redshift 
(i.e., objects with secondary objects within 1 arcsec, $|\Delta z| < 0.1$, and with $\Delta m_{\mathrm{F160W}} \leq 1$, 
as well as objects with disturbed/suspicious line or spatial profiles, based on visual inspection),
as these systems may have disturbed kinematics (see, e.g., \citealt{Shapiro08}). 
Finally, any object that has a line width smaller than the instrumental resolution is excluded from this analysis.

The final kinematic sample includes 681 galaxies. Of these galaxies, kinematics are measured 
using  \Halpha, $\OIII 5007\AA$, and \Hbeta for 481, 195, and 5 galaxies, respectively 
(with some objects having two or more observations of the best-available line from different masks).
Within our sample, both \Halpha and \Hbeta are detected and have reliable fluxes in 320 galaxies 
(i.e., Balmer-decrement -orrected \Halpha SFRs are adopted), 
one Balmer line is detected with reliable flux for 283 galaxies 
(i.e., \Halpha or \Hbeta SFRs are estimated using $A_V$ from SED fitting), 
and no Balmer line is detected for 78 galaxies (i.e., SED SFRs are adopted).

The stellar masses, SFRs, and effective radii of our final kinematics sample are shown in 
Figure~\ref{fig:sample}, colored by redshift range. 
For comparison, we show the best-fit stellar mass-SFR relation at $z\sim1.4-2.6$ by 
\citet{Shivaei15} (left panel) and the best-fit size-stellar mass relations for star-forming galaxies 
at $z=1.75$, $z=2.25$, and $z=2.75$ by \citet{vanderWel14a} (right panel). 
The low- and medium-redshift galaxies in our sample are in excellent agreement with the best-fit 
relation of \citet{Shivaei15}, and overall the sample shows the expected trend of 
higher SFRs at higher redshifts. 
The galaxies also generally follow the size-mass relations measured at similar redshifts, 
though the galaxies at $z\sim3$ are generally smaller than the 
relation measured at $z=2.75$, as our highest sample redshift range ($2.9\leq{}z\leq 3.8$) 
probes higher redshifts than presented by \hbox{\citet{vanderWel14a}.}

\subsection{Sample Spatial Resolution}
\label{sec:spatial_res}

We next measure the point-spread function (PSF) corrected spatial extent of the highest-S/N emission line for each galaxy, 
as 2D kinematic fitting will only yield meaningful constraints if the emission line is resolved. 
By measuring the intrinsic emission line sizes from the spectra, 
we directly account both for size variations between a galaxy's 
continuum and emission and for projection and seeing effects that 
impact how much of a galaxy's light falls within the slit. 
Intrinsic emission line half-light diameters are measured 
following a similar procedure to that used in \citet{Simons16}.

First, we mask skylines and low-S/N cross-dispersion spatial rows 
within the 2D spectrum of each galaxy. 
Any continuum is then subtracted from the 2D emission line. 
We measure the continuum slope from a weighted linear fit to the 1D spectrum 
and then fit the continuum row by row using weighted linear fits where only the intercept is variable. 
The observed emission line FWHM is measured from a weighted fit to a Gaussian profile. 
The intrinsic half-light diameter, $D_{\mathrm{50,emis}}$, is then determined by subtracting 
the PSF FWHM in quadrature from the observed line FWHM. 
For galaxies with observed FWHMs or unmasked spatial ranges less than the PSF size, 
we set $D_{\mathrm{50,emis}} = 0$.\footnote{Note that $D_{\mathrm{50,emis}}$ is only used 
for the spatial resolution classification and not for analysis, 
so for simplicity we do not adopt an upper limit.}
The distribution of $D_{\mathrm{50,emis}}/\mathrm{FWHM_{PSF}}$ 
versus the galaxy photometric major axis-slit misalignment ($|\deltPA|$) for our sample 
is shown in Figure~\ref{fig:sampleressplit}a. 

Galaxies are then classified as spatially resolved if they 
satisfy $D_{\mathrm{50,emis}} \geq 0.8\times\mathrm{FWHM_{PSF}}$. 
\citet{Simons16} demonstrate that kinematics can be accurately recovered 
down to this limit for the typical central $\mathrm{S/N}\sim\!15$ per pixel 
for their sample, and the MOSDEF observations have similar S/N values. 
Their finding is consistent with the results of our kinematic recovery 
tests presented in Appendix~\ref{sec:appendixA} (see Figure~\ref{fig:MISFITtest}), 
reinforcing the adoption of this resolution criterion.

\vfill \eject 

\section{Kinematic Measurements}
\label{sec:kin_meas}

The kinematic properties and dynamical masses of our sample are measured 
from emission lines (\Halpha, \OIII, or \Hbeta), 
together with the structural parameters derived from the \HST/F160W imaging, 
using the 3D models and methods developed in \citet{Price16}. 
In this analysis,  we measure and analyze both ``resolved'' and ``unresolved'' kinematics, 
as discussed below. For clarity, definitions of key kinematic variables are listed in Table~\ref{tab:kinvar}.

\subsection{Resolved Kinematics}
\label{sec:kin_res}

For spatially resolved galaxies, the rotation velocity and velocity dispersion can be 
measured from 2D spectra. 
However, the MOSDEF galaxies were observed with random 
misalignments between the galaxy major axis and the slit, 
resulting in reduced rotation signatures since part of the galaxy falls outside of the slit. 
We thus use the 3D models and methods developed by 
\citet{Price16} to fit the rotation and velocity dispersion from the 2D MOSFIRE spectra, 
incorporating the spatial information from the \HST/F160W imaging to 
account for the degree of slit-axis misalignment.

Full details of the kinematic models and the fitting code (\MISFIT; \textit{MIS}aligned-slit kinematic \textit{FIT}ting) 
are given in Appendix~A of \citet{Price16}. 
In brief, the models have three free kinematic parameters: the asymptotic velocity ($V_a$) and 
turnover radius ($r_t$) of an arctan rotation curve model, and a 
constant intrinsic velocity dispersion (\sigmavint). The models also account for inclination 
(assuming an intrinsic disk axis ratio $(b/a)_0=0.19$, following \citealt{Miller11}) 
and the galaxy sizes, brightness profiles, slit misalignments and kinematic aperture losses, 
seeing conditions, and instrumental resolution. 
Specifically, each model is calculated by first constructing 
a galaxy profile from the intrinsic kinematic and structural parameters, 
convolving with the PSF and instrument line-spread function, 
and finally extracting the kinematics within a misaligned slit.
As the \MISFIT models are constructed by forward modeling, the impact of beam smearing on 
the recovered kinematics is explicitly included. Thus, the fitting is able to directly constrain 
uncertainties from the range of intrinsic kinematics that match the observed data 
after applying beam smearing and other observational effects.

We subtract the continuum emission and mask the low-S/N rows and columns of the 
2D emission lines similar to the procedure of \citet{Price16}, 
while adopting appropriate wavelength ranges that 
exclude neighboring features for the different emission lines (i.e., \Halpha, \OIII, \Hbeta). 
Any emission lines from serendipitous objects are masked, if they fall within the fitting regions. 
The masked, continuum-subtracted 2D emission lines are then fit to the models using the 
\texttt{python} Markov Chain Monte Carlo (MCMC) 
package \texttt{emcee} \citep{Foreman-Mackey13}. 
The MCMC sampling was conducted for 100 steps with 2000 walkers,  
with a burn-in phase of 50 steps. 
For every object, the final acceptance fraction is between 0.2 and 0.5 
and the chain was run for longer than 10 times 
the maximum estimated parameter autocorrelation time. 
We then marginalize over the posterior distributions to determine 
the best-fit values and confidence intervals for the intrinsic $V(\re)$,  $\Vtt = V(r=2.2r_s)$ 
(where $r_s = \re/1.676$ is the scale length of an exponential disk), and $\sigmavint$.

While the \MISFIT 3D models directly account for the misalignment between the galaxy 
kinematic major axis and the slit, \deltPA, 
we lack independent measurements of the kinematic major-axis position angles. 
We therefore assume that the photometric and kinematic major axes are aligned. 
This assumption is generally reasonable, as \citet{Wisnioski15} 
find that these axes are within $\deltPA\leq30^{\circ}$ for most of their sample. 
Moreover, even with direct modeling of slit misalignments, 
objects that are very misaligned will suffer from degeneracies between 
the blended rotation signatures and the intrinsic velocity dispersion. 
In Appendix~\ref{sec:appendixA}, 
we use a suite of simulated galaxies and find that kinematics are recovered less accurately 
and have larger scatter if $|\deltPA|>45^{\circ}$ (Figure~\ref{fig:MISFITtest}, fourth column). 
Thus, we restrict the sample for which we fit the 2D kinematics to only galaxies that 
are spatially resolved ($D_{\mathrm{50,emis}} \geq 0.8 \times \mathrm{FWHM_{PSF}}$; 
Section~\ref{sec:spatial_res}) and relatively aligned with the slit ($|\deltPA|\leq45^{\circ}$). 
The spatial resolution and slit misalignment classification of our sample 
is shown in Figure~\ref{fig:sampleressplit}a.

Using the 2D fitting results, we then determine which spatially resolved/aligned galaxies have robust 
rotation detections. A galaxy is classified as having detected rotation if the best-fit rotation velocity 
at the effective radius, $V(\re)$, is 
\textit{(a)} nonzero with 68\% confidence (\mbox{$|V(\re)|/\langle err_{V(\re)}\rangle \geq 1$}) 
\mbox{and $\ $ 
\textit{(b)} $\ $ larger $\ \; $ than $\ $ the $\ \; $ instrument $\ $ resolution $\ \; $ limit} 
\mbox{($2V(\re)/\sigma_{\mathrm{inst.\ res}}\geq1$).}
We show these rotation resolution criteria for the spatially resolved/aligned galaxies 
in Figure~\ref{fig:sampleressplit}b.

\begin{figure*}
\vglue -8pt
\centering
\includegraphics[width=0.8\textwidth]{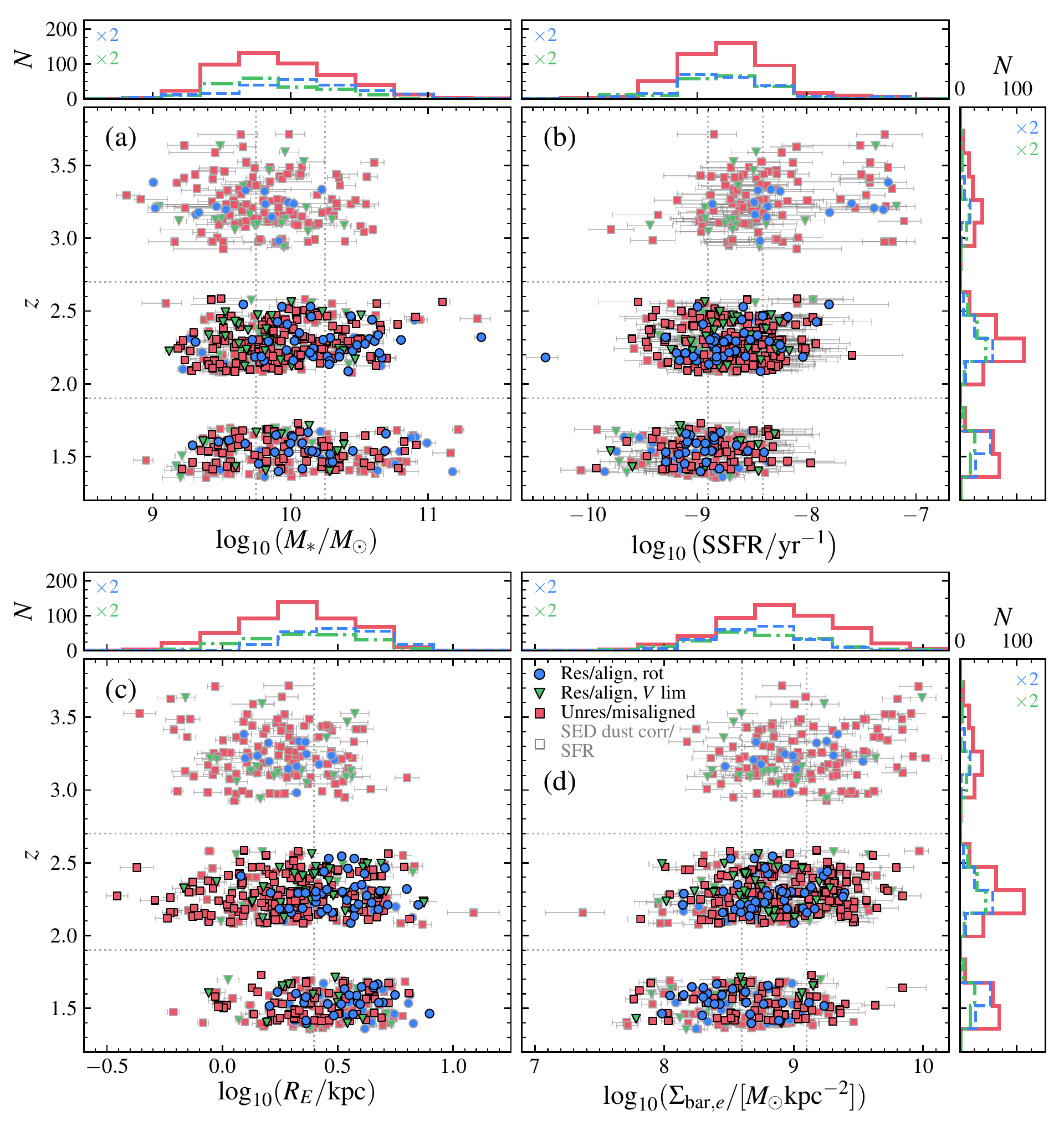}\vglue -10pt
\caption{
Comparison of redshift versus 
\textit{(a)} stellar mass, \textit{(b)} specific star formation rate (sSFR), \textit{(c)} effective radius (\re), 
and \textit{(d)} baryonic mass surface density (\Sigmabar) 
for resolved/aligned galaxies with detected rotation (blue circles) or with velocity limits (green triangles), 
and for unresolved/misaligned galaxies (red squares). 
Histograms of $z$, $\log_{10}(\Mstar/\Msun)$, $\log_{10}(\mathrm{sSFR}/\mathrm{yr}^{-1})$, 
$\log_{10}(\re/\mathrm{kpc})$, and $\log_{10}(\Sigmabar/[\Msun \mathrm{kpc}^{-2}])$ show the distribution of 
the resolved/aligned, rotation-detected (blue dashed), rotation-limit (green dashed-dotted) and unresolved/misaligned 
(solid red) galaxies 
(with the number of aligned/resolved rotation and velocity-limit galaxies multiplied by 2 for clarity). 
As in Figure~\ref{fig:sample}, galaxies without Balmer-decrement dust-corrected SFRs are 
marked with gray outlines. 
Bin boundaries in $z$, $\log_{10}(\Mstar/\Msun)$, $\log_{10}(\mathrm{sSFR}/\mathrm{yr}^{-1})$, 
$\log_{10}(\re/\mathrm{kpc})$, and $\log_{10}(\Sigmabar/[\Msun \mathrm{kpc}^{-2}])$ 
used for later analysis (see Section~\ref{sec:results}) are marked with gray dashed lines. 
The resolved/aligned galaxies with detected rotation tend to be slightly more massive and larger  
than the rotation-limit and unresolved/misaligned galaxies at a given redshift, but the subsamples have similar sSFRs. 
The baryonic surface densities of the resolved/aligned detected rotation and rotation-limit galaxies tend to be similar, and 
these galaxies generally are less dense than the unresolved/misaligned galaxies. 
}
\label{fig:samplebins}
\end{figure*}

\begin{table}
\scriptsize
 \centering \caption{Kinematic variables}
 \setlength{\tabcolsep}{0.06in} 
\label{tab:kinvar}
\begin{tabular*}{0.895\columnwidth}{l @{\extracolsep{\fill}}  l}
\hline
\hline  
& \vspace{-8pt} \\
\Vtt; $V(\re)$ & Intrinsic rotation velocity at $r_{2.2} = 2.2 r_s \approx 1.3 \re$; \re \\
\sigmavint & Intrinsic galaxy velocity dispersion (assumed to be const.)  \vspace{6pt} \\ 
\sigmaonedobs & Directly observed 1D integrated velocity dispersion \\
\sigmavobs & Integrated 1D velocity dispersion, corrected for inst. res. \\
\sigmavmodel & Kinematic model integrated 1D velocity dispersion  \vspace{3pt} \\
\hline
\end{tabular*}
\end{table}

Based on these sample classification criteria, 
we find that 200 galaxies are spatially resolved and aligned with the slit (``resolved/aligned''). 
Of the resolved/aligned galaxies, rotation is detected for 105 (``resolved/aligned, rotation''), 
and we place limits on the rotation for the other 95 galaxies (``resolved/aligned, rotation-limit''). 
The remaining 481 galaxies are classified as ``unresolved/misaligned''. 
For the galaxies with duplicate observations of the best-available line, we perform kinematic fits 
(either resolved or unresolved; Section~\ref{sec:kin_unres}) 
on the separate observations and then classify the object following its best-case kinematic observation(s) 
(resolved rotation, then rotation-limit, then unresolved/misaligned). 
The kinematics used in the rest of the analysis is taken to be 
the average of any duplicate best-case kinematics observations, weighted by the emission line S/N, 
or the single best-case observation, if the other observation(s) are less spatially/kinematically resolved.

The distribution of the galaxies within these three categories over redshift, stellar mass, sSFR,  
effective radius (\re), and baryon mass surface density (\Sigmabar) is shown in 
Figure~\ref{fig:samplebins}. Overall, resolved/aligned galaxies with detected rotation tend to have 
slightly higher masses and effective radii than those galaxies with rotation limits or that are 
unresolved/misaligned, but all categories have similar sSFRs. Additionally, we detect rotation in 
relatively few galaxies at $z\sim3$ compared to the galaxies at $z\sim1.5$ and $z\sim2$.  
This dearth of rotation-detected objects at $z\sim3$ is likely driven by 
evolution of galaxy half-light sizes, resulting in fewer resolution elements at 
higher redshifts on average (e.g., see Fig~\ref{fig:samplebins}c).

\subsection{Unresolved Kinematics}
\label{sec:kin_unres}

Kinematics for galaxies classified as ``unresolved/misaligned'' and ``resolved/aligned, rotation-limit'' 
(Section~\ref{sec:kin_res}) are measured from the integrated 1D spectra. 
However, while these two categories of galaxies have no detected rotation, 
we might expect that they have at least partial rotational support, 
which has likely been masked by the seeing-limited conditions (e.g., \citealt{Newman13}), 
together with the misalignment of the galaxies and slits. 
Therefore, as in \citet{Price16}, we model the integrated, unresolved kinematics 
using the same 3D models while assuming a fixed value of $\vtosigre = V(\re)/\sigmavint$.

Briefly, we first measure the integrated velocity dispersions $\sigmaonedobs$ 
by fitting the emission line (\Halpha, \OIII, or \Hbeta), any neighboring lines, and 
the continuum simultaneously using Gaussian profiles and a linear component, assuming that the 
emission lines have the same width in velocity space.
 The measured integrated velocity dispersion $(\sigma_{V,obs})$ 
is then corrected for instrumental resolution by subtracting 
$\sigma_{V,\mathrm{sky}}=\sigma_{{inst.\ res}}$  
in quadrature from $\sigmaonedobs$, which is measured from the skylines in each spectrum. 
Errors on \sigmavobs are estimated by repeating the fitting and correction procedure 
on 500 random, error-perturbed copies of each spectrum.

We then use the same \MISFIT 3D kinematic models to convert the observed 
velocity dispersions $\sigma_{V,\mathrm{obs}}$ to an intrinsic rms 
velocity 
$\Vrms (R_E) = \sqrt{ V(R_E)^2 + \sigma_{V,0}^2 } $, 
as described in Appendix~B of \citet{Price16}. 
In summary, each galaxy is modeled as an inclined disk using the best-fit \textsc{Galfit} structural parameters  
and is offset from the slit by the measured $\deltPA$. 
Rotation and velocity dispersion are included by assuming a fixed ratio of $\vtosigre$ 
(see Section~\ref{sec:dyn_masses_unres}) 
and assuming an arctan velocity curve where $r_t = 0.4 \, r_s = 0.4 \, (\re/1.676)$ \citep{Miller11}. 
We then compute the luminosity-weighted, seeing-convolved integrated velocity dispersion 
($\sigma_{V,\mathrm{model}}$) and rms velocity at \re ($\Vrms(\re)_{\mathrm{model}}$) 
from the galaxy model. 
Finally, we calculate the composite $\Vrms(\re)_{\mathrm{1D,corr}}$ as in Equation~(2) of \citet{Price16}, 
using the measured instrument resolution corrected integrated velocity dispersion ($\sigma_{V,\mathrm{obs}}$) 
together with the model-integrated velocity dispersion ($\sigma_{V,\mathrm{model}}$) and the model 
rms velocity ($\Vrms(\re)_{\mathrm{model}}$).


\subsection{Measuring Dynamical Masses and \vtosig}
\label{sec:dyn_masses}

We determine the dynamical masses (\Mdyn) for the galaxies in our sample by combining their 
measured kinematics and structural information. Here we give the 
calculation of \Mdyn from both 2D and 1D spectra, detail how \vtosig is constrained for 
galaxies without resolved and detected rotation, 
and compare the dynamical masses measured using both methods.

\subsubsection{Resolved Rotation from 2D Spectra}
\label{sec:dyn_masses_res}

For galaxies with resolved and detected rotation measured from 2D spectra, 
the dynamical masses are determined from their inferred rotation velocities. 
Furthermore, we apply an asymmetric drift correction to account for the 
nonnegligible pressure support in these galaxies (e.g., 
\citealt{Epinat09}, \citealt{Newman13}, \citealt{Wuyts16}). 
The pressure-corrected circular velocity is 
\begin{equation}
\Vcirc(r) = \sqrt{V(r)^2 + 2\left( r/r_s \right) \sigmavint^2  }, 
\label{eq:vcirc}
\end{equation}
where $r_s$ is the disk scale length (e.g., \citealt{Burkert10}), 
yielding $\Vcirc(\re) = \sqrt{V(\re)^2 + 3.35 \sigmavint^2}$ at \re.
We then calculate the total dynamical masses as
\begin{equation}
\Mdyn = \ktot(\re) \frac{\Vcirc(\re)^2\re}{G},
\label{eq:Mdyn}
\end{equation}
where $G$ is the gravitational constant. 
We adopt a virial coefficient of $\ktot(\re) = 2.128$, 
which corresponds to a flattened S\'ersic mass profile for a system with 
intrinsic axis ratio $q=0.4$ 
(to approximate the composite oblateness of the galaxy disk and halo; 
see \citealt{Miller11}, \citealt{Dutton11b}) 
and $n=1$ (exponential; typical for star-forming galaxies; \citealt{Wuyts11b}), 
following the calculations given by \citet{Noordermeer08}.\footnote{We explicitly 
apply an asymmetric drift correction and a constant virial coefficient $\ktot(\re)$ 
for consistency with other work (e.g., \citealt{Wuyts16}, \citealt{Ubler17}), 
as opposed to using $\Vrms(\re)$ and defining a composite $\keff(\re)$ as in \citet{Price16}. 
We note that this alternative approach would not \mbox{impact} the results of our previous study, 
with only a $\Delta\log_{10}\Mdyn = -0.006\unit{dex}$
 mass difference between these methods.  }
The impact of the assumed virial coefficient is discussed in Section~\ref{sec:reconcile_masses}.

For the resolved/aligned galaxies with detected rotation, we use the best-fit values of $V(\re)$ and $\sigmavint$ to 
directly determine $\Vcirc(\re)$ and \Mdyn using Equations~\ref{eq:vcirc} and \ref{eq:Mdyn}.

\subsubsection{Unresolved Kinematics from 1D Spectra}
\label{sec:dyn_masses_unres}

For galaxies without robustly detected rotation or that are 
unresolved/misaligned, we cannot simultaneously constrain $V(\re)$ and \sigmavint, 
as we measure their kinematics from integrated 1D spectra. 
We must instead assume a fixed value of \vtosigre 
to determine $\Vrms(\re)_{\mathrm{1D,corr}}$. 
From this, we can then calculate $\Vcirc(\re)|_{(\vtosig)}$ (from Eq.~\ref{eq:vcirc}): 
\begin{equation}
\Vcirc(\re)|_{(\vtosig)}=\sqrt{\left[\frac{\vtosigre^2+3.35}{\vtosigre^2+1}\right]}\Vrms(\re)_{\mathrm{1D,corr}},\!
\label{eq:vcirc_unres}
\end{equation}
assuming an exponential disk profile, where $\re=1.676\,r_s$. 
Finally, we calculate \Mdyn for this fixed \vtosigre using Equation~\ref{eq:Mdyn}.

\begin{figure*}[t!]
\centering
\vglue 10pt
\includegraphics[width=0.8925\textwidth]{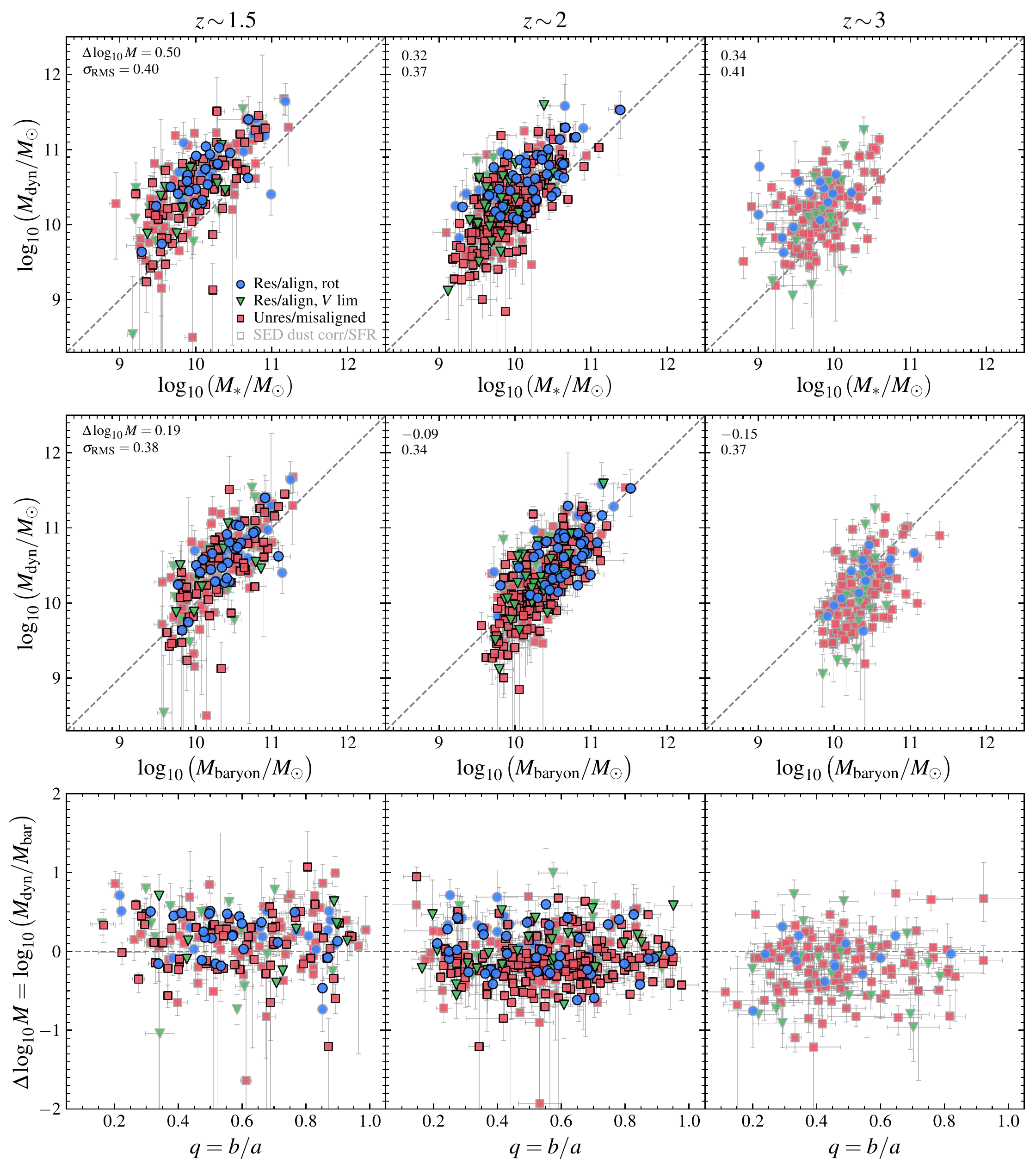}
\caption{
Comparison of dynamical and 
stellar \textit{(top row)} and 
baryonic (stellar and gas; \textit{middle row}) masses for galaxies at 
$z\sim1.5$ \textit{(left)}, $z\sim2$ \textit{(middle)}, and $z\sim3$ \textit{(right)}. 
The resolved/aligned detected rotation, resolved/aligned rotation-limit, 
and misaligned/unresolved galaxies are shown as 
blue circles, green triangles, and red squares, respectively.
The average \vtosigre used to calculate \Mdyn for the unresolved/misaligned and rotation-limit galaxies 
is fit in bins of $z$ and \Mstar (see Figure~\ref{fig:samplebins}a). 
Objects without Balmer-decrement dust-corrected SFRs are denoted with gray outlines. 
The gray dashed lines show $\Mdyn=\Mstar$ {(top row)} and $\Mdyn=\Mbar$ {(middle row)}, 
and the median offset \deltlogM and rms scatter are marked in each panel. 
\textit{(Bottom row)} Comparison of ratio between dynamical and baryonic masses 
versus galaxy axis ratio, $b/a$, for the three redshift ranges. 
The gray dashed horizontal line denotes no offset. 
In general, all galaxies have good agreement between their 
dynamical and baryonic masses, and we see no differences between 
\Mdyn/\Mbar versus axis ratio (i.e., inclination) for the three subsample classifications. 
}
\vglue 30pt
\label{fig:dyn_mass}
\end{figure*}

However, we $\,\!$ can $\,\!$ statistically $\,\!$ constrain $\,\!$ the $\,\!$ average 
\hbox{\vtosigre{} for these galaxies by examining the} 
correlation between the difference between dynamical and baryonic 
($\Mbar = \Mstar + \Mgas$) masses as a function of galaxy axis ratio. 
For disks, low inclination reduces the observed rotation signal while having no impact on an isotropic dispersion. 
For example, if \vtosigre is overestimated, $V(\re)$ will be overestimated and thus the composite $\Vrms(\re)$ will 
be overcorrected for inclination effects, leading to an overestimate of \Mdyn. 
For this case the degree to which \Mdyn is overestimated should depend on axis ratio; 
edge-on ($b/a\sim0.19$) galaxies have no inclination correction, 
while the correction is very high for face-on galaxies ($b/a\sim1$). 
Since we expect no intrinsic trend between \Mdyn or \Mbar and inclination (a random orientation projection effect), 
we would see this inclination overcorrection trend imprinted as a positive trend between 
$\Delta \log_{10}M = \log_{10}(\Mdyn/\Mbar)$ and $b/a$. 
Thus, we can infer the best-fit \vtosig as the value that best removes 
any dependence of $\Delta \log_{10}M$ on $b/a$. 
We note that this approach of modeling galaxies as inclined disks does not prevent 
us from finding very low to no ensemble-average rotation.

We constrain the ensemble average \vtosigre using the trend of $\Delta \log_{10}M$ versus $b/a$ 
as introduced in \citet{Price16}, but following a slightly different procedure. 
First, we calculate the dynamical masses for the galaxies 
without detected rotation and those that are unresolved/misaligned 
over a range of \vtosigre values (from $\vtosigre=0$ to 10). 
We then measure the slope between $b/a$ and \deltlogM 
for these galaxies at every value of \vtosigre, and determine the slope error 
by generating 500 bootstrap samples with replacement and measuring the slope for each realization. 
The best-fit \vtosigre is then estimated by finding the value that removes 
the trend of \deltlogM with $b/a$ (i.e., a slope of zero). 
The confidence interval is taken to be the range of \vtosigre where the slopes are 
consistent with zero within the uncertainties. 
Average \vtosigre for the unresolved/misaligned or undetected rotation galaxies are measured using this method 
in bins of redshift, stellar mass, sSFR, effective radius, and baryon surface density. 
The set of bin boundaries used in this work are shown with gray dashed lines in Figure~\ref{fig:samplebins}.

We then use the best-fit \vtosigre in each bin 
to calculate $\Vrms(\re)_{\mathrm{1D,corr}}$ (Sec.~\ref{sec:kin_unres}), 
$\Vcirc(\re)|_{(\vtosig)}$ (Eq.~\ref{eq:vcirc_unres}), and finally \Mdyn (Eq.~\ref{eq:Mdyn}) 
for the nonresolved, non-rotation-detection galaxies. 
We additionally determine \vtosigtt (see Sec~\ref{sec:kin_trends_vtosig}) 
from the best-fit \vtosigre, using the assumed arctan curve (Sec~\ref{sec:kin_unres}). 
The process is repeated for each of the binning parameter spaces 
($z$, $z$-\Mstar, $z$-sSFR, $z$-\re, and $z$-\Sigmabar), 
allowing us to use consistently measured \vtosigtt and \Mdyn for the non-rotation-detected 
galaxies when examining trends in these parameter spaces.

\subsubsection{Full-sample Dynamical Masses}
\label{sec:dyn_masses_comp}

The dynamical, baryonic, and stellar masses of the galaxies in our sample 
are shown in Figure~\ref{fig:dyn_mass}. 
For the galaxies that are spatially unresolved/misaligned or have unresolved kinematics, 
the average \vtosigre is determined in bins of $z$ and \Mstar (see Figure~\ref{fig:samplebins}a). 
Overall, as expected, the dynamical masses exceed the stellar masses for most galaxies 
in all three redshift ranges, 
with a slightly higher offset at $z\sim1.5$ relative to the higher redshift bins. 
The dynamical and baryonic masses are also reasonably consistent, with 
the median \Mdyn-\Mbar offset decreasing with increasing redshift. 
All galaxies (resolved/aligned galaxies with and without detected rotation,  
as well as the unresolved/misaligned galaxies) follow the same \Mstar-\Mdyn and \Mbar-\Mdyn relations. 
The three subsamples also have similar distributions of 
$\Mdyn/\Mbar$ versus axis ratio (i.e., inclination). 
We do note that our star-forming galaxy sample 
is incomplete below $\log_{10}(\Mbar/\Msun)\lesssim 9.5-9.8$. 
The impact of this incompleteness is discussed further in Section~\ref{sec:caveats}.

\section{Results}
\label{sec:results}

We use these kinematic and structural observations for galaxies in the MOSDEF survey 
to explore connections between kinematics and other properties and
to constrain the evolution of the dynamical-baryonic mass offset and 
the inferred baryonic and dark matter fractions in galaxies between $z\sim1.4-3.8$.  
Previous studies with more detailed kinematic observations 
have examined these properties, but the large sample size, three redshift intervals, and wide range of 
galaxy masses within each epoch of our sample provide a unique opportunity to investigate the kinematics 
and mass budgets for the star-forming galaxy population at $z\sim1-3$.


\subsection{Comparison between Internal Kinematics\\ 
and Galaxy Properties}
\label{sec:kin_trends}

We begin by examining how the kinematics of our sample vary with stellar mass (\Mstar), 
sSFR (which correlates with gas fraction; \citealt{Tacconi13}), 
and baryon surface density (\Sigmabar), in order to probe how galaxy structure and turbulence are set.

\begin{figure*}[t!]
\centering
\vglue -5.5pt
\includegraphics[width=0.7\textwidth]{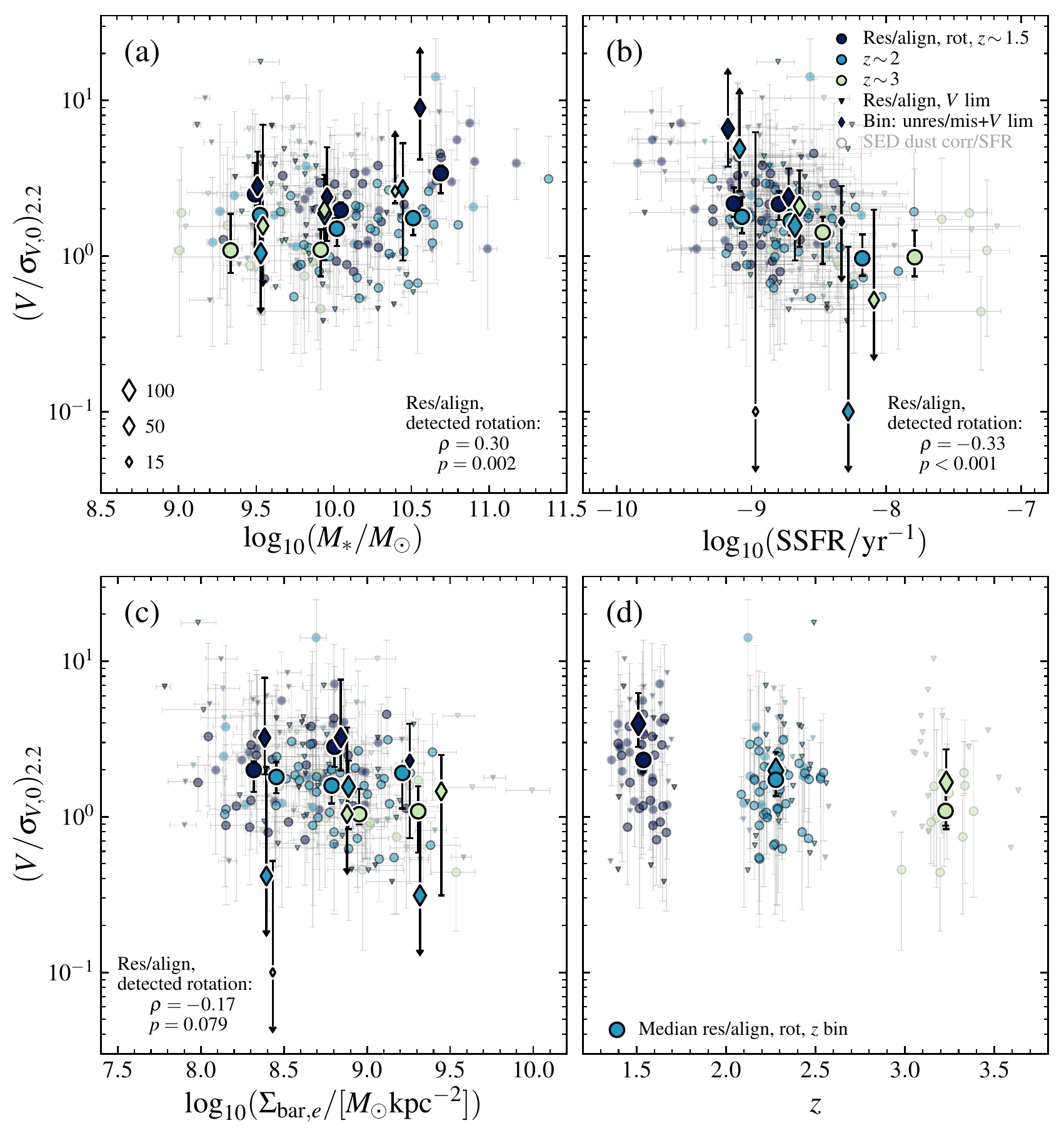}
\vglue -10pt
\caption{
\vtosigtt vs. \textit{(a)} stellar mass, \textit{(b)} sSFR, \textit{(c)} baryon surface density (\Sigmabar), and \textit{(d)} 
redshift for the galaxies in our sample, 
colored by redshift range as in Figure~\ref{fig:sample}.  
Resolved/aligned galaxies with detected rotation are shown as circles, 
and the upper limits on \vtosigtt for rotation-limit galaxies are marked with triangles. 
Gray outlines denote galaxies without Balmer-decrement-corrected SFRs. 
Median \vtosigtt for the resolved/aligned rotation galaxies binned by parameter and redshift 
are shown as large circles (if $N_{\mathrm{bin}}\geq5$). 
Additionally, the measured median \vtosigtt for binned unresolved/misaligned and resolved/aligned velocity-limit galaxies 
are shown with diamonds, with the size denoting the number of galaxies in each bin. 
The bin boundaries for the ensemble \vtosigtt fits in panels \textit{(a)}, \textit{(b)}, and \textit{(c)} 
are as defined in Figure~\ref{fig:samplebins}(a), (b), and (d), respectively, 
and panel \textit{(d)} is binned by redshift (also as in Figure~\ref{fig:samplebins}). 
Spearman correlation coefficients and $p$-values between 
the parameters for the resolved/aligned galaxies with detected rotation (circles) are listed in the panels. 
We find a positive correlation between \Mstar and \vtosigtt 
and a negative correlation between sSFR and \vtosigtt for these rotation-detected galaxies, 
but see no significant correspondence between \Sigmabar and \vtosigtt. 
These trends may reflect marginal disk stability, where galaxies with lower sSFR (and gas fraction) 
and higher stellar mass have naturally have lower turbulence and thus more support from ordered motions.
The trends also suggest that surface density has little impact on internal kinematic structure.
Furthermore, we observe an average increase in \vtosig toward lower redshifts, 
but this is consistent with being entirely driven by 
the mass differences between redshift bins (Figure~\ref{fig:samplebins}) together with
the average decrease in galaxy gas fractions and sSFRs over time
(i.e., reflecting the \vtosig-\Mstar and \vtosig-sSFR trends; panels \textit{(a)} and \textit{(b)}.
}
\label{fig:vtosig}
\end{figure*}


\begin{figure*}
\vglue -5pt 
\hglue -6pt
\includegraphics[width=1.02225\textwidth]{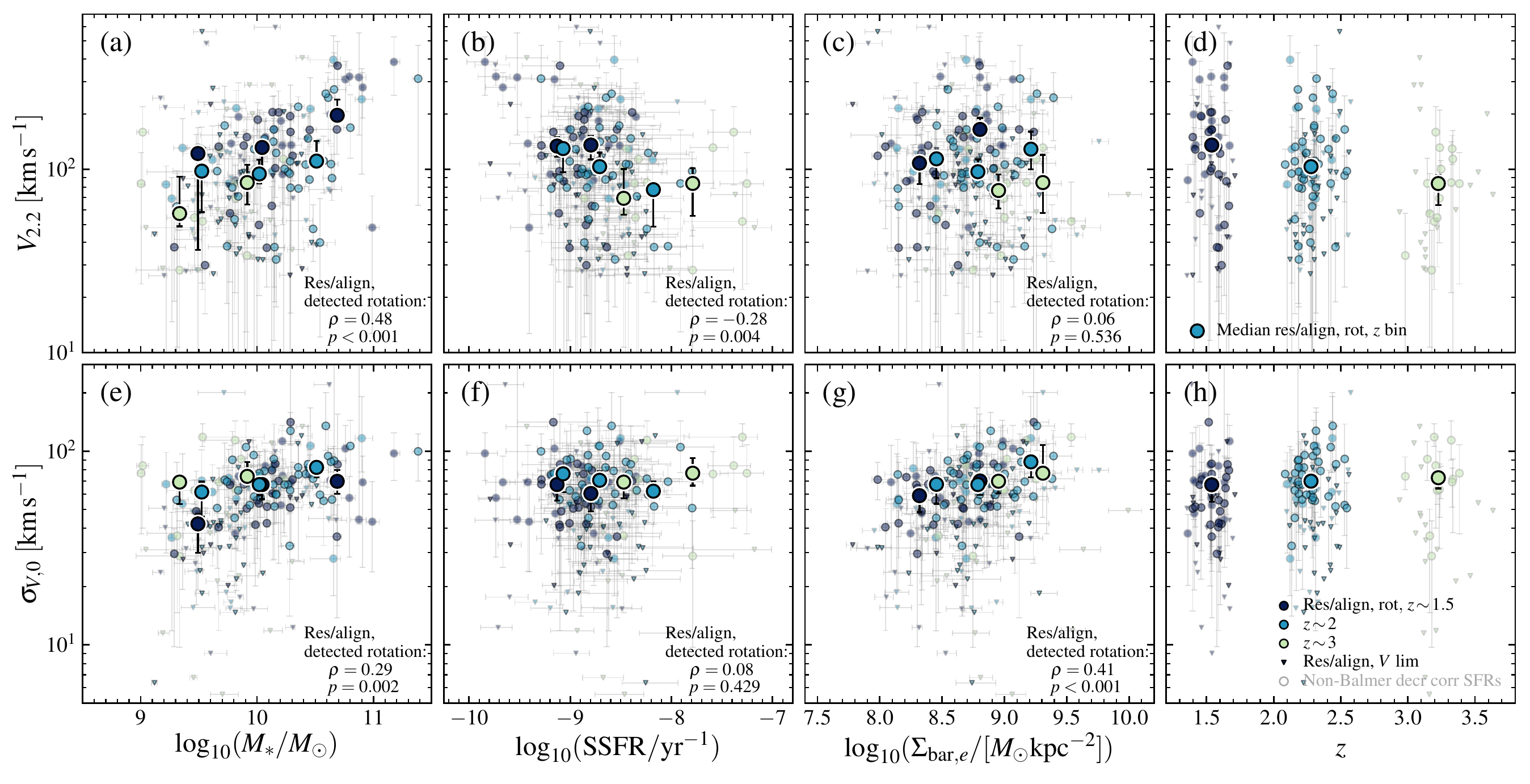}
\vglue -10pt
\caption{
Comparison of \Vtt \textit{(top)} and $\sigmavint$ \textit{(bottom)} versus 
stellar mass \textit{(first column)}, sSFR \textit{(second column)}, \Sigmabar \textit{(third column)}, and redshift \textit{(fourth column)} for 
aligned/resolved galaxies colored by redshift range as in Figure~\ref{fig:sample}. 
The point definitions are the same as in Figure~\ref{fig:vtosig}. 
For galaxies with $V$ limits (triangles), the upper limit of \Vtt is shown. 
There are positive correlations between \Vtt and \Mstar and between \sigmavint and \Sigmabar, 
with weaker, less significant correlations between \sigmavint and \Mstar (positive) and \Vtt and sSFR (negative), 
and no discernible correlation of \sigmavint and sSFR or \Vtt and \Sigmabar. 
The strong relation between velocity and mass is expected, since kinematics trace a system's potential. 
However, the trends with dispersion suggest that surface density, rather than 
integrated properties, is more important in setting turbulence in high-redshift star-forming galaxies. 
}
\label{fig:v_sig}
\end{figure*}


\subsubsection{Trends with \vtosig}
\label{sec:kin_trends_vtosig}

The amount of kinematic support from ordered versus random motions (\vtosig) provides information 
about the internal structures of galaxies. In particular, low values of \vtosig may indicate that 
a galaxy has a thick disk and high gas turbulence, while galaxies with high \vtosig tend to have 
ordered, thin disks. We thus investigate the relationship between \vtosig and other properties 
to constrain what processes drive the internal structures of star-forming galaxies at high redshifts.

In Figure~\ref{fig:vtosig}, we present  $\vtosigtt = V(2.2r_s)/\sigmavint$\footnote{We consider 
$V(r=2.2r_s)$, the radius where an exponential rotation curve peaks, 
for the analysis of the ratio of ordered-to-disordered motions (\vtosig), 
to provide reasonable comparisons with existing measurements, 
as we do not directly constrain the turn-over or flattening for the rotation curves of our galaxies.
}
versus stellar mass (Figure~\ref{fig:vtosig}a), sSFR (Figure~\ref{fig:vtosig}b), 
baryon mass surface density, $\Sigmabar$ 
(measured within the deprojected, major-axis effective radius \re; Figure~\ref{fig:vtosig}c), 
and redshift (Figure~\ref{fig:vtosig}d) for the resolved/aligned galaxies. 
Median \vtosigtt for the rotation-detected galaxies (large circles) are measured 
within bins of stellar mass, sSFR, and \Sigmabar split by redshift, and also binned by redshift 
alone, following the bin boundaries shown in Figure~\ref{fig:samplebins}. 
The uncertainties on the medians are estimated by perturbing \vtosigtt for each rotation-detected 
galaxy by its error over 500 realizations. 
We also show the ensemble average \vtosigtt of the 
unresolved/misaligned and velocity-limit galaxies (diamonds). 
These ensemble-average values and their uncertainties are determined following the 
procedure of Section~\ref{sec:dyn_masses}, and are separately measured within 
each of the above bins (again, see Figure~\ref{fig:samplebins}). 
For clarity, the markers for best-fit median bins with no rotation ($\vtosigtt=0$) are 
displayed at $\vtosigtt=0.1$, given the log plot scale. 
Uncertainties that are consistent with the lower or upper ensemble fit boundary 
($\vtosigtt=0$ or $10$) are marked with arrows. 
For reference, the median \vtosigre and \vtosigtt for the resolved/aligned, rotation-detected galaxies 
and the unresolved/misaligned and rotation-limit galaxies within all bins are given in 
Tables~\ref{tab:kinprop_res} and \ref{tab:kinprop_unres}, respectively.

We find a moderate positive correlation (Spearman rank correlation coefficient 
$\rho=0.30$ with $\sim\!\!2.9\sigma$ significance) 
between stellar mass and \vtosigtt for our entire sample of galaxies with detected rotation. 
We observe a similar trend in the \vtosigtt values measured in bins of $z$ and \Mstar for 
the unresolved/misaligned and rotation-limit galaxies. 
Additionally, there is a moderate negative correlation 
($\rho=-0.33$ with $>\!\!3\sigma$ significance)
between \vtosigtt and sSFR for the detected rotation galaxies, with a similar trend for the bins of 
unresolved/misaligned and rotation-limit galaxies. 
In contrast, we find a very weak, insignificant negative trend between \Sigmabar and \vtosigtt.

These trends of \vtosigtt with \Mstar and sSFR are in fairly good agreement with the findings of 
\citet{Newman13} at $z\sim1-2.5$, 
\citet{Wisnioski15} at $z\sim1$ and 2, \citet{Simons16} and  \citet{Alcorn18} at $z\sim2$, 
and \citet{Turner17} at $z\sim3.5$. 
We note that \citeauthor{Wisnioski15} find a larger trend of \vtosig with redshift at fixed stellar mass than we 
do, but the redshift intervals differ between the two studies ($z\sim1$ and 2 versus $z\sim1.5$, 2, and 3 used here). 
Additionally, the \citeauthor{Wisnioski15} sample excludes galaxies with $\vtosig<1$, while ours does not.

\citet{Wisnioski15} suggest that the trend of \vtosig with sSFR is indicative of 
galaxies tending to have more ordered motions when they have lower gas fractions, 
as predicted by the Toomre disk stability criterion (\citealt{Toomre64}; 
see also \citealt{Genzel11}, \citealt{Wisnioski15}). 
As sSFR correlates with gas fraction \citep{Tacconi13}, 
we would thus expect to see a negative correlation between \vtosig and sSFR.

Since sSFR and stellar mass are 
also correlated, we would expect to observe both a decreasing trend of 
sSFR and an increasing trend of stellar mass toward higher \vtosig. 
The slightly stronger correlation of \vtosig with sSFR that we observe suggests that 
gas fractions may be more important than stellar mass in setting a galaxy's dynamical structure. 
Furthermore, the lack of correlation with \Sigmabar suggests that surface density has little impact 
on the relative amounts of ordered and unordered motions in these high-redshift galaxies.

The overall trends of \vtosig with \Mstar and sSFR also appear to hold 
for galaxies at $z\sim1.5$ and $z\sim2$ separately. 
The trends with \vtosig at $z\sim3$ are inconclusive due to the larger errors 
caused by small numbers of galaxies in some bins and the lack of dynamic range. 
Thus, while massive galaxies at $z\sim1-2.5$ have thicker, more turbulent disks due to higher gas fractions 
than their local counterparts, the underlying physics regulating marginal disk stability may be the same 
and in place by $z\sim2$.

For general reference, we also show the individual and median \vtosigtt 
versus redshift (Figure~\ref{fig:vtosig}d) for all galaxies, 
without controlling for any sample differences between the redshift bins. 
When we consider galaxies at fixed stellar mass, sSFR, or baryonic mass surface density, 
there are offsets that are suggestive of a decrease in \vtosigtt with increasing redshift. 
However, given the measurement uncertainties, 
these offsets are also consistent with no evolution with redshift. 
The apparent decrease of \vtosigtt with increasing redshift in 
Figure~\ref{fig:vtosig}d thus reflects the average increase in sSFR (i.e., increase in 
gas fraction) and the average decrease in stellar mass between the redshift ranges 
(see Figure~\ref{fig:sample} and Tables~\ref{tab:kinprop_res},~\ref{tab:kinprop_unres}) 
and should not be interpreted as a pure redshift evolution. 
While our results show no conclusive redshift trends of \vtosigtt with fixed galaxy 
mass, sSFR, or surface density, we cannot definitively rule out any \vtosig evolution over time.

Our median \vtosig values are in good agreement with the trend of previous results out to $z\sim3.5$ 
(e.g., \citealt{Epinat08}, \citealt{ForsterSchreiber09}, \citealt{Law09},  
\citealt{Gnerucci11}, \citealt{Epinat12}, \citealt{Green14}, 
\citealt{Wisnioski15}, \citealt{Harrison17},  \citealt{Swinbank17}, 
\citealt{Turner17}; see Fig.~7 of \citealt{Turner17}). 
As discussed above, our values for \vtosigtt at $z\sim2$ are somewhat lower compared to the \KMOSTD values 
because we do not apply a \vtosig cut. 
Additionally, our $z\sim2$ sample extends to lower masses than the \citeauthor{Wisnioski15} $z\sim2$ sample, 
which could further explain our lower \vtosig values.


\subsubsection{Trends with Rotation and Intrinsic Dispersion for Galaxies with Robustly Detected Rotation}
\label{sec:kin_trends_v_sig}

We also separately investigate correlations between rotation velocity, velocity dispersion, and 
other galaxy properties (as motivated in the beginning of Section~\ref{sec:kin_trends}) 
to directly examine how disk rotation velocity and turbulence evolve. 
Figure~\ref{fig:v_sig} shows measurements of $\Vtt=V(r=2.2r_s)$ and \sigmavint 
versus stellar mass, sSFR, baryon surface density, and redshift for 
the resolved/aligned galaxies with detected rotation. 
Median values within bins of redshift, stellar mass, sSFR, and \Sigmabar are also shown 
and are listed in Table~\ref{tab:kinprop_res}.

We find the strongest correlation for our resolved/aligned rotation-detected sample 
between \Vtt and \Mstar ($\rho=0.48$ at $>\!\!3\sigma$). 
Additionally, there is fairly strong trend between \sigmavint and \Sigmabar ($\rho=0.41$ at $>\!\!3\sigma$). 
There are also weaker, less significant correlations between \sigmavint and \Mstar ($\rho=0.29$ at $\sim\!2.8\sigma$)
and \Vtt and sSFR ($\rho=-0.28$ at $\sim\!2.7\sigma$).
We do not find any correlation between \sigmavint and sSFR or 
between \Vtt and \Sigmabar for the rotation-detected sample.

These trends (or lack thereof) of \Vtt and \sigmavint are in qualitative 
agreement with previous work on massive star-forming galaxies at high redshifts. 
The observed correlation between \Vtt and \Mstar in the redshift slices agrees generally with the results presented by 
\citet{Harrison17} at $z\sim1$, \citet{Simons16}, \citet{Straatman17}, and \citet{Alcorn18} at $z\sim2$, 
and \citet{Turner17} at $z\sim3.5$. 
Similarly, the weak trend between \sigmavint and \Mstar split by redshift is consistent with 
the findings of \citet{Stott16} at $z\sim1$, 
\citet{Wisnioski15} at $z\sim1-2$, 
\citet{Alcorn18} at $z\sim2$, and \citet{Turner17} at $z\sim3.5$. 
In contrast, \citet{Simons17} report no significant difference in dispersion with mass at redshifts between 
$z\sim0.2-2$. However, we note that our median \sigmavint values in bins of \Mstar are consistent with 
the \citeauthor{Simons17} median values at both $z\sim1.5$ and 2. 
Finally, the lack of correlation between \sigmavint and sSFR that we see is similar to what 
\citet{Wisnioski15} find.

The strong correspondence between rotation velocity and stellar mass is 
expected, as kinematics trace the total potential of a system. 
From the Toomre disk stability criterion, we would also expect the intrinsic dispersion to scale 
with both rotation velocity and gas fraction (e.g., \citealt{Wisnioski15}). 
Additionally, the equilibrium or regulator galaxy growth model -- where 
star-forming galaxies are in more or less steady equilibrium between gas inflow, star formation, 
and outflows (e.g., \citealt{Bouche10}, \citealt{Dave12}, \citealt{Lilly13}, \citealt{Dekel14}) -- 
would also predict a relation between intrinsic galaxy dispersion and gas fraction.
Our finding of a slight correlation between \sigmavint and stellar mass 
is in line with expectations from disk stability theory. 
However, we do not see a correlation between dispersion and sSFR 
(as a proxy for global gas fraction; e.g., \citealt{Tacconi13}), as we would expect from the equilibrium model. 
Instead, we find that the intrinsic dispersion \sigmavint is most strongly correlated with baryon surface density. 
The stronger correlation of dispersion with surface density, rather than sSFR or stellar mass, suggests 
that density may be more important in setting the amount of random motion than a galaxy's global properties 
(e.g., total gas fraction or rotation velocity/total mass). 
However, given the uncertainties in our kinematic and sSFR measurements (and 
the inherent selection effects in our ``rotation-detected'' sample), 
our results cannot definitively rule out a relation between 
intrinsic velocity dispersion and gas fraction.

\begin{figure*}[t]
\vglue -4.5pt 
\centering
\hglue -5pt 
\includegraphics[width=1.02\textwidth]{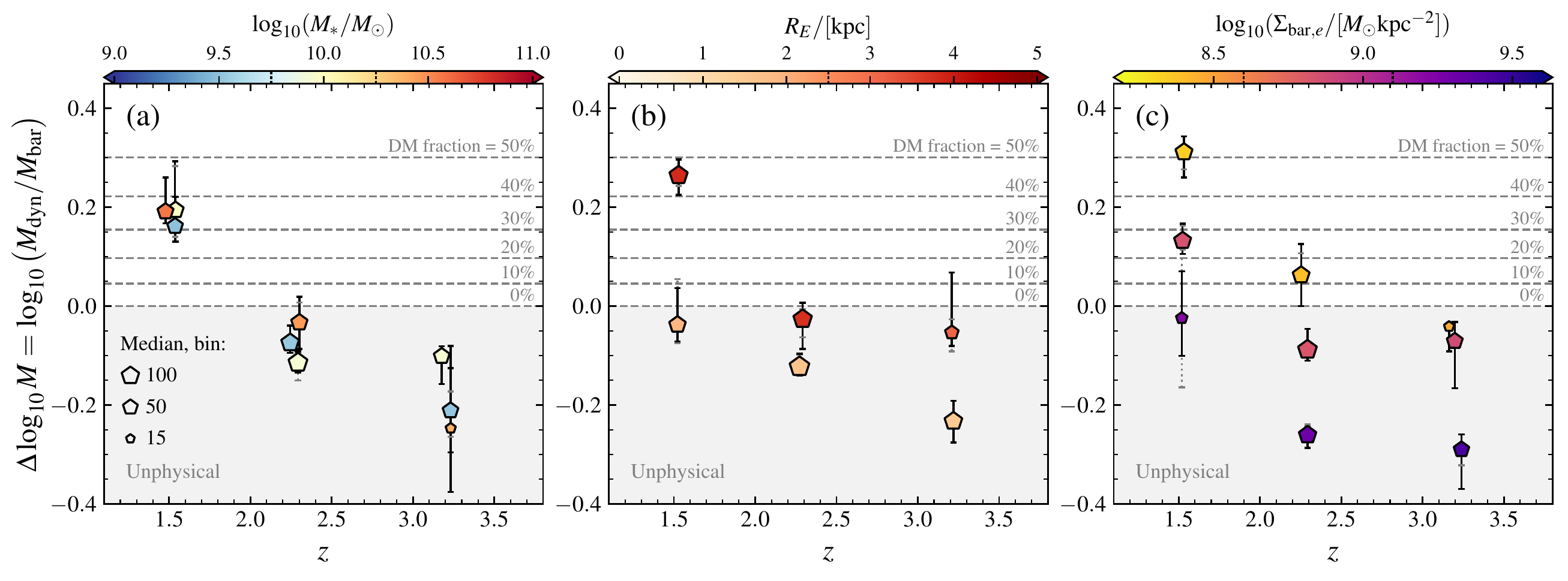} 
\vglue -8pt
\caption{
Trend of \Mdyn/\Mbar and inferred dark matter fraction with redshift, in bins of 
\textit{(a)} stellar mass, \textit{(b)} effective radius, and \textit{(c)} baryon surface density. 
The median mass offsets $\Delta \log_{10}M = \log_{10}(\Mdyn/\Mbar)$ for all galaxies 
(including all three kinematic subsample classifications) 
in bins of $z$ and \Mstar, \re, or \Sigmabar (Figures~\ref{fig:samplebins}a,c,d) are shown as pentagons, 
scaled by the number of galaxies in each bin and colored by the 
median \Mstar, \re, and \Sigmabar, respectively. 
The average \vtosigre for the unresolved/misaligned and rotation-limit galaxies 
is fit within these respective bins of $z$ and \Mstar, \re, and \Sigmabar. 
Black bars denote the bootstrapped errors on the median for $\Delta \log_{10}M$. 
Dotted gray bars demonstrate how the median $\Delta \log_{10}M$ changes when using 
the $\pm1\sigma$ range of median \vtosig measured for the rotation-limit 
and unresolved/misaligned galaxies in each bin. 
Constant dark matter fractions are shown as gray dashed lines, 
and the unphysical region where $\Mdyn < \Mbar$ is shaded gray. 
We observe a decrease in $\Delta \log_{10}M$ and the inferred dark matter fraction toward higher redshifts. 
This decrease is seen even at fixed mass, suggesting that the average growth of 
galaxy masses with time is not responsible for this trend. 
At fixed redshift, we find that the median dark matter fraction within the effective radius 
is relatively constant with stellar mass but is generally higher for galaxies with larger sizes 
or lower surface densities. 
For the higher redshift ranges, the median \deltlogM for most bins lies within the unphysical 
region where $\Mdyn<\Mbar$. These negative \deltlogM values suggest that 
one, if not more, of the assumptions and methods used to derive dynamical and baryonic masses 
may not be valid, 
in particular 
at higher redshifts or smaller sizes.  
}
\label{fig:DMfrac}
\end{figure*}

There is a general trend of  \Vtt increasing toward later times. 
This trend may primarily reflect the differences in stellar mass (and sSFR) in each redshift range 
(see Figure~\ref{fig:samplebins} and Table~\ref{tab:kinprop_res}). 
When examining galaxies at 
fixed stellar mass, sSFR, and \Sigmabar, we see do
suggestive trends of increasing median \Vtt over time. 
However, these offsets are also generally consistent with no redshift evolution, 
given the uncertainties and small number of ``rotation-detected'' galaxies. 
Nonetheless, considering the mass differences between redshift slices, 
our observed trends of \Vtt with $z$ are generally consistent with the 
evolution from $z\sim0.2$ to $z\sim2$ found using DEEP2 and SIGMA data 
(\citealt{Kassin12}, \citealt{Simons17}) 
and agree with the overall trend of results from $z\sim0$ out to $z\sim3.5$ (\citealt{Turner17}, and the 
literature values presented therein).

In comparison, we find suggestive trends of \sigmavint decreasing from 
$z\sim3.5$ to $z\sim1.5$ when fixing stellar mass, sSFR, or \Sigmabar, but 
these trends when holding galaxy properties fixed 
are also broadly consistent with no evolution given the uncertainties. 
Our observed values of \sigmavint at $z\sim1.4-3.3$ are in excellent agreement 
with the findings of DEEP2/SIGMA (\citealt{Kassin12}, \citealt{Simons17}) and 
are generally consistent with a range of studies at $z\sim0-3.5$ (e.g., \citealt{Newman13}; 
see \citealt{Wisnioski15}, \citealt{Turner17} and included literature references). 
Even though we do not definitively detect a redshift trend in \sigmavint, 
the combination of our data and these previous measurements 
agrees fairly well with the expected evolution 
for marginally stable disks (e.g., \citealt{Toomre64}, \citealt{ForsterSchreiber06}, \citealt{Genzel11}; 
see Fig.~8 of \citealt{Wisnioski15}, Fig.~5 of \citealt{Turner17}).

\subsection{Evolution of the Dynamical-Baryonic Mass Offset} 
\label{sec:dmfrac}

Next, by comparing the dynamical and baryonic masses for the galaxies in our sample, 
we investigate how the inferred baryonic and dark matter fractions vary with galaxy properties and 
evolve over time. 
Here we examine trends with stellar mass (\Mstar), 
effective radius (\re), and baryon mass surface density (\Sigmabar). 
These parameters enable us to examine how baryon and dark matter fractions 
vary over redshift as a function of galaxy mass and size (as a proxy for the extent into the halo). 
We will also consider whether there is a redshift-independent relation between baryon 
surface density and total baryon fraction (as suggested by \citealt{Wuyts16}).

The sample is split into bins of redshift with stellar mass, effective radius, and baryon mass surface density 
as shown in Figure~\ref{fig:samplebins}. 
The ensemble average \vtosigre for the galaxies without detected rotation and that are 
unresolved/misaligned 
are measured within these bins following the procedure described in Section~\ref{sec:dyn_masses}. 
We then combine all galaxies within each bin 
(all of the detected rotation, rotation-limit, and unresolved/misaligned galaxies) 
and calculate the median offset between dynamical and baryonic masses, 
$\deltlogM=\log_{10}(\Mdyn/\Mbar)$, and determine errors through bootstrap resampling.

These mass offsets are shown versus redshift in Figure~\ref{fig:DMfrac}, where each bin is colored by 
the median \Mstar, \re, and \Sigmabar (left to right, respectively). 
We find that the mass difference $\deltlogM$ decreases toward higher redshifts, 
where galaxies have higher baryon fractions ($\fbar = \Mbar/\Mdyn$). 
This overall trend of increasing mass offset with time holds even in fixed bins of 
stellar mass, effective radius, and surface density, 
suggesting that this general result is not solely 
driven by differences in these properties for our sample within different redshift ranges. 
The offsets within these bins are also given in Table~\ref{tab:massoffsets}.
These results agree well with previous work finding that massive galaxies are 
typically baryon dominated within the galaxy scale at high redshift 
(e.g., \citealt{ForsterSchreiber09}, \citealt{Wuyts16}, \citealt{Burkert16}, \citealt{Stott16}, 
\citealt{Alcorn16}, \citealt{Genzel17}, \citealt{Lang17}), 
though these studies did not probe the mass offsets 
as a joint function of redshift and other galaxy properties. 
Other studies of high-redshift galaxies find much higher dark matter fractions within much larger radii 
(e.g., $\gtrsim60\%$ within $6r_s$; \citealt{Tiley18}). 
While we do not have the S/N in the outskirts of our galaxies to constrain the dark matter 
fractions within such large radii (here we only constrain the values within $\re$), 
these large-radius results are not necessarily inconsistent with the smaller-radius results, as 
the dark matter fraction should naturally be higher when probing apertures farther out into the halo.

This observed baryon fraction trend agrees fairly well with theoretical work. 
Results from hydrodynamical simulations also show an average 
increase in the baryon fraction of star-forming or disk galaxies within their half-light (or half-mass) radii 
toward higher redshifts (e.g., \citealt{Lovell18}; $z\sim0-4$, \citealt{Teklu18}; $z\sim2$). 
However, theory and observations begin to have tension at $z\gtrsim2$, where the 
observed very high or unphysical \fbar are higher than the predicted baryon fractions from simulations 
(\citealt{Wuyts16}, \citealt{Lovell18}, \citealt{Teklu18}).

We find little to no variation in mass offset with \Mstar at fixed redshift (Figure~\ref{fig:DMfrac}a). 
At $z\sim1.5$, the median mass offset of all \Mstar bins corresponds to 
baryon fractions within the effective radius of $\fbar\sim\!65\%$, 
or dark matter fractions of $\fdm\sim\!35\%$ 
(where $\fdm=1-\fbar$). 
For all stellar mass bins at $z\gtrsim2$, the mass offsets are negative and fall within 
the unphysical region where $\Mdyn<\Mbar$. 
However, we note that the highest stellar mass bin ($\log_{10}(\Mstar/\Msun) > 10.25$) at $z\sim2.3$ 
is consistent with up to $\fdm\sim\!5\%$ (or $\fbar\sim\!95\%$) within the uncertainties. 
Our results at $z\sim1.5$ and $z\sim2.3$ for the highest stellar mass bin are in relatively good agreement 
with the results by \citet{Wuyts16} for similar redshifts and masses. 
The relative lack of variation of \fbar with stellar mass observed in this work is in contrast to predictions 
from theoretical work. Results from the Illustris TNG simulations at $z\sim2$ predict an average 
increase in the galaxy-scale (e.g., $r_{\mathrm{half}}$) baryon fraction for 
disk galaxies with increasing stellar mass \citep{Lovell18}. 
Furthermore, \citet{Lovell18} predict very little baryon or dark matter fraction evolution in 
galaxies with $\log_{10}(\Mstar/\Msun)=9$ from $z\sim4$ to the present day, in contrast 
to the results we find for our lowest mass bin ($\log_{10}(\Mstar/\Msun)\sim9.5$).

When controlling for galaxy size, we find larger galaxies have 
higher $\deltlogM$ and lower baryon fractions (Figure~\ref{fig:DMfrac}b). 
For galaxies at $z\sim1.5$, galaxies with $\re\geq2.5\unit{kpc}$ have 
$\fbar\sim55\%$ 
within their effective radii, while smaller galaxies with $\re<2.5\unit{kpc}$ have 
$\fbar>100\%$ and no inferred dark matter 
(but are consistent with up to a $\sim\!10\%$ 
dark matter fraction within the uncertainties). 
At $z\sim2.3$ and $z\sim3.3$, the baryon fractions for galaxies with $\re\geq2.5\unit{kpc}$ 
are unphysical ($\fbar\gtrsim100\%$) 
but are consistent with small dark matter fractions, 
while the smaller galaxies fall fully within the unphysical regime.

Finally, we examine the mass offset as a function of 
baryon surface density at fixed redshift (Figure~\ref{fig:DMfrac}c). 
We find that denser galaxies have lower \deltlogM and higher inferred baryon fractions. 
At $z\sim1.5$, the lowest- and medium-density galaxies 
($\langle\log_{10}(\Sigmabar/[\Msun\unit{kpc}^{-2}])\rangle\sim8.4,8.8$) 
have 
$\fbar\sim50\%, 75\%$ within \re, respectively, 
while the highest-density galaxies 
($\langle\log_{10}(\Sigmabar/[\Msun\unit{kpc}^{-2}])\rangle\sim9.2$) have 
$\fbar>100\%$, but are consistent with {$\fdm\sim15-25\%$ within the uncertainties. 
The mass offsets decrease to higher redshifts, 
with the lowest-density galaxies having increased $\fbar\sim90\%$
at $z\sim2.3$ and $\fbar>100\%$ (no dark matter) at $z\sim3.3$, 
and with he medium- and high-density galaxy bins all having $\fbar>100\%$ at $z\gtrsim2$.

While our results suggest that \fbar depends little on mass, 
the extent of a galaxy within its halo could explain variations of baryon fraction with 
size, density, and redshift (\citealt{Price16}, \citealt{Wuyts16}, \citealt{Ubler17}). 
Larger or less dense galaxies may extend farther into their halos, resulting in 
lower \fbar within their effective radii. The average size growth of galaxies over time 
would likewise be expected to lead to lower baryon densities at lower redshifts.

Our observed trends of increasing \fbar with increasing density (\Sigmabar), 
decreasing radius (\re), and increasing redshift support this postulate that 
galaxy extent within its halo impacts the observed \fbar within \re. 
This finding is in excellent agreement with the full-sample results of \citet{Wuyts16} ($z\sim0.6-2.6$) and 
generally agrees with the correlation between \fbar and \Sigmabar seen 
for galaxies from the Illustris simulation matched to the \KMOSTD sample \citep{Wuyts16}.  
However, our results also show that there is redshift evolution in the 
baryon fraction-density (\fbar-\Sigmabar) and -size (\fbar-\re) trends, 
suggesting that neither size nor density alone sets a galaxy's baryon fraction.

A key factor we have not included is how the dark matter halo mass profiles evolve 
and how galaxy and halo properties are related. 
Exact theoretical predictions of central halo profiles and their evolution are uncertain, given the 
potential impact of adiabatic halo contraction (or expansion) and the response of the halo to a baryonic disk 
(e.g., see discussion in \citealt{Duffy10}, \citealt{Dutton14}, \citealt{Velliscig14}, \citealt{Courteau15}, \citealt{Dutton16}). 
The particular details of how the halo and galaxy coevolve could strongly impact 
the interpretation of 
our derived central baryon/dark matter fraction results. 
For instance, if $\re/R_{\mathrm{halo}}$ changes with redshift, holding \re fixed would also result in \deltlogM 
probing halo evolution within this fixed physical radius (e.g., concentration). 
Likewise, evolution in $\Mstar/M_{\mathrm{halo}}$, as well as halo mass profile evolution, could 
also impact the redshift trends of \deltlogM at fixed \Mstar and \Sigmabar. 
Taken together, these results suggest 
that the evolution of galaxies' baryon fractions over time reflects 
the complex interplay of galaxy mass-size growth, global gas fraction, and halo growth and evolution, 
and thus it is not surprising that none of the individual examined properties are 
responsible for a universal, redshift-independent \fbar relation.

The negative offsets \deltlogM (i.e., \mbox{$\fbar\!>\!100\%$}) observed at $z\gtrsim2$, 
especially for smaller and denser galaxies, 
imply that at least some of the assumptions used to measure the dynamical 
and baryonic masses may be invalid for these galaxies. 
At minimum, we would expect the dynamical mass to account for all of the observed baryonic mass. 
This tension, as well as potential ways the masses might be reconciled, 
is further discussed in Section~\ref{sec:reconcile_masses}. 
Nonetheless, the observed redshift evolution of \fbar is 
likely not entirely driven by these mass measurement uncertainties. 
We would expect potential underestimates of dynamical masses or 
overestimates of gas masses to be the least problematic for  
larger, less dense galaxies (as they mostly do not have unphysical baryon fractions), 
and we do see time evolution of \fbar for these galaxies. 
In contrast, the exact details of the evolution of \fbar/\fdm for smaller, denser 
galaxies will require more detailed future studies.

\section{Discussion}
\label{sec:discussion}

\subsection{Reconciling Baryonic and Dynamical Masses\\at High Redshifts}
\label{sec:reconcile_masses_main}

While our results support a decreasing dark matter fraction within \re toward higher redshifts, 
there is tension between the dynamical and baryonic masses for 
a large fraction ($54\%$) of our sample, particularly in the higher redshift bins.  
On average, galaxies at $z\sim1.5$ and for the highest mass bin at $z\sim2$ are consistent with 
nonzero dark matter fractions. 
The $z\sim1.5$ median mass offsets leave room for a more bottom-heavy IMF (e.g., \citealt{Salpeter55}), 
but other studies suggest that a \citet{Chabrier03} IMF is more appropriate for star-forming galaxies 
(e.g., \citealt{Bell01}, \citealt{Tacconi08}, \citealt{Dutton11a}, \citealt{Brewer12}). 
However, even for a \citet{Chabrier03} IMF, the dynamical masses on average are lower than 
the inferred baryonic masses for all stellar masses at $z\gtrsim2$. 
Similarly, when splitting the sample by \re or \Sigmabar, small and dense galaxies generally 
have lower inferred dynamical than baryonic masses. 
The unphysical offsets where $\Mdyn < \Mbar$ suggest that some of the assumptions used 
to derive masses are invalid for these galaxies. 
We discuss and explore possible solutions to these tensions below.

\subsubsection{Impact of Higher Star Formation Efficiencies or Virial Coefficients}
\label{sec:reconcile_masses}

Two potential causes of the tension between the dynamical and baryonic masses (or possibly both) 
are systematic overestimates of the baryonic masses and systematic underestimates of the dynamical masses. 
Given that we do not directly measure molecular gas masses, nor do we have the 
detailed, high-S/N observations required to fully model the dynamical masses of our galaxies, 
we have adopted prescriptions to infer the molecular gas masses from measured SFRs and 
to convert measured velocities and dispersions into dynamical masses. Both of our adopted prescriptions 
could therefore contribute to our observed mass tension.

To start, if we overestimate the galaxies' gas masses, we would also overestimate their baryonic masses.
In this study, we have used the SFR-gas mass relation of \citet{Kennicutt98} for star-forming galaxies in the 
local universe to convert observed, dust-corrected SFRs into gas masses. 
However, this relation may not hold for galaxies at higher redshifts, 
particularly for higher surface density galaxies. 
Work on galaxies at $z\sim1-3$ finds Kennicutt-Schmidt relation slopes that vary 
from $N=1.28$ \citep{Genzel10} to 
$N=1.7$ \citep{Bouche07}, which bracket the local value of $N=1.4$ by \citet{Kennicutt98}. 
If the true slope for our galaxies is higher than the local relation (i.e., closer to $N=1.7$), our inferred 
gas masses would overestimate the true values, which would ease the tension between 
the dynamical and baryonic masses.


\begin{figure*}
\centering
\vglue -7pt 
\includegraphics[width=0.85\textwidth]{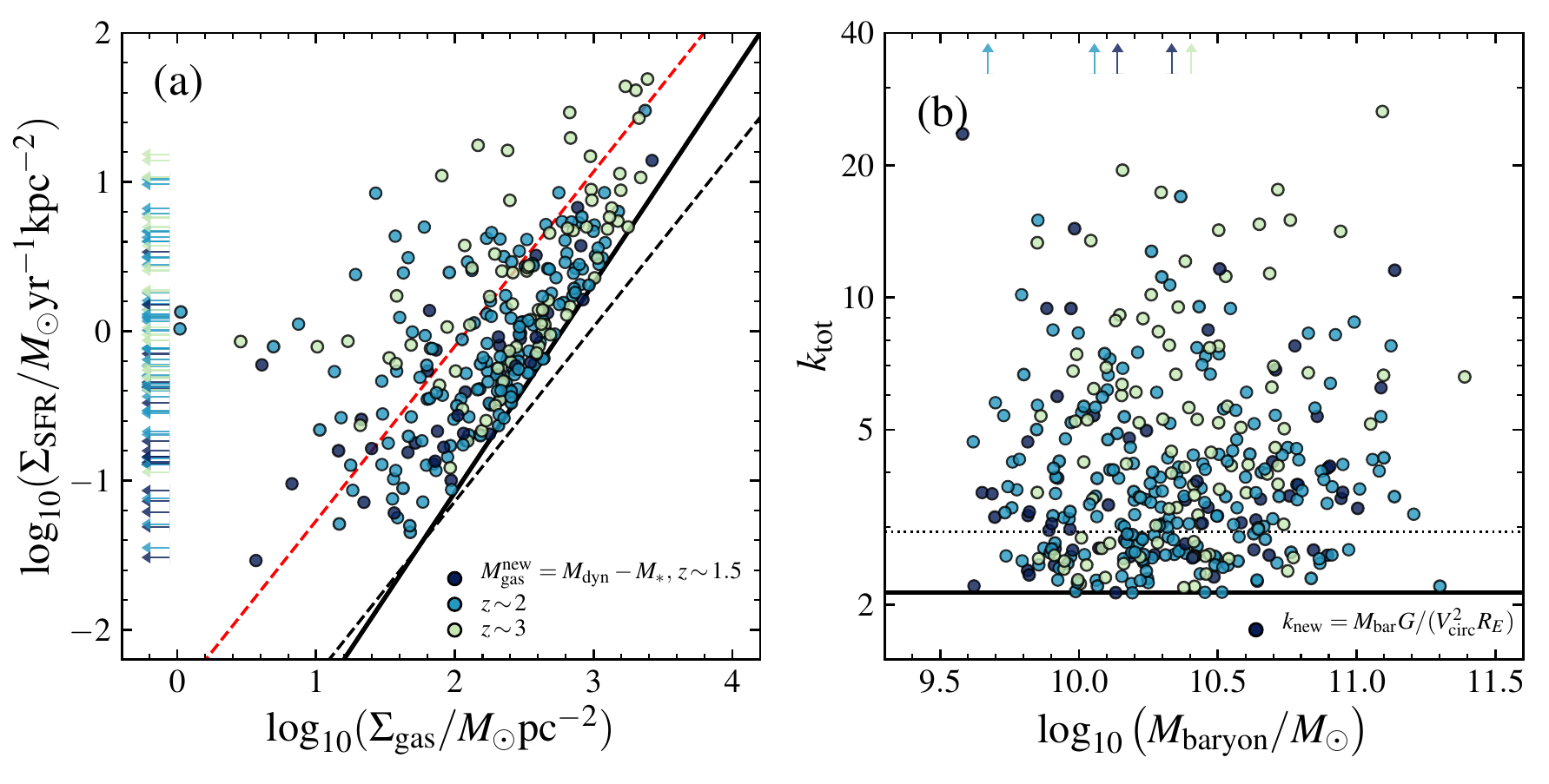}
\vglue -10pt
\caption{
Calculation changes required to reconcile unphysical ($\Mdyn\!<\!\Mbar$) mass offsets. 
We explore how (\textit{a}) the gas surface density and (\textit{b}) adopted virial coefficient (\ktot) 
would need to change so that baryonic and dynamical masses agree for galaxies falling in the 
unphysical regime. 
Here we use the average \vtosig for the unresolved/misaligned and rotation-limit galaxies 
measured in bins of $z$ and \Mstar  (see Figure~\ref{fig:samplebins}a). 
In the left panel, we show the Kennicutt-Schmidt (K-S) relation by \citet{Kennicutt98}, 
which we use to infer \Mgas for our sample (solid black line). 
Also shown are the K-S relations for ``normal'' star-forming and ULIRG/SMG-like galaxies 
(black and red dashed lines, respectively) by \citet{Genzel10}. 
For the galaxies that are unphysical under our default assumptions (i.e., $\Mdyn<\Mbar^{\mathrm{orig}}$), 
we use the limiting case of $\Mgas^{\mathrm{new}}=\Mdyn\!-\!\Mstar$ to determine the maximum 
consistent \Sigmagas.  
We show the shifted $\Sigmagas^{\mathrm{new}}$ for these objects as circles, colored by redshift 
(as in Figure~\ref{fig:sample}). 
Galaxies with very low $\Sigmagas^{\mathrm{new}}$ (or where $\Mstar\!>\!\Mdyn$) are marked with arrows 
at the far left of the panel. 
We find that decreasing \Sigmagas for the ``unphysical'' galaxies down to the ULIRG/SMG regime 
can improve the total fraction of galaxies with consistent masses 
($\Mdyn \geq \Mbar$) from $46\%$ to $76\%$. 
The remainder of the objects 
would require even more extreme SFEs than the ULIRG/SMG relation. 
In the right panel, the fiducial virial coefficient, $\ktot=2.128$, is shown with a thick black line. 
Also shown is the coefficient for a spherical exponential distribution ($\ktot=2.933$; dotted black line). 
We use the limiting case of $k_{\mathrm{new}}=\Mbar G/(\Vcirc^2\re)$ to calculate the minimum \ktot required 
to bring the masses of the ``unphysical'' galaxies into agreement. 
The point definitions are the same as the left panel, and objects with very high $k_{\mathrm{new}}$ 
are marked with arrows at the top of the panel. 
Increasing \ktot up to the value for a spherical mass distribution 
can increase the consistent fraction to $62\%$.
However, some objects would require very high, unrealistic virial coefficients ($\ktot\gtrsim10$) 
to make \Mdyn consistent with \Mbar. 
Using either (or both) of these calculation modifications can thus help reduce the 
baryon-dynamical mass tension, but further work is needed to fully 
constrain the sources of measurement uncertainties and biases. 
(Note that for both panels, only the shifted values for the initially unphysical objects are shown. 
The fiducial values lie on the \citealt{Kennicutt98} and $\ktot=2.128$ lines.)
}
\vglue 2pt
\label{fig:nonhomology}
\end{figure*}

In Figure~\ref{fig:nonhomology}a, we explore the possibility that the some gas masses may be overestimated 
by examining what gas surface densities (i.e., modified star formation efficiencies [SFEs]) 
would be needed to have consistent masses (i.e., $\Mdyn\geq\Mbar$). 
We show the relation of \citet{Kennicutt98} with a thick black line, which we 
use to infer gas masses (through \Sigmagas) from the SFR surface densities of our sample.
We then calculate the maximum gas surface density $\Sigmagas^{\mathrm{new}}$ 
the ``unphysical'' ($\Mdyn<\Mbar$) galaxies could have and still have physically consistent masses, 
using $\Mgas^{\mathrm{new}}=\Mdyn-\Mstar$ (i.e., no dark matter). 
With the \citet{Kennicutt98}-derived gas masses, 54\% of our sample has unphysical masses, 
with $\Mdyn<\Mbar$. 
When calculating the maximum physically allowable gas masses for these unphysical baryon fraction galaxies 
(i.e., higher SFE and decreased gas surface densities), 
we find that 76\% of our sample (up from 46\% for our fiducial calculation)
falls between the ``normal'' star-forming galaxy and ULIRG/SMG galaxy 
star-formation surface density to gas surface density regimes by \citet{Genzel10} (marked with 
black and red dashed lines, respectively). 
Such an increase in the SFE for these galaxies may be reasonable, 
given observations showing a decrease in gas depletion time (i.e., increased SFE) 
toward higher redshifts (e.g., \citealt{Tacconi18}). 
Even more extreme SFEs than the ULIRG/SMG relation by \citet{Genzel10} would be needed to 
bring the masses into agreement for the remaining galaxies with $\Mdyn\geq\Mstar$ (10\% of total sample). 
SFE changes cannot be invoked for the galaxies with $\Mstar>\Mdyn$ (14\% of total sample), 
though given the uncertainties, we may expect some scatter even into this region. 
Overall, in agreement with a similar test by \citet{Wuyts16}, this check suggests that 
moderately increased SFEs relative to the \citet{Kennicutt98} Schmidt relation could explain the mass tension 
for the majority of our sample, but further work and observations are 
needed to fully quantify the Schmidt relation in high-redshift galaxies.

Alternatively (or additionally), our measurements and prescriptions may underestimate the 
true dynamical masses of our sample.
First, if the ionized gas kinematics do not fully trace the total system potential, 
then the dynamical masses inferred from the kinematics will underestimate the true total masses. 
Second, we have assumed an underlying oblate intrinsic mass profile and a constant mass-to-light ratio in order to 
infer total dynamical masses from the measured kinematics at \re, through the virial coefficient \ktot. 
If a galaxy's true mass profile is more spherical than the assumed oblate mass profile, 
using the adopted \ktot can underestimate the total galaxy mass.

Similar to the test with gas masses, we explore what virial coefficient \ktot 
would need to be adopted to reconcile the masses in Figure~\ref{fig:nonhomology}b. 
We show the fiducial $\ktot=2.128$ as a thick black horizontal line. 
For objects with $\Mdyn<\Mbar$, 
we calculate the minimum modified $\ktot$ that would be necessary for physical masses, 
using the limiting case of $\Mdyn=\Mbar$, or taking $k_{\mathrm{new}}=\Mbar G/(\Vcirc^2\re)$. 
After determining $k_{\mathrm{new}}$ for the unphysical baryon fraction galaxies, 
62\% of the galaxies (up from 46\%) would have virial coefficients falling between the fiducial $\ktot$ 
for a $q=0.4$ oblate exponential profile and the value for a spherical exponential potential 
($\ktot=2.933$, dotted black horizontal line). 
The remainder would need even 
higher virial coefficients (corresponding to prolate potentials), up to very extreme, unrealistic values ($\ktot\gtrsim10$). 
\citet{vanderWel14} have argued that a large fraction of massive galaxies at $z\gtrsim2$ 
have spheroidal/disk geometries, 
while the majority of low-mass galaxies at $z\gtrsim2$ are elongated and prolate, 
using an analysis of the axis ratio distribution of galaxies out to $z\sim2.5$, 
which might suggest virial coefficients higher than the oblate value we have used in this work.  
We do note that very few galaxies in our kinematics sample at $z\sim3$ have 
axis ratios $b/a>0.8$, as expected for prolate geometries, but this may only reflect the 
relatively small sample size at $z\sim3$. Nonetheless, much more detailed kinematic observations of 
small galaxies at $z\sim2-3$ would be necessary to fully constrain their internal geometries. 

To further quantify the impact of a higher SFE or virial coefficient, 
we also determine how the median dynamical-to-baryonic mass ratio, \deltlogM, 
changes if we adopted the ULIRG/SMG relation from \citet{Genzel10} or the spherical \ktot for our full sample. 
We find that these modifications would increase \deltlogM by 
$\sim\!0.27$ and  $\sim\!0.14\unit{dex}$, respectively. 
Figure~\ref{fig:caveats} shows these offsets, along with other systematic uncertainties 
discussed throughout Section~\ref{sec:discussion}.


\subsubsection{Alternative Gas Mass Estimates: Scaling Relations}
\label{sec:scaling}

We next examine how our mass fraction results (e.g., \fbar) would change --- 
and whether any of the dynamical-baryonic mass tension could be alleviated --- 
if we measure gas masses for our full sample following high-redshift gas mass scaling relations instead 
of using the \citet{Kennicutt98} Schmidt relation. 
For this test, we estimate gas masses using the best-fit molecular gas scaling relation of 
\citet{Tacconi18}, using the \citet{Whitaker14a} main-sequence prescription. 
We also adopt the ``best'' SFRs (and matching sSFRs) 
derived following the ladder technique of \citet{Wuyts11a} in place of 
our primarily \Halpha- and \Hbeta-based SFRs, 
as these SFRs match those used to derive the scaling relations. 
We then remeasure \vtosigre for the galaxies without detected rotation 
using the scaling-relation baryonic masses (following Section~\ref{sec:dyn_masses_unres}), 
and redetermine the dynamical-baryonic mass offset as in Section~\ref{sec:dmfrac}.

Our primary results are unchanged if scaling-relation gas masses are used. 
On average the scaling-relation gas masses are slightly higher $\sim0.13\unit{dex}$) than the 
Schmidt-relation gas masses, resulting in even higher average \fbar 
(the median \deltlogM decreases by $\sim\!-0.07\unit{dex}$; see Fig.~\ref{fig:caveats}). 
Nonetheless, the general trend of increasing \fbar toward higher redshifts 
at fixed mass, size, or surface density is unchanged, and the dynamical-baryonic mass tension at 
$z\gtrsim2$ is still present when adopting these alternative gas masses. 
When using the scaling-relation gas masses, 
there is a suggestive trend 
in our lowest redshift bin ($z\sim1.5$) where higher stellar mass galaxies have slightly higher \fbar. 
However, the current data uncertainties hamper a firm conclusion regarding any definitive differences between this 
finding and our fiducial results where \fbar does not depend on \Mstar at a given redshift. 
Overall, we conclude that our results are not greatly impacted by the choice of 
adopting gas masses estimated by inverting the \citet{Kennicutt98} relation 
instead of masses derived using scaling relations. 


\begin{figure}
\centering
\vglue -4.5pt
\hglue -4pt 
\includegraphics[width=0.488\textwidth]{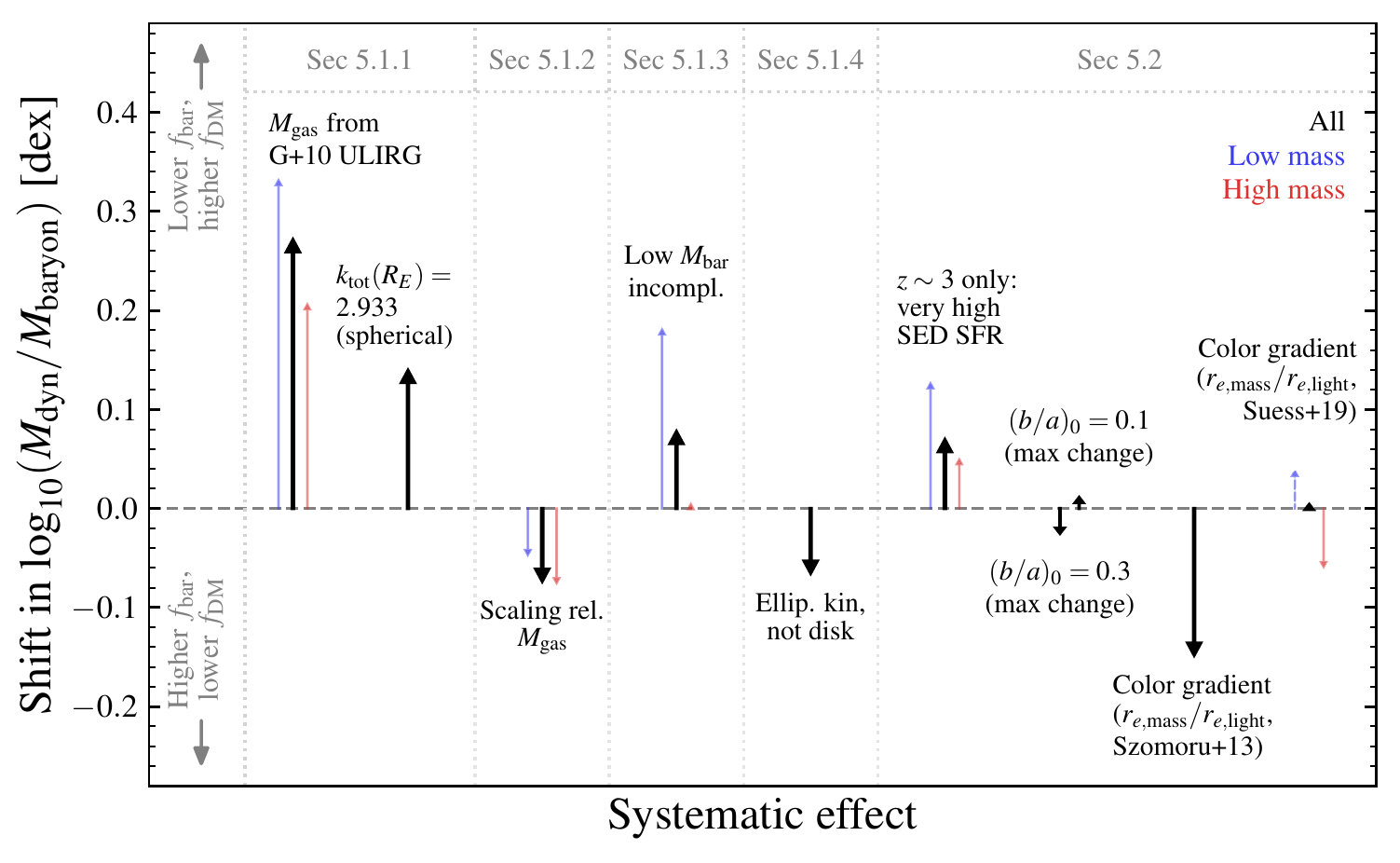}
\vglue -5pt
\caption{
Overview of the impact of systematic uncertainties, as well as 
alternative analysis assumptions (as discussed in Sec.~\ref{sec:discussion}), 
on the difference between dynamical and baryonic masses (e.g., shown in Fig.~\ref{fig:DMfrac}). 
The median impact for the full sample (or a subset) and for low- and high-mass (as in Fig.~\ref{fig:samplebins}) 
galaxies in each case are shown with black, blue, and red arrows, respectively.
}
\label{fig:caveats}
\end{figure}

\subsubsection{Impact of Incompleteness at Low Baryonic Masses}
\label{sec:incompleteness}

Another concern regarding our dynamical-baryonic mass comparison 
comes from the incompleteness of our sample at the lowest baryonic masses 
($\log_{10}(\Mbar/\Msun)\lesssim9.5-9.8$; see Figure~\ref{fig:dyn_mass}). 
This incompleteness can lead to underestimates of \deltlogM, 
particularly at the lowest masses, as objects with lower \Mbar at fixed \Mdyn near 
the completeness limit are missing from this analysis. 
This underestimate of \deltlogM is equivalent to an overestimate of \fbar, 
especially at lower stellar masses and higher redshifts, as the median stellar mass 
of our sample decreases toward higher redshifts. 
We examine the impact of this incompleteness by repeating the analysis of 
Section~\ref{sec:dmfrac}, but including only galaxies with $\log_{10}(\Mdyn/\Msun)\geq9.8$.
We do not refit the unresolved/misaligned/rotation nondetection galaxies \vtosigre, 
but simply apply the cut to the fiducial \Mdyn.

We find increases of 
\hbox{$\sim\!0.17$},  
\hbox{$\sim\!0.19$}, and
\hbox{$\sim\!0.15\unit{dex}$}
in \deltlogM for the lowest stellar mass bin ($\log_{10}(\Mstar/\Msun)<9.75$)
at $z\sim1.5$, $z\sim2$, and $z\sim3$, 
respectively,  when applying this dynamical mass cut.
Altogether, the full-sample median mass offset increases by $\sim\!0.08\unit{dex}$
(see Fig.~\ref{fig:caveats}). However, the greater impact of incompleteness at 
lower masses could be driving part of the tension with theoretical results. 
Based on the shifts from this dynamical mass cut, accounting for incompleteness could result in lower baryon fractions 
for the lower-mass galaxies relative to the higher-mass galaxies, as found by \citet{Lovell18}.
Nonetheless, our primary results remain unchanged: \fbar increases toward higher redshifts, and most galaxies 
still have $\Mdyn<\Mbar$ at $z\gtrsim2$. We conclude that, on the whole, 
incompleteness does not solely drive our results 
but could be partially responsible for the lack of offset in \fbar between stellar mass bins.

\subsubsection{Unresolved Kinematics: Early-type Galaxies?}
\label{sec:dyn_masses_early}

For this analysis, we have modeled the galaxies that are 
unresolved, are misaligned, or have no detected rotation as inclined thick disks, 
but this may not hold for all of these objects. 
Instead, it is possible that the dynamics of at least some of these systems are 
better described by early-type, random-motion-dominated structures. 
While our models and adopted dynamical mass calculation 
do allow for no ensemble rotation ($\vtosigre=0$), resulting in inclination-independent values 
and an ``effective'' virial coefficient of $\ktot=7.13$, 
there may still be differences in our fiducial \Mdyn and masses 
determined following prescriptions for early-type galaxies. 
As we cannot directly distinguish between inclined thick disks and early-type dynamics in 
our integrated 1D velocity dispersion measurements, we thus also calculate the dynamical masses of all the unresolved/misaligned and rotation-limit 
galaxies under the assumption that they are early-type galaxies (following \citealt{Cappellari06}; 
see Section 5.2 of \citealt{Price16}), to examine how the assumed 
kinematic structure of these galaxies impacts our results.

To determine early-type dynamical masses, we 
calculate aperture-corrected integrated dispersions $\sigma_{e,\mathrm{corr}}$ from the 
observed, instrument-corrected 1D velocity dispersions $\sigmavobs$. 
We then use circularized radii to calculate both the dynamical and gas masses for consistency. 
As we found in our previous study, the dynamical and baryonic masses for the full sample 
are still in reasonable agreement ($\deltlogM=0.06,-0.18,-0.23$ at $z\sim1.5,2,3$, respectively; 
the full-sample median offset is $\sim\!-0.07\unit{dex}$, see Fig.~\ref{fig:caveats}), 
and there is no strong residual trend of $\deltlogM$ with $b/a$. 
However, as we found before, under the early-type assumption, 
the unresolved/misaligned/{\allowbreak}
\makebox[0.47\textwidth][s]{rotation-undetected galaxies have systematically lower}\\
$\deltlogM$ than the rotation-detected galaxies, with even higher inferred baryon fractions. 
This increased dynamical-baryonic mass tension 
suggests that, on average, these galaxies have at least a modest amount of rotational support, 
supporting the adoption of the best-fit ensemble average \vtosigre and the corresponding  
\Mdyn as derived in Section~\ref{sec:dyn_masses_unres} for this study. 
However, systematic uncertainties (see Section~\ref{sec:caveats}) make it difficult to 
definitively conclude whether or not these galaxies have rotational support.

\subsection{Other Analysis Caveats and the Path Forward}
\label{sec:caveats}

In addition to the caveats discussed in the previous section, 
other assumptions made in this work could impact our results. 
First, many of the caveats discussed in \citet{Price16} also apply to this analysis. 
Specifically, the accuracy of the \textsc{Galfit} structural parameters is not fully accounted for in 
our kinematic measurements. These structural parameters are integral to both the 1D and 2D kinematic modeling, 
but no structural parameter errors are included in the kinematic fitting. 
Our analysis also will have over- and underestimates in the measured kinematics 
if the photometric and kinematic major axes are misaligned, resulting in different position angle misalignments 
$\Delta\mathrm{PA}$ for the slit and the photometric and kinematic major axes 
and introducing extra scatter into our measurements. 
Similarly, the uncertainty in the measured galaxy axis ratios impacts the inclination correction, 
particularly for smaller galaxies (e.g., $z\sim3$ galaxies). 
If the observed ratios are larger or smaller than the true values, then this will lead to over- and underestimates 
in the kinematics and dynamical masses, respectively, which will also increase the scatter of our measurements. 
Additionally, we assume a fixed intrinsic disk thickness of $(b/a)_0=0.19$ to derive galaxy inclinations. 
Variations in the true intrinsic thicknesses (e.g., thicker or thinner) will lead to over- and underestimates (respectively) 
of the inclination correction for our galaxies, which could increase the scatter 
and may also introduce systematic offsets in the derived dynamical properties.
For thicker $(b/a)_0=0.3$ or thinner $(b/a)_0=0.1$, the derived dynamical masses 
could be up to $\sim-0.025\unit{dex}$ lower and $\sim0.011\unit{dex}$ higher than the fiducial 
dynamical masses, respectively (see Fig.~\ref{fig:caveats}).

Furthermore, we have assumed fixed forms for the kinematic profiles of our galaxies. 
We adopt an arctan rotation curve model (e.g., following \citealt{Weiner06a}, \citealt{Miller11}). 
Some work suggests that particularly massive high-redshift star-forming galaxies have 
falling rotation curves, such as a \citet{Freeman70} exponential disk model 
(e.g., \citealt{Wisnioski15}, \citealt{vanDokkum15}, \citealt{Genzel17}, \citealt{Lang17}). 
However, for our objects, we only reliably probe the kinematics out to $\sim2.2r_s$. 
Arctan model profiles are similar to the other rotation curve models over this range, 
so our choice of rotation profile should not impact our results very strongly. 
We also assume a constant intrinsic velocity dispersion profile. 
If the true velocity dispersion instead decreases with increasing radius, 
this would result in higher median \vtosigre measured for the galaxies without resolved rotation 
(also see discussion in Section~5.7 of \citealt{Price16}). 
Higher spatial resolution and higher-S/N observations are needed 
to properly constrain the form of the velocity dispersion profiles.

Second, beyond the assumption of a SFR-gas mass relation (e.g., from \citealt{Kennicutt98}, 
as discussed in Section~\ref{sec:reconcile_masses}), 
we also use a mix of SFR indicators, as Balmer-decrement-corrected \Halpha SFRs 
are not available for the entire sample. 
Mixing SFR indicators could lead to systematic differences in the inferred gas masses. 
However, in Appendix~\ref{sec:appendixB}, we compare SFR indicators for galaxies at $z\sim2$ with 
Balmer-decrement-corrected \Halpha SFRs and find no bias between the \Halpha/\Hbeta+\Av indicators 
and \Halpha SFRs, though the SED SFRs underestimate the \Halpha SFRs (as also found by 
\citealt{Reddy15} for galaxies with high SFRs). This offset would lead to even larger inferred 
gas masses, increasing the tension between the dynamical and baryonic mass measurements. 
If we use the median offset between the other SFR indicators and the \Halpha SFRs 
(as measured in Appendix~\ref{sec:appendixB}) to correct all non-Balmer-decrement \Halpha SFRs, 
we find the same median dynamical-to-baryonic mass ratio for the full sample (or at low or high masses), 
because the SED SFRs are only used for $\sim\!11\%$ of the sample.

However, some $z\sim3$ galaxies with very high sSFRs may have the opposite problem, 
where the SED and \Hbeta+\Av SFR indicators may overestimate the intrinsic \Halpha SFRs 
(as discussed in Appendix~\ref{sec:appendixB}). 
As directly measured Balmer-decrement \Halpha SFRs for $z\sim3$ galaxies are currently unavailable, 
we instead test how \deltlogM changes at $z\sim3$ if we only include $z\sim3$ galaxies with 
$\log_{10}(\mathrm{sSFR}/\unit{yr}^{-1}) \leq -7.8$. 
When performing this cut, we find a $\sim0.07\unit{dex}$ increase in the mass offset at $z\sim3$ 
(see Fig.~\ref{fig:caveats}), 
which is insufficient to move the median mass offsets at $z\sim3$ into the physical regime.

Third, we have assumed that the galaxies have constant mass-to-light profiles (i.e., equal half-light and half-mass radii). 
If half-mass radii are smaller than rest-frame optical \re, as found by 
\citet{Szomoru13}, we would overestimate the true dynamical masses, which would result in 
increased tension between the dynamical and baryonic masses. 
Specifically, for the average ratio $r_{1/2,\mathrm{mass}}/\re\sim0.75$ found by \citet{Szomoru13}, 
dynamical masses measured using half-mass radii would be lower than the 
fiducial \Mdyn by $\sim\!-0.15\unit{dex}$ (see Fig.~\ref{fig:caveats}). 
Recently, \citet{Suess19} found that the median ratio $R_{E,\mathrm{mass}}/R_{E,\mathrm{light}}$ 
for star-forming galaxies decreases from $\sim1$ at $z\gtrsim2$ down to $\sim0.75-0.8$ at $z\sim1.5$, 
so this reduction in \Mdyn would likely impact our lowest redshift bin more than our higher redshift bins. 
\citeauthor{Suess19} also find that $R_{E,\mathrm{mass}}/R_{E,\mathrm{light}}$ decreases with stellar mass, 
which would imply that at high \Mstar we would find 
lower \deltlogM (higher \fbar), with less offset for lower masses. 
To quantify the impact variable mass-to-light profiles may have on our results, 
we use the redshift- and mass-dependent ratios between $R_{E,\mathrm{mass}}/R_{E,\mathrm{light}}$ 
from \citet{Suess19} to calculate how \Mdyn changes. We assume $r_t = 0.4 r_s = 0.4\re/1.676$ 
and use either the measured $V(\re)$ or inferred $V(\re)$ (from \vtosigre fit in bins of redshift and stellar mass), 
and we also apply the extrapolated relations to stellar masses below the \citeauthor{Suess19} sample completeness limits. 
We find essentially no change 
in the full-sample median dynamical-to-baryonic mass ratio (formally a $\sim\!0.004\unit{dex}$ increase; 
see Fig.~\ref{fig:caveats}), though for the mass and redshift bins where \citeauthor{Suess19} find ratios of 
 $R_{E,\mathrm{mass}}/R_{E,\mathrm{light}}\sim0.75-0.8$, the offset within those bins will be close to the 
 $\sim\!-0.15\unit{dex}$ shift for the \citeauthor{Szomoru13} case.

This problem of smaller half-mass than half-light radii would be further compounded if emission-line 
half-light radii were adopted, since the emission line profiles 
generally extend farther out than the stellar light profiles \citep{Nelson16}. 
In this case, the half-emission-line radii would be even larger than the half-mass radii. 
Since our dynamical mass derivation requires kinematics measured at the half-mass radius, 
if kinematics and sizes were instead measured at the on-average larger $R_{E,\Halpha}$, 
we would overestimate the dynamical masses of our sample.

However, in \citet{Price16} we found that the seeing-matched \Halpha and 
stellar light profiles for the resolved rotation galaxies were similar, 
which suggests that using stellar light profiles to model and measure galaxy kinematics 
is a reasonable assumption for most of our sample. 
In contrast, if the half-mass radii are larger than \re for some galaxies, 
the presented values would be underestimates of the total dynamical masses. 
Mock observations of multiple lines of sight to the same galaxy show that the specific viewing 
direction to a galaxy can strongly impact the observed half-light radius, 
leading to additional variations between the half-light and half-mass radii (including 
cases with half-light radii smaller than the half-mass radius; \citealt{Price17}). 
This potential line-of-sight size variation would introduce additional scatter into the dynamical mass measurements.

Finally, as mentioned in Section~\ref{sec:reconcile_masses}, 
if the ionized gas kinematics do not fully trace the potential of these galaxies 
(at least out to the half-mass radius), the kinematics and dynamical masses will underestimate the true values. 
This concern would particularly impact lower-S/N objects and small galaxies close to the seeing limit, 
where we determine the kinematics from integrated 1D dispersion measurements.

Despite these analysis caveats, the following conclusions are fairly robust.
First, while the precise baryon and dark matter fractions are somewhat uncertain, there is strong evidence that 
massive star-forming galaxies are highly baryon dominated at high redshifts. 
Furthermore, small and compact galaxies at high redshifts are more baryon dominated than more extended 
galaxies, suggesting that the extent of a galaxy within its halo plays a key role in setting the 
dark matter fraction within the half-light radius. 
Second, the trends of ordered to random motion (\vtosigtt) with stellar mass and sSFR (as a proxy for gas fraction) 
suggest that, on average, massive galaxies have marginally stable disks \citep{Toomre64} by $z\sim2$. 
The higher gas fractions in galaxies at earlier epochs would naturally lead to more turbulent, thick disks under the 
Toomre disk stability criterion. This average gas fraction evolution may be enough to explain the observed structural evolution, where disks over similar parameter space are thicker at earlier times.

To disentangle the potential biases in this kinematics analysis 
and fully understand the source of the dynamical-baryonic mass tension for the higher-redshift galaxies, 
it is crucial to both directly measure molecular gas masses and have more detailed kinematic constraints. 
We will begin to address these open areas by examining molecular gas masses from the 
PHIBSS survey \citep{Tacconi18} for the MOSDEF galaxies that fall within both samples. 
Additionally, we will compare  
MOSDEF masses and kinematics with measurements of the same galaxies 
from the \KMOSTD survey \citep{Wisnioski15}, to understand the 
limitations of the slit kinematic observations. 
Beyond these comparisons, future deep and high spatial resolution 
kinematic observations are also key to fully characterize the internal structures of high-redshift galaxies, 
particularly for small or low-mass galaxies, where the current kinematic constraints are most uncertain.

\section{Conclusions}
\label{sec:conclusions}

In this paper, we use spectra from the MOSDEF survey together with 
CANDELS \HST/F160W imaging to study 
the kinematics and dynamical masses of 681 galaxies at $1.34 \leq z \leq 3.8$, 
with stellar masses ranging from $\Mstar \sim 10^{9} \Msun$ to $\Mstar \sim 10^{11.5} \Msun$. 
In addition to kinematics and structural parameters, 
we use stellar masses derived from multiwavelength photometry 
and infer gas masses from either dust-corrected \Halpha or \Hbeta SFRs or 
SED SFRs if Balmer lines are unavailable. 

We use the 3D kinematic models (\MISFIT) developed in \citet{Price16} to 
measure the galaxy kinematics from the misaligned galaxy-slit MOSFIRE observations. 
We use these models to measure both rotation and velocity dispersions from the 2D spectra 
for the 105 galaxies that have robust rotation detections. 
For the remaining 576 galaxies, we measure galaxy-integrated 1D velocity dispersions and use the 
kinematic models to convert the observed velocity dispersions into combined kinematic 
rms velocities and dynamical masses, assuming a fixed ratio \vtosigre. 
These \vtosigre are derived as ensemble averages for the galaxies without detected rotation 
within bins of redshift, stellar mass, sSFR, \re, and \Sigmabar by removing any trend between 
$\log_{10}(\Mdyn/\Mbar)$ and axis ratio $b/a$ (i.e., inclination).

We explore the relation between the ratio of rotation to velocity dispersion, $\vtosigtt$, 
as a function of stellar mass, sSFR,  \Sigmabar, and redshift. 
We find that \vtosigtt increases with increasing stellar mass and decreases with increasing sSFR. 
These trends may indicate 
that these galaxies are marginally stable \citep{Toomre64},
where galaxies with lower gas fractions (e.g., lower sSFR or higher stellar mass) will 
naturally have less turbulent motions.  
We additionally examine the relation between \Vtt and \sigmavint and stellar mass, 
sSFR, \Sigmabar, and redshift for the galaxies with detected rotation. 
We find that \Vtt is most correlated with \Mstar, and that 
\sigmavint is most correlated with \Sigmabar. 
There are weak, less significant trends between \sigmavint-\Mstar and \Vtt-sSFR, 
and no correlation between \sigmavint and sSFR.  
At fixed stellar mass, we see suggestive trends of \Vtt increasing and \sigmavint decreasing 
over time, but these trends are also generally consistent with no evolution 
given the uncertainties.

Using the mass measurements, we find that the median offset   
between dynamical and baryonic masses, $\deltlogM=\log_{10}(\Mdyn/\Mbar)$, 
decreases with increasing redshift. The offset is relatively constant with stellar mass at fixed redshift. 
In contrast, we find that both larger galaxies and galaxies with lower surface densities 
tend to have higher mass offsets (i.e., lower \fbar). 
The observed mass offset evolution implies an evolving dark matter fraction, where 
galaxies at $z\gtrsim2$ are very strongly baryon dominated within their effective radii.
The evolution of \fbar does not appear to be controlled by a single 
galaxy property, but instead reflects the intertwined effects of 
galaxy mass-size growth, gas fraction, and halo growth and evolution.

However, we find tension between the dynamical and baryonic masses at $z\gtrsim2$, 
particularly for galaxies with small sizes or high densities. 
For these galaxies, the measured baryonic masses exceed the estimated dynamical masses 
(i.e., $\Mdyn < \Mbar$). 
This mass discrepancy could be explained for a number of these galaxies having an offset 
Schmidt relation with higher SFEs (i.e., a smaller \Sigmagas can sustain the same $\Sigma_{\mathrm{SFR}}$),  
or if the galaxies have a higher virial coefficient \ktot (i.e., more spherical mass distributions).  
Nonetheless, our conclusions  that galaxies become more baryon dominated toward higher redshifts 
and that massive galaxies generally have marginally stable disks by 
$z\sim2$ are fairly robust even in the face of these measurement tensions.

The approach of using multiplexing, seeing-limited NIR spectrographs to constrain the average 
properties of high-redshift galaxies allows us to study kinematics for one of the largest 
samples of star-forming galaxies at $z\sim1.5-3$, 
using a homogenous data set to extend down to 
much lower stellar masses than other surveys over a large range in redshift. 
However, further observations are necessary to reconcile the tension between 
the dynamical and baryonic masses at high redshifts and small sizes. 
In particular, direct observations of molecular gas masses are necessary to accurately 
measure baryonic masses. 
Detailed follow-up observations with adaptive-optics-assisted integral field unit (IFU) 
spectrographs are also crucial to better constrain the dynamical structures of these galaxies 
with unphysical baryon fractions.




\vglue 18pt
\acknowledgements

\vglue -18pt 
We acknowledge useful discussions with 
T.~T.~Shimizu, S.~Wuyts, E.~J.~Nelson, 
R.~Genzel,  L.~Tacconi, S.~Belli, K.~Suess, and R.~Trainor. 
We are grateful to the anonymous

\vfill\eject
\vfill\eject

\noindent
referee for their valuable comments. 
We thank the 3D-HST Collaboration, which provided the 
spectroscopic and photometric catalogs used 
to select the MOSDEF targets and derive stellar population parameters. 
This paper is based upon work supported by the National Science Foundation Graduate
Research Fellowship Program under Grant No. DGE 1106400 (S.H.P.). 
We acknowledge support from NSF AAG grants AST-1312780, 1312547, 1312764, and 1313171; 
archival grant AR-13907 provided by NASA through the Space Telescope Science Institute; 
and grant NNX16AF54G from the NASA ADAP program. 
The data presented in this paper were obtained at the W. M. Keck Observatory, which is operated 
as a scientific partnership among the California Institute of Technology, the University of California, 
and the National Aeronautics and Space Administration. The Observatory was made possible by 
the generous financial support of the W. M. Keck Foundation. 
We extend special thanks to those of Hawaiian ancestry on whose sacred mountain we are privileged to be guests. 
We are most fortunate to have the opportunity to conduct observations from this mountain. 
This work is also based on observations made with the NASA/ESA {\it Hubble Space Telescope}, 
which is operated by the Association of Universities for Research in Astronomy, Inc., under NASA 
contract NAS 5-26555. Observations associated with the following GO and GTO programs were 
used: 12063, 12440, 12442, 12443, 12444, 12445, 12060, 12061, 12062, 12064 (PI: Faber); 
12177 and 12328 (PI: van Dokkum); 12461 and 12099 (PI: Riess); 11600 (PI: Weiner); 
9425 and 9583 (PI: Giavalisco); 12190 (PI: Koekemoer); 11359 and 11360 (PI: O'Connell); 11563 (PI: Illingworth).

This research made use of Astropy,\footnote{http://www.astropy.org} a community-developed core Python package for Astronomy \citep{astropy:2013, astropy:2018}.

\vglue 5pt

\hglue -9.825cm
\rule{\textwidth}{1pt}

\vglue 5pt

\appendix

\begin{figure*}[t]
\centering
\vglue -4pt
\hglue -3pt 
\includegraphics[width=1.012\textwidth]{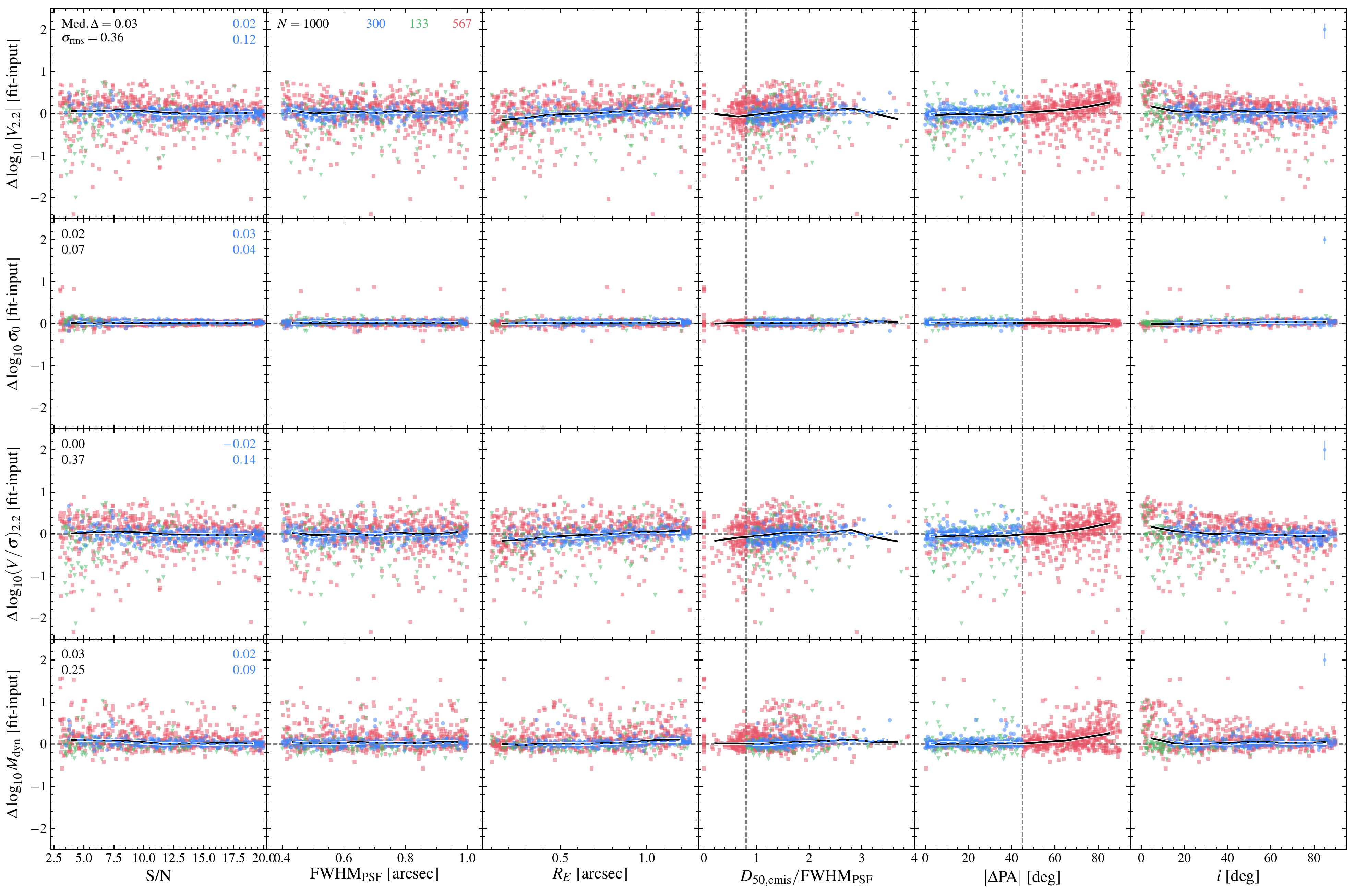}
\vglue -6pt
\caption{
Test of \MISFIT kinematic measurements for 2D spectra under a variety of observational conditions. 
We show the recovery of \Vtt, \sigmavint, \vtosigtt, and \Mdyn (\textit{top to bottom}) 
as a function of central S/N per pixel, PSF FWHM, galaxy effective radius (\re, arcsec), 
emission line spatial resolution ($D_{\mathrm{50,emis}}/\mathrm{FWHM_{PSF}}$),  
galaxy major axis-slit misalignment ($|\deltPA|$), and inclination (\textit{left to right}). 
Mock observations are marked as spatially resolved/aligned with detected rotation (blue circles) 
or with rotation limits (green triangles), or as unresolved/misaligned (red squares).
For reference, the resolved/aligned criteria are shown as vertical gray dashed lines in the 
$D_{\mathrm{50,emis}}/\mathrm{FWHM_{PSF}}$ and $|\deltPA|$ panels. 
The running median value offsets are shown in each panel 
for both the full mock sample and for only the resolved/aligned objects with detected rotation 
(black solid and blue dashed lines, respectively). The global median offset and rms scatter are also listed 
for the full and resolved/aligned, rotation-detected samples. 
Median uncertainties for the resolved/aligned, rotation-detected sample are shown in 
the upper right corner of the rightmost panels. 
Overall, the kinematic parameters are recovered fairly well on average, especially for 
mock objects that are spatially resolved/aligned and have detected rotation. 
}
\label{fig:MISFITtest}
\end{figure*}

\section{Kinematic Recovery Tests with \MISFIT}
\label{sec:appendixA}

Kinematics derived from slit spectra where the 
galaxy major axis and slit are misaligned suffer from a number of 
observational challenges. 
Any rotation signal will be blurred, as velocities from multiple radii are sampled within a given spatial slice, 
and some kinematic signal will be lost as portions of the galaxy fall outside of the slit. 
We account for the effects of galaxy-slit misalignments by fitting the kinematics using 
the forward models of \MISFIT, which we developed and first presented in \citet{Price16}
(see also Section~\ref{sec:kin_res}). 
However, the combined effects of galaxy size, slit misalignment, inclination, PSF FWHM, and S/N 
complicate the recovery of the intrinsic kinematics 
and thus may introduce scatter or bias in population-wide trends. 
In this appendix, we fit mock galaxy spectra with \MISFIT and 
examine how well kinematics are recovered under a variety of observational conditions, 
in order to quantify the scatter and bias in our kinematic measurements. 
Fixed kinematic and structural parameters (with the exception of \re) are selected for this test, 
to focus on the impact of the observational parameters 
(and to avoid prohibitively large mock data sets with long computation times).

We first generate a set of 1000 mock galaxies at $z=2$, 
each with $V_a=100\unit{km/s}$, $\sigmavint=50\unit{km/s}$, 
turnover radius $r_t=0.\arcsec2$, S\'ersic index $n=1.8$ 
(typical for emission line profiles at $z\sim0.7-1.5$; \citealt{Nelson16}), 
and intrinsic axis ratio $q_0=0.19$. 
We assume an instrument resolution of $\sigma_{\mathrm{inst}}=40\unit{km/s}$, which is 
typical for MOSFIRE in the \textit{K} band (observing \Halpha at $z\sim2$). 
The following parameters are then randomly drawn from uniform ranges: 
galaxy-slit misalignment $\deltPA \in [-90^{\circ}, 90^{\circ}]$, inclination $i\in[0^{\circ},90^{\circ}]$ (face-on to edge-on), 
effective radius $\re \in [0.\arcsec1, 1.\arcsec25]$, PSF FWHM $\in [0.\arcsec4,1.\arcsec0]$, 
and central S/N per pixel $\in [3,20]$. 
The kinematic parameters are chosen to be fairly typical for our sample. 
The free parameter intervals (S/N, $\mathrm{FWHM_{PSF}}$, \re, \deltPA, and $i$)  
are selected to cover the general range of values seen in the MOSDEF sample.

For each mock galaxy, we construct a model spectrum 
using the given fixed and randomly drawn structural and kinematic parameters, 
assuming that the PSF is Gaussian. 
Random noise is then added so that the central S/N per pixel equals the randomly selected S/N. 
Next, we use \MISFIT to analyze the kinematics for each mock realization, 
using the same procedure used in the main analysis (see Section~\ref{sec:kin_res}). 
From these results, we derive the best-fit \Vtt, \sigmavint, and \vtosigtt, and 
calculate \Mdyn (using $V(\re)$, \sigmavint, and \re) as in Section~\ref{sec:dyn_masses}. 
Each mock observation is then classified as spatially resolved/aligned with detected rotation 
or with rotation limits, or as unresolved/misaligned following the criteria given in Section~\ref{sec:kin_res}.

We show the offset between the input and recovered values of \Vtt, \sigmavint, \vtosigtt, and \Mdyn (top to bottom) 
versus central S/N per pixel, PSF FWHM, emission line resolution $D_{\mathrm{50,emis}}/\mathrm{FWHM_{PSF}}$, 
galaxy-slit misalignment $|\deltPA|$, and inclination $i$ (left to right) in Figure~\ref{fig:MISFITtest}. 
Overall, the parameters are recovered well on average, with little bias between 
the input and recovered parameters for both the full sample and 
for the subsample of resolved/aligned and rotation-detected mock observations. 
While the values are recovered well on average, there are relatively large scatters 
in the recovered \Vtt, \vtosigtt, and \Mdyn for the full mock set ($\sim0.25-0.4\unit{dex}$), 
though there is a lower scatter in the recovered \sigmavint ($0.07\unit{dex}$). 
The scatter in the recovered parameters is lower when considering only the subset of 
mock observations that meet the resolution/alignment and detection cuts ($\sim0.1-0.15\unit{dex}$ 
for \Vtt, \vtosigtt, and \Mdyn, and $0.04\unit{dex}$ for \sigmavint).

We see no systematic trends in the parameter recovery with central S/N per pixel or PSF FWHM. 
We see a slight trend of lower \Vtt and \vtosigtt when the emission line is barely resolved or unresolved 
(i.e., low $D_{\mathrm{50,emis}}/\mathrm{FWHM_{PSF}}$),  
so we thus restrict 2D kinematic fitting to galaxies with $D_{\mathrm{50,emis}}/\mathrm{FWHM_{PSF}}\geq0.8$ 
(similar to the recovery results of \citealt{Simons16}). 
There is a noticeable trend where \Vtt, \vtosigtt, and \Mdyn are overestimated for large slit-galaxy misalignments, 
justifying the alignment cut where we do not fit 2D kinematics when $|\deltPA|>45^{\circ}$. 
We note that we still see a slight trend of lower \Vtt and \vtosigtt for the objects in the resolved/aligned 
and rotation-detected subsample with the smallest \re (maximum of $\sim\!0.1\unit{dex}$), 
but that there are no trends in the recovered \Mdyn or \sigmavint.
Finally, we see slight overestimates of \Vtt, \vtosigtt, and \Mdyn for more face-on mock objects. 
However, this trend is not seen for the resolved/aligned and rotation-detected sample, as 
the kinematic quality cut appears to remove the objects most impacted by the face-on uncertainties.

Overall, this test demonstrates that while kinematics of individual objects may not be precisely recovered 
(resulting in scatter that may impact the recovery of trends when using individual measurements), 
the ensemble sample kinematic properties are well recovered with \MISFIT, and thus our 
results do not suffer greatly from measurement biases.

\begin{figure*}[t] 
\vglue -5pt
\hglue -4pt
   \includegraphics[width=1.02\textwidth]{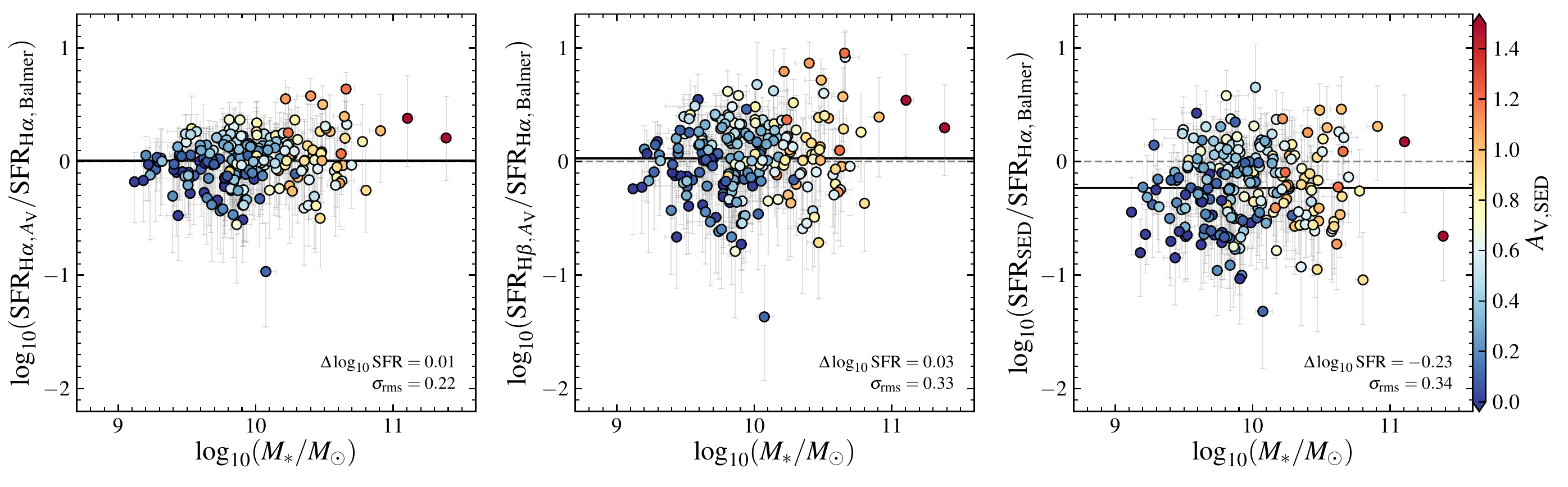}
\vglue -10pt
\caption{
Comparison of different SFR measurement methods for galaxies at $z\sim2$. 
The ratios of \Halpha, \Av dust-corrected (\textit{left}), 
\Hbeta, \Av dust-corrected (\textit{middle}), and 
SED (\textit{right}) SFRs 
to \Halpha, Balmer-decrement-corrected SFRs are shown versus stellar mass.
Galaxy colors indicate the SED-derived \Av values. 
The median offset for each comparison is shown with a solid black line, and no offset is marked with dashed gray lines. 
On average the \Halpha, \Av and \Hbeta, \Av SFRs agree well with the Balmer-decrement-corrected \Halpha SFRs, 
though we do observe a fair amount of scatter ($\sigma_{\mathrm{rms}}\!\sim0.2-0.3\unit{dex}$). 
The SED SFRs tend to be lower than the Balmer-decrement-corrected \Halpha SFRs 
($\Delta\log_{10}\mathrm{SFR}=-0.23\unit{dex}$, with $\sigma_{\mathrm{rms}}=0.34\unit{dex}$). 
This offset implies that, on average, the inferred \Mgas and \Mbar would be higher 
if Balmer-decrement-corrected \Halpha SFRs could be used instead of SED SFRs.
This change would amplify, not reduce, the tension between \Mbar and \Mdyn. 
}
\vglue -6pt
\label{fig:sfrmethodcompare}
\end{figure*}

\section{Comparison of Mixed SFR Indicators and Kinematics from Different Emission Lines}
\label{sec:appendixB}

Successfully comparing trends of structure, kinematics, and matter content between different cosmic epochs 
requires either using the same measurement tracers or characterizing any bias between the tracers. 
In this appendix, we investigate whether biases arise in our analysis from using a combination 
of different SFR indicators and kinematics from different emission lines.

In this analysis, we use a ``ladder'' of SFR indicators,  preferring Balmer-decrement-corrected 
\Halpha SFRs when available (320 galaxies), 
but otherwise using (in order of preference) \Av-corrected \Halpha SFRs (176 galaxies), 
\Av-corrected \Hbeta SFRs (107 galaxies), or SED SFRs (78 galaxies), 
with $\Av=A_{\mathrm{V,\,neb}}$ inferred from SED fit $A_{\mathrm{V,\,cont}}$ values. 
To test the accuracy of the lower-priority SFR indicators relative to the best-available Balmer-decrement-corrected 
\Halpha SFRs, we select galaxies at $z\sim2$ that have secure ($\mathrm{S/N}\geq3$) 
\Halpha and \Hbeta detections. We restrict this test to only galaxies at $z\sim2$ to provide the closest-available 
analogs to galaxies at $z\sim3$, where only \Av-corrected \Hbeta and SED SFRs are available. 
We then alternatively derive the \Av-corrected \Halpha and \Hbeta SFRs, as well as the SED SFRs, using 
the FAST SED fit results (see Section~\ref{sec:mosdef}). 
The offset between each of these lower-priority SFR indicators and the Balmer-decrement-corrected \Halpha SFRs 
is shown in Figure~\ref{fig:sfrmethodcompare}.

We find excellent agreement between the \Av-corrected \Halpha and \Hbeta SFRs 
and the Balmer-decrement-corrected \Halpha SFRs ($\Delta\log_{10}\mathrm{SFR}\sim0.01-0.03\unit{dex}$), 
though there is a fair amount of scatter between the indicators ($\sigma_{\rms}\sim0.2-0.3\unit{dex}$). 
However, we find that the SED SFRs are lower than the Balmer-decrement-corrected \Halpha SFRs 
by $-0.23\unit{dex}$, with a moderately large scatter of $0.34\unit{dex}$. 
This offset implies that the SED SFR-inferred \Mgas and \Mbar would underestimate the 
values derived from Balmer-decrement-corrected \Halpha SFRs. 
Thus, accounting for this offset would lead to higher \Mbar, increasing 
the tension between \Mbar and \Mdyn. 
Therefore, this bias cannot explain the observed tension between the dynamical and baryonic masses, but 
would rather amplify the problem. 
We do note that intrinsic differences between galaxies at $z\sim2$ and $z\sim3$ (e.g., metallicity, star-formation history shape), 
as well as differences in the available rest-frame photometry, 
may impact the accuracy of SED modeling, 
which could potentially lead to cases at $z\sim3$ where SED SFRs actually overestimate \Halpha SFRs 
(i.e., the very high sSFR objects at $z\sim3$; Figure~\ref{fig:sample}). 
If objects had higher SED than \Halpha SFRs, \Mbar and the baryon 
fraction would be overestimated when using SED SFRs, which could potentially explain some of the 
dynamical-baryon mass tension. 
Also, the scatter between these indicators could potentially affect 
the quantification of trends involving individual galaxies and not just median values.

\begin{figure*}[t] 
\vglue -5pt
\centering
   \includegraphics[width=0.7\textwidth]{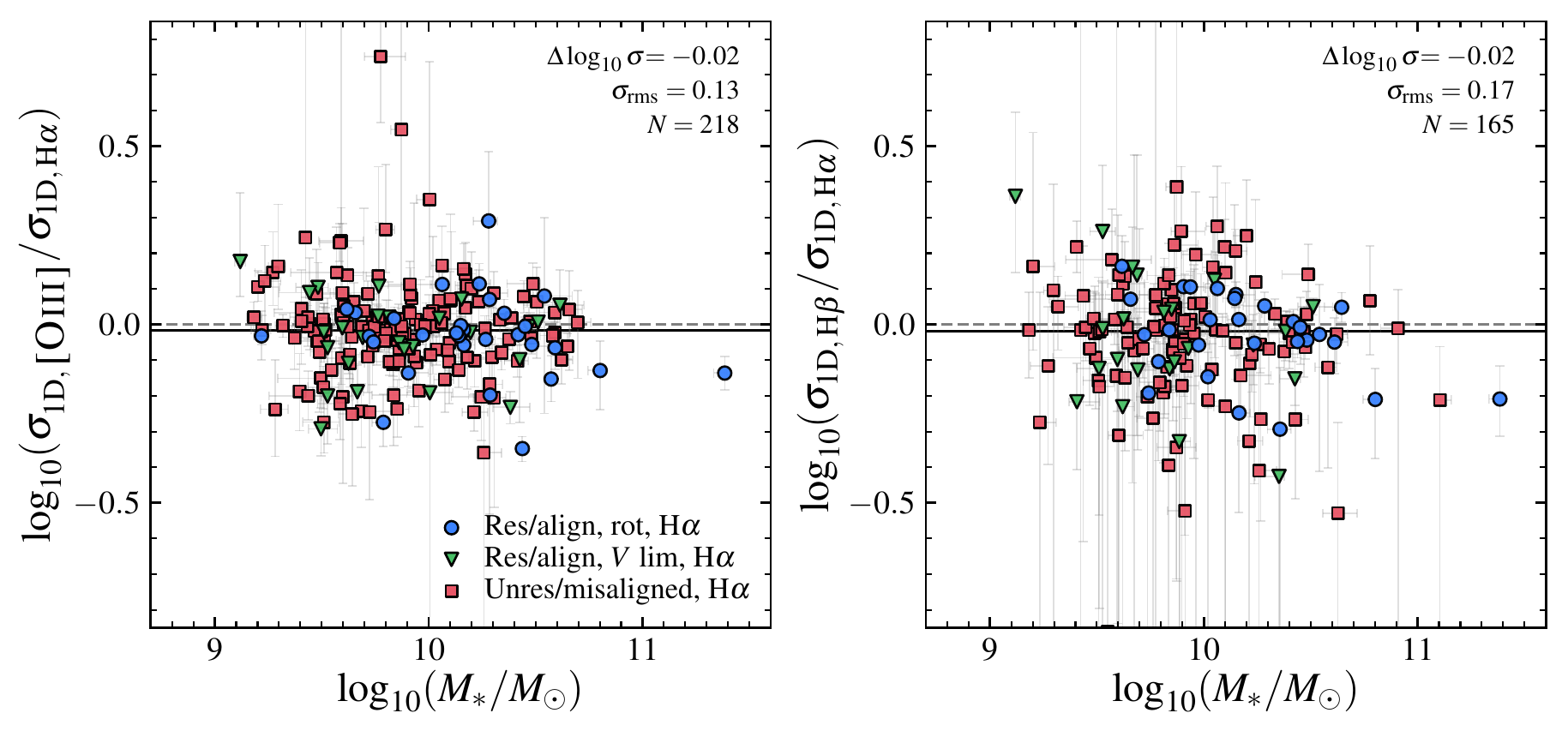}
\vglue -10pt
\caption{
Comparison between \sigmavobs measured from 
\OIII (\textit{left}) or \Hbeta (\textit{right}) and \Halpha for galaxies at $z\sim2$. 
The symbols show the \Halpha kinematic classification, 
as defined in Figures~\ref{fig:sampleressplit} and \ref{fig:samplebins}. 
Solid black lines mark the median offsets, and no offset is shown as a dashed gray line. 
The \OIII and \Hbeta 1D velocity dispersions are in excellent agreement with the \Halpha values, 
with relatively little scatter ($\sigma_{\mathrm{rms}}\!\sim0.1-0.2\unit{dex}$). 
}
\label{fig:kincomp1D}
\end{figure*}

Next, we turn to potential biases from kinematic measurements with different emission lines. 
In our analysis, we use a mix of 2D and 1D measurements from 
\Halpha (481 galaxies), \OIII (195 galaxies), and \Hbeta (5 galaxies). 
Thus, we select objects for which we have measurements of both \OIII and \Halpha or 
\Hbeta and \Halpha. As with the SFR method comparison, we select galaxies at $z\sim2$ 
to provide the best analogs to the $z\sim3$ subsample. 
For this test, we do not require all three lines to be detected for each object, 
as the spectral coverage and skyline contamination of our spectra 
make clean kinematic measurements of all three lines in a single object relatively rare.

We first compare the 1D kinematics from different lines, as shown in Figure~\ref{fig:kincomp1D}. 
The \sigmavobs values from \Halpha, \OIII, and \Hbeta agree very well on average, 
with offsets of only $-0.02\unit{dex}$ between \OIII or \Hbeta and \Halpha. 
We do find some scatter ($\sigma_{\mathrm{rms}}\!\sim0.1\unit{dex}$ and $\sim0.2\unit{dex}$ 
for \OIII vs. \Halpha and \Hbeta vs. \Halpha, respectively), 
but overall mixing 1D dispersions measured from different emission 
lines likely does not introduce any bias into our analysis.

Finally, we compare the 2D kinematics measured from \Halpha, \OIII, and \Hbeta. 
In Figure~\ref{fig:kincomp2D}, we compare the values of 
\Vtt, \sigmavint, and \Mdyn measured from 2D fitting of \OIII or \Hbeta with those from \Halpha. 
The values measured from \Halpha and the other lines are generally in good agreement, with 
at most a $\sim0.13\unit{dex}$ offset. 
We do find a fair amount of scatter in the recovered values, 
but this is generally smaller than the uncertainty in the offsets.  
Nonetheless, as with the kinematic recovery and SFR indicator tests, the scatter between 
the different emission lines' 2D kinematics could possibly influence the investigated galaxy property trends. 
Although the 2D comparison sample is small, this test 
suggests that the results derived from different emission lines are compatible, 
so that mixing of kinematic tracers does not systematically impact our results.

\begin{figure*}[t] 
\vglue -5pt
\centering
\hglue -2pt 
   \includegraphics[width=1.01\textwidth]{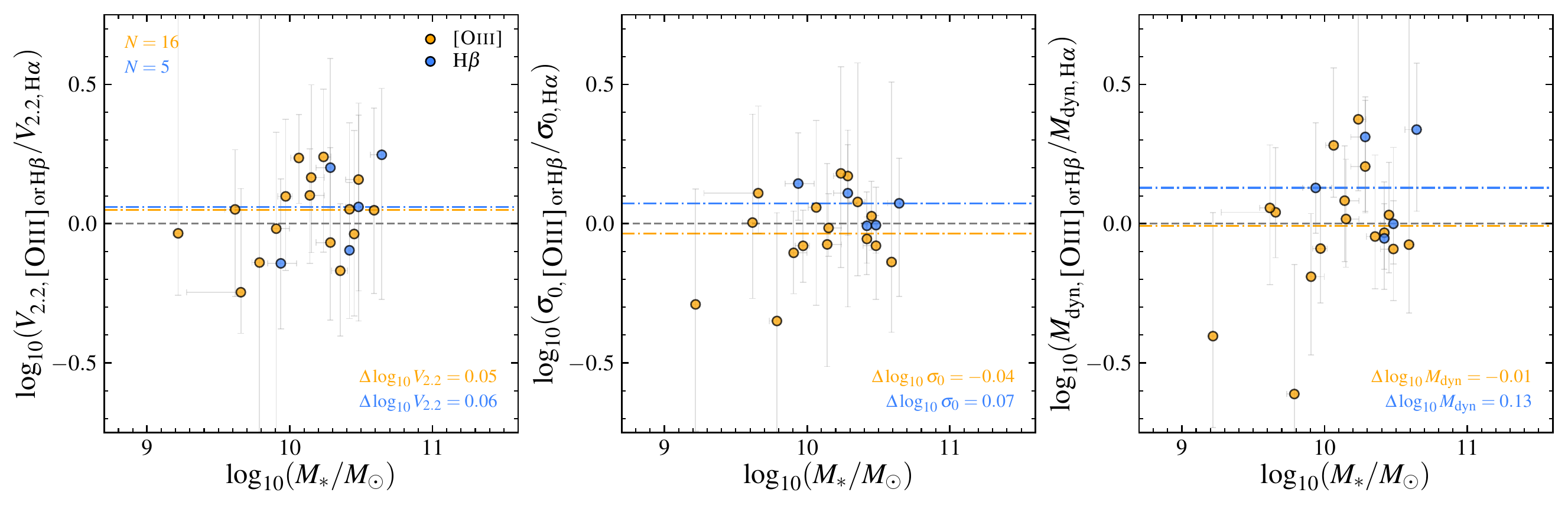}
\vglue -10pt
\caption{
Comparison of \Vtt (\textit{left}), \sigmavint (\textit{middle}), and \Mdyn (\textit{right}) measured with 
different lines for resolved/aligned galaxies with detected rotation at $z\sim2$. 
Offsets between \OIII or \Hbeta and \Halpha are shown as orange and blue circles, respectively. 
The median offsets are marked with dashed-dotted lines of the same colors, 
and the dashed gray lines mark no offset. 
The offsets are generally small (at most, $\sim\!0.13\unit{dex}$). 
}
\vglue 10pt 
\label{fig:kincomp2D}
\end{figure*}

\begin{deluxetable*}{ccccccccccc}
\setlength{\tabcolsep}{0.04in}
\tabletypesize{\scriptsize}\tablenum{2}
\tablecaption{Median kinematic properties, Resolved/Aligned galaxies with detected rotation}\label{tab:kinprop_res}
\tablehead{\colhead{ } & \colhead{ } & \colhead{} & 
\multicolumn{8}{c}{Median Properties} \\ 
\cline{4-11} 
\colhead{Redshift} &\colhead{Bin} &  
\colhead{$\ \ N\ \ $} &  
\colhead{$\langle z \rangle$} & \colhead{$\langle\log_{10}(M_*)\rangle$}  
& \colhead{$\langle\log_{10}(\mathrm{sSFR})\rangle$}  
& \colhead{$\langle\log_{10}(R_E)\rangle$}  
& \colhead{$\langle\vtosigre\rangle$}  
& \colhead{$\langle\vtosigtt\rangle$}  
& \colhead{$\langle{}|V_{2.2}|\rangle$} & \colhead{$\langle\sigma_{V,0}\rangle$} 
 \vspace{-5pt} \\ 
\colhead{} & \colhead{} & \colhead{} &  
\colhead{---} & \colhead{$[\log_{10}(M_{\odot})]$}  
& \colhead{$[\log_{10}(\mathrm{yr^{-1}})]$}  
& \colhead{$[\log_{10}(\mathrm{kpc})]$}  
& \colhead{---}  
& \colhead{---}  
& \colhead{$[\mathrm{km/s}]$}  
& \colhead{$[\mathrm{km/s}]$}  
}  
\startdata 
\vspace{2pt}
  & $\qquad\quad\;\,\,[8, 9.75]$  & 5 & 1.59 & 9.49 & -9.09 & 0.32 & $1.9_{-1.1}^{+0.8}$ & $2.5_{-1.5}^{+1.6}$ & $122_{-86}^{+1}$ & $42_{-12}^{+5}$   \\
  & $\log_{10}(M_*)=[9.75, 10.25]$  & 22 & 1.54 & 10.04 & -8.90 & 0.54 & $1.6_{-0.3}^{+0.2}$ & $2.0_{-0.5}^{+0.2}$ & $131_{-29}^{+4}$ & $67_{-11}^{+2}$   \\
  & $\qquad\qquad\;\,\,[10.25, 12]$  & 14 & 1.53 & 10.69 & -9.16 & 0.66 & $2.8_{-0.7}^{+0.4}$ & $3.4_{-0.9}^{+0.5}$ & $197_{-16}^{+41}$ & $70_{-10}^{+9}$  
\hglue -2pt \vspace{2pt} \\  \cline{2-11} \vspace{2pt}
  & $\qquad\qquad\qquad\;[-11, -8.9]$  & 24 & 1.53 & 10.15 & -9.13 & 0.56 & $1.7_{-0.3}^{+0.5}$ & $2.2_{-0.4}^{+0.6}$ & $134_{-14}^{+17}$ & $67_{-11}^{+2}$   \\
$1.3\leq z \leq 1.8$ & $\log_{10}(\mathrm{sSFR})=[-8.9, -8.4]$  & 15 & 1.54 & 10.13 & -8.80 & 0.53 & $1.8_{-0.4}^{+0.4}$ & $2.1_{-0.4}^{+0.5}$ & $135_{-21}^{+13}$ & $60_{-12}^{+3}$   \\
  & $\qquad\qquad\qquad\;[-8.4, -7.0]$  & 2 & 1.50 & 10.01 & -8.37 & 0.67 & $2.3_{-0.7}^{+4.0}$ & $2.8_{-0.9}^{+5.2}$ & $199_{-58}^{+60}$ & $72_{-30}^{+16}$  
\hglue -2pt \vspace{2pt} \\  \cline{2-11} \vspace{2pt}
  & $\qquad\qquad\qquad[6.0, 8.6]$  & 20 & 1.54 & 9.94 & -9.08 & 0.59 & $1.6_{-0.4}^{+0.3}$ & $2.0_{-0.5}^{+0.3}$ & $108_{-27}^{+7}$ & $59_{-10}^{+3}$   \\
  & $\log_{10}(\Sigma_\mathrm{bar})=[8.6, 9.1]$  & 19 & 1.54 & 10.29 & -8.92 & 0.53 & $2.3_{-0.6}^{+0.3}$ & $2.8_{-0.8}^{+0.3}$ & $165_{-13}^{+26}$ & $70_{-11}^{+9}$   \\
  & $\qquad\qquad\qquad\;[9.1, 13.0]$  & 2 & 1.53 & 10.49 & -8.85 & 0.40 & $2.3_{-0.8}^{+3.8}$ & $2.8_{-1.2}^{+4.3}$ & $162_{-86}^{+63}$ & $89_{-21}^{+87}$  
\hglue -2pt \vspace{2pt} \\  \cline{2-11} \vspace{2pt}
  & Full redshift bin  & 41 & 1.54 & 10.13 & -8.97 & 0.57 & $1.9_{-0.3}^{+0.2}$  & $2.3_{-0.4}^{+0.2}$ & $135_{-10}^{+14}$ & $67_{-11}^{+-0}$  
\hglue -2pt  \\  \hline \vspace{2pt} 
  & $\qquad\quad\;\,\,[8, 9.75]$  & 9 & 2.29 & 9.52 & -8.45 & 0.39 & $1.4_{-0.6}^{+0.6}$ & $1.8_{-0.8}^{+0.7}$ & $98_{-38}^{+6}$ & $62_{-15}^{+9}$   \\
  & $\log_{10}(M_*)=[9.75, 10.25]$  & 19 & 2.28 & 10.02 & -8.81 & 0.46 & $1.0_{-0.1}^{+0.4}$ & $1.5_{-0.3}^{+0.2}$ & $94_{-10}^{+19}$ & $67_{-8}^{+5}$   \\
  & $\qquad\qquad\;\,\,[10.25, 12]$  & 22 & 2.27 & 10.51 & -8.72 & 0.63 & $1.4_{-0.3}^{+0.1}$ & $1.7_{-0.4}^{+0.1}$ & $111_{-2}^{+30}$ & $82_{-10}^{+3}$  
\hglue -2pt \vspace{2pt} \\  \cline{2-11} \vspace{2pt}
  & $\qquad\qquad\qquad\;[-11, -8.9]$  & 10 & 2.21 & 10.21 & -9.07 & 0.44 & $1.5_{-0.3}^{+0.5}$ & $1.8_{-0.4}^{+0.6}$ & $130_{-32}^{+7}$ & $76_{-13}^{+5}$   \\
$2.0\leq z \leq 2.6$ & $\log_{10}(\mathrm{sSFR})=[-8.9, -8.4]$  & 32 & 2.29 & 10.25 & -8.71 & 0.56 & $1.4_{-0.3}^{+0.1}$ & $1.7_{-0.3}^{+0.2}$ & $103_{-5}^{+20}$ & $71_{-4}^{+7}$   \\
  & $\qquad\qquad\qquad\;[-8.4, -7.0]$  & 8 & 2.36 & 9.87 & -8.18 & 0.48 & $0.8_{-0.2}^{+0.3}$ & $1.0_{-0.2}^{+0.4}$ & $77_{-29}^{+9}$ & $62_{-4}^{+8}$  
\hglue -2pt \vspace{2pt} \\  \cline{2-11} \vspace{2pt}
  & $\qquad\qquad\qquad[6.0, 8.6]$  & 14 & 2.21 & 9.88 & -8.78 & 0.55 & $1.5_{-0.4}^{+0.3}$ & $1.8_{-0.4}^{+0.5}$ & $114_{-22}^{+6}$ & $67_{-13}^{+5}$   \\
  & $\log_{10}(\Sigma_\mathrm{bar})=[8.6, 9.1]$  & 27 & 2.28 & 10.24 & -8.72 & 0.54 & $1.3_{-0.3}^{+0.1}$ & $1.6_{-0.4}^{+0.1}$ & $97_{-9}^{+19}$ & $67_{-4}^{+7}$   \\
  & $\qquad\qquad\qquad\;[9.1, 13.0]$  & 9 & 2.32 & 10.59 & -8.59 & 0.41 & $1.5_{-0.5}^{+0.1}$ & $1.9_{-0.8}^{+0.1}$ & $129_{-31}^{+28}$ & $88_{-15}^{+2}$  
\hglue -2pt \vspace{2pt} \\  \cline{2-11} \vspace{2pt}
  & Full redshift bin  & 50 & 2.28 & 10.16 & -8.71 & 0.52 & $1.4_{-0.3}^{+0.0}$  & $1.7_{-0.4}^{+0.0}$ & $103_{-7}^{+11}$ & $70_{-4}^{+5}$  
\hglue -2pt  \\  \hline \vspace{2pt} 
  & $\qquad\quad\;\,\,[8, 9.75]$  & 7 & 3.21 & 9.33 & -7.79 & 0.32 & $0.8_{-0.2}^{+0.6}$ & $1.1_{-0.3}^{+0.7}$ & $57_{-8}^{+35}$ & $69_{-18}^{+3}$   \\
  & $\log_{10}(M_*)=[9.75, 10.25]$  & 7 & 3.24 & 9.92 & -8.36 & 0.32 & $0.8_{-0.3}^{+0.3}$ & $1.1_{-0.3}^{+0.3}$ & $85_{-24}^{+20}$ & $74_{-6}^{+15}$   \\
  & $\qquad\qquad\;\,\,[10.25, 12]$ & 0 & --- & --- & --- & --- & --- & --- & --- & ---
\hglue -2pt \vspace{2pt} \\  \cline{2-11} \vspace{2pt}
  & $\qquad\qquad\qquad\;[-11, -8.9]$ & 0 & --- & --- & --- & --- & --- & --- & --- & --- \\
$2.9\leq z \leq 3.8$ & $\log_{10}(\mathrm{sSFR})=[-8.9, -8.4]$  & 5 & 3.23 & 9.82 & -8.47 & 0.32 & $1.1_{-0.3}^{+0.3}$ & $1.4_{-0.5}^{+0.4}$ & $69_{-16}^{+28}$ & $69_{-14}^{+4}$   \\
  & $\qquad\qquad\qquad\;[-8.4, -7.0]$  & 9 & 3.22 & 9.53 & -7.79 & 0.29 & $0.7_{-0.2}^{+0.4}$ & $1.0_{-0.2}^{+0.5}$ & $84_{-29}^{+17}$ & $77_{-11}^{+13}$  
\hglue -2pt \vspace{2pt} \\  \cline{2-11} \vspace{2pt}
  & $\qquad\qquad\qquad[6.0, 8.6]$  & 1 & 3.16 & 9.31 & -8.47 & 0.32 & $1.2_{-0.8}^{+2.9}$ & $1.6_{-1.1}^{+3.4}$ & $57_{-47}^{+36}$ & $37_{-22}^{+12}$   \\
  & $\log_{10}(\Sigma_\mathrm{bar})=[8.6, 9.1]$  & 8 & 3.21 & 9.75 & -8.40 & 0.34 & $0.8_{-0.1}^{+0.4}$ & $1.0_{-0.1}^{+0.5}$ & $77_{-15}^{+16}$ & $70_{-8}^{+5}$   \\
  & $\qquad\qquad\qquad\;[9.1, 13.0]$  & 5 & 3.32 & 9.81 & -7.59 & 0.20 & $0.6_{-0.2}^{+0.6}$ & $1.1_{-0.5}^{+0.5}$ & $85_{-32}^{+31}$ & $77_{-1}^{+29}$  
\hglue -2pt \vspace{2pt} \\  \cline{2-11} \vspace{2pt}
  & Full redshift bin  & 14 & 3.23 & 9.74 & -8.34 & 0.32 & $0.8_{-0.1}^{+0.3}$  & $1.1_{-0.2}^{+0.4}$ & $84_{-20}^{+7}$ & $73_{-8}^{+6}$  
\hglue -2pt \vspace{2pt} \\   \enddata 
\end{deluxetable*}

\begin{deluxetable*}{ccccccccc}
\setlength{\tabcolsep}{0.04in}
\tabletypesize{\scriptsize}\tablenum{3}
\tablecaption{Median kinematic properties, Unresolved/Misaligned and Undetected Rotation Galaxies}\label{tab:kinprop_unres}
\tablehead{\colhead{ } & \colhead{ } & \colhead{} & 
\multicolumn{6}{c}{Median Properties} \\ 
\cline{4-9} 
\colhead{Redshift} &\colhead{Bin} &  
\colhead{$\ \ N\ \ $} &  
\colhead{$\langle z \rangle$} & \colhead{$\langle\log_{10}(M_*)\rangle$}  
& \colhead{$\langle\log_{10}(\mathrm{sSFR})\rangle$}  
& \colhead{$\langle\log_{10}(R_E)\rangle$}  
& \colhead{$\langle\vtosigre\rangle$\tablenotemark{a}}  
& \colhead{$\langle\vtosigtt\rangle$\tablenotemark{a,b}}  
 \vspace{-5pt} \\ 
 \colhead{} & \colhead{} & \colhead{} &  
\colhead{---} & \colhead{$[\log_{10}(M_{\odot})]$}  
& \colhead{$[\log_{10}(\mathrm{yr^{-1}})]$}  
& \colhead{$[\log_{10}(\mathrm{kpc})]$}  
& \colhead{---}  
& \colhead{---}  
}  
\startdata 
\vspace{2pt}
  & $\qquad\quad\;\,\,[8, 9.75]$  & 58 & 1.53 & 9.50 & -8.78 & 0.34 & $2.7_{-1.1}^{+1.8}$ & $2.8_{-1.1}^{+1.9}$  \\
  & $\log_{10}(M_*)=[9.75, 10.25]$  & 51 & 1.54 & 9.95 & -8.95 & 0.40 & $2.3_{-1.1}^{+2.5}$ & $2.4_{-1.1}^{+2.6}$  \\
  & $\qquad\qquad\;\,\,[10.25, 12]$  & 54 & 1.47 & 10.55 & -9.13 & 0.58 & $8.6_{-4.6}^{+99}$ & $9.0_{-4.8}^{+99}$ 
\hglue -2pt \vspace{2pt} \\  \cline{2-9} \vspace{2pt}
  & $\qquad\qquad\qquad\;[-11, -8.9]$  & 90 & 1.49 & 10.15 & -9.17 & 0.51 & $6.3_{-2.7}^{+99}$ & $6.6_{-2.8}^{+99}$  \\
$1.3\leq z \leq 1.8$ & $\log_{10}(\mathrm{sSFR})=[-8.9, -8.4]$  & 60 & 1.53 & 9.68 & -8.70 & 0.38 & $2.3_{-0.8}^{+1.2}$ & $2.4_{-0.8}^{+1.2}$  \\
  & $\qquad\qquad\qquad\;[-8.4, -7.0]$  & 13 & 1.57 & 9.83 & -8.32 & 0.30 & $1.6_{-1.6}^{+1.1}$ & $1.7_{-1.7}^{+1.1}$ 
\hglue -2pt \vspace{2pt} \\  \cline{2-9} \vspace{2pt}
  & $\qquad\qquad\qquad[6.0, 8.6]$  & 70 & 1.53 & 9.64 & -9.06 & 0.52 & $3.1_{-1.3}^{+4.4}$ & $3.2_{-1.4}^{+4.6}$  \\
  & $\log_{10}(\Sigma_\mathrm{bar})=[8.6, 9.1]$  & 71 & 1.50 & 10.23 & -8.86 & 0.44 & $3.1_{-1.2}^{+4.2}$ & $3.2_{-1.2}^{+4.4}$  \\
  & $\qquad\qquad\qquad\;[9.1, 13.0]$  & 22 & 1.52 & 10.23 & -8.79 & 0.20 & $2.2_{-1.5}^{+1.6}$ & $2.3_{-1.6}^{+1.7}$ 
\hglue -2pt \vspace{2pt} \\  \cline{2-9} \vspace{2pt}
  & Full redshift bin  & 163 & 1.52 & 9.93 & -8.94 & 0.44 & $3.8_{-1.1}^{+2.2}$ & $4.0_{-1.1}^{+2.3}$  
\hglue -2pt  \\  \hline \vspace{2pt} 
  & $\qquad\quad\;\,\,[8, 9.75]$  & 99 & 2.24 & 9.53 & -8.65 & 0.21 & $1.0_{-1.0}^{+0.9}$ & $1.0_{-1.0}^{+0.9}$  \\
  & $\log_{10}(M_*)=[9.75, 10.25]$  & 133 & 2.29 & 9.93 & -8.76 & 0.31 & $1.8_{-0.8}^{+1.2}$ & $1.9_{-0.8}^{+1.2}$  \\
  & $\qquad\qquad\;\,\,[10.25, 12]$  & 51 & 2.31 & 10.44 & -8.76 & 0.47 & $2.6_{-1.7}^{+2.5}$ & $2.7_{-1.8}^{+2.6}$ 
\hglue -2pt \vspace{2pt} \\  \cline{2-9} \vspace{2pt}
  & $\qquad\qquad\qquad\;[-11, -8.9]$  & 77 & 2.29 & 9.95 & -9.04 & 0.35 & $4.7_{-2.2}^{+99}$ & $4.9_{-2.3}^{+99}$  \\
$2.0\leq z \leq 2.6$ & $\log_{10}(\mathrm{sSFR})=[-8.9, -8.4]$  & 147 & 2.28 & 9.85 & -8.67 & 0.33 & $1.5_{-0.6}^{+0.7}$ & $1.6_{-0.6}^{+0.7}$  \\
  & $\qquad\qquad\qquad\;[-8.4, -7.0]$  & 59 & 2.28 & 9.81 & -8.28 & 0.19 & $0.0_{-0.0}^{+1.1}$ & $0.0_{-0.0}^{+1.1}$ 
\hglue -2pt \vspace{2pt} \\  \cline{2-9} \vspace{2pt}
  & $\qquad\qquad\qquad[6.0, 8.6]$  & 72 & 2.26 & 9.77 & -8.84 & 0.55 & $0.4_{-0.4}^{+1.6}$ & $0.4_{-0.4}^{+1.7}$  \\
  & $\log_{10}(\Sigma_\mathrm{bar})=[8.6, 9.1]$  & 122 & 2.29 & 9.85 & -8.70 & 0.32 & $1.5_{-0.7}^{+0.7}$ & $1.6_{-0.7}^{+0.7}$  \\
  & $\qquad\qquad\qquad\;[9.1, 13.0]$  & 89 & 2.29 & 9.91 & -8.45 & 0.10 & $0.3_{-0.3}^{+0.7}$ & $0.3_{-0.3}^{+0.7}$ 
\hglue -2pt \vspace{2pt} \\  \cline{2-9} \vspace{2pt}
  & Full redshift bin  & 283 & 2.29 & 9.86 & -8.70 & 0.31 & $1.9_{-0.6}^{+0.6}$ & $2.0_{-0.6}^{+0.6}$  
\hglue -2pt  \\  \hline \vspace{2pt} 
  & $\qquad\quad\;\,\,[8, 9.75]$  & 55 & 3.23 & 9.53 & -8.42 & 0.20 & $1.5_{-1.5}^{+5.2}$ & $1.6_{-1.6}^{+5.4}$  \\
  & $\log_{10}(M_*)=[9.75, 10.25]$  & 56 & 3.17 & 9.94 & -8.65 & 0.29 & $1.9_{-1.0}^{+1.3}$ & $2.0_{-1.0}^{+1.4}$  \\
  & $\qquad\qquad\;\,\,[10.25, 12]$  & 19 & 3.23 & 10.39 & -8.68 & 0.35 & $2.5_{-0.4}^{+99}$ & $2.6_{-0.4}^{+99}$ 
\hglue -2pt \vspace{2pt} \\  \cline{2-9} \vspace{2pt}
  & $\qquad\qquad\qquad\;[-11, -8.9]$  & 16 & 3.13 & 9.92 & -8.97 & 0.29 & $0.0_{-0.0}^{+6.0}$ & $0.0_{-0.0}^{+6.2}$  \\
$2.9\leq z \leq 3.8$ & $\log_{10}(\mathrm{sSFR})=[-8.9, -8.4]$  & 71 & 3.19 & 9.91 & -8.65 & 0.29 & $2.0_{-0.9}^{+1.4}$ & $2.1_{-0.9}^{+1.5}$  \\
  & $\qquad\qquad\qquad\;[-8.4, -7.0]$  & 43 & 3.30 & 9.66 & -8.08 & 0.23 & $0.5_{-0.5}^{+1.4}$ & $0.5_{-0.5}^{+1.5}$ 
\hglue -2pt \vspace{2pt} \\  \cline{2-9} \vspace{2pt}
  & $\qquad\qquad\qquad[6.0, 8.6]$  & 18 & 3.17 & 9.69 & -8.63 & 0.53 & $0.0_{-0.0}^{+0.5}$ & $0.0_{-0.0}^{+0.5}$  \\
  & $\log_{10}(\Sigma_\mathrm{bar})=[8.6, 9.1]$  & 58 & 3.19 & 9.90 & -8.66 & 0.30 & $1.0_{-1.0}^{+2.6}$ & $1.0_{-1.0}^{+2.7}$  \\
  & $\qquad\qquad\qquad\;[9.1, 13.0]$  & 54 & 3.24 & 9.82 & -8.41 & 0.06 & $1.4_{-1.1}^{+1.0}$ & $1.5_{-1.1}^{+1.0}$ 
\hglue -2pt \vspace{2pt} \\  \cline{2-9} \vspace{2pt}
  & Full redshift bin  & 130 & 3.21 & 9.84 & -8.53 & 0.26 & $1.6_{-0.8}^{+1.0}$ & $1.7_{-0.8}^{+1.0}$  
\hglue -2pt \vspace{2pt} \\   \enddata 
\tablenotetext{a}{$\vtosig$ for the unresolved/misaligned and velocity-limit galaxies 
without constrained upper uncertainties 
(e.g., fit slope is consistent with zero at maximum grid $\vtosigre=10$) 
have upper uncertainties marked as 99.}
 \vspace{-6pt} 
\tablenotetext{b}{Calculated assuming $r_t = 0.4 R_{s} = 0.4/1.676 \, R_E = 0.4/2.2 R_{2.2}$
 for the unresolved/misaligned and velocity-limit galaxies.}
 \vspace{-20pt} 
\end{deluxetable*}

\begin{deluxetable*}{ccccccccc}
\tabletypesize{\scriptsize}\tablenum{4}
\tablecaption{Median mass offsets\tablenotemark{a}}\label{tab:massoffsets}
\tablehead{\colhead{} & \colhead{} & \colhead{} & 
\colhead{$\langle z \rangle$} & \colhead{$\langle\log_{10}(M_*)\rangle$} 
& \colhead{$\langle\log_{10}(\mathrm{sSFR})\rangle$} 
& \colhead{$\langle\log_{10}(R_E)\rangle$} 
& \colhead{$\langle\log_{10}(\Sigma_{\mathrm{bar}})\rangle$} 
& \colhead{$\langle\Delta \log_{10}M\rangle$} 
\vspace{-5pt} \\
\colhead{Redshift} & \colhead{Bin} & \colhead{$N$} & 
\colhead{---} & \colhead{$[\log_{10}(M_{\odot})]$} 
& \colhead{$[\log_{10}(\mathrm{yr^{-1}})]$} 
& \colhead{$[\log_{10}(\mathrm{kpc})]$} 
& \colhead{$[\log_{10}(M_{\odot}\mathrm{kpc}^{-2})]$} 
& \colhead{[dex]} } 
\startdata \vspace{2pt}
  & $\qquad\quad\;\,\,[8, 9.75]$ & 63 & 1.54 & 9.50 & -8.81 & 0.33 & 8.40 & $0.16_{-0.03}^{+0.13}$\\
  & $\log_{10}(M_*)=[9.75, 10.25]$ & 73 & 1.54 & 9.97 & -8.94 & 0.45 & 8.61 & $0.19_{-0.06}^{+0.03}$\\
  & $\qquad\qquad\;\,\,[10.25, 12]$ & 68 & 1.48 & 10.56 & -9.13 & 0.61 & 8.84 & $0.19_{-0.02}^{+0.07}$
\hglue -2pt \vspace{2pt} \\  \cline{2-9} \vspace{2pt}
  & $\qquad\qquad\qquad\;[-11, -8.9]$ & 114 & 1.50 & 10.15 & -9.16 & 0.53 & 8.58 & $0.24_{-0.05}^{+0.03}$\\
  & $\log_{10}(\mathrm{sSFR})=[-8.9, -8.4]$ & 75 & 1.54 & 9.82 & -8.72 & 0.39 & 8.73 & $0.10_{-0.02}^{+0.03}$\\
$1.3\leq z \leq 1.8$ & $\qquad\qquad\qquad\;[-8.4, -7.0]$ & 15 & 1.53 & 9.84 & -8.34 & 0.31 & 8.77 & $0.15_{-0.20}^{+0.09}$
\hglue -2pt \vspace{2pt} \\  \cline{2-9} \vspace{2pt}
  & $\quad\qquad R_E =\ [0.0, 2.5]$ & 78 & 1.52 & 9.67 & -8.81 & 0.29 & 8.82 & $-0.04_{-0.03}^{+0.07}$\\
  & $\qquad\qquad\qquad\;[2.5, 15.0]$ & 126 & 1.53 & 10.22 & -9.09 & 0.58 & 8.58 & $0.26_{-0.04}^{+0.03}$
\hglue -2pt \vspace{2pt} \\  \cline{2-9} \vspace{2pt}
  & $\qquad\qquad\qquad\,[6.0, 8.6]$ & 90 & 1.53 & 9.77 & -9.06 & 0.54 & 8.34 & $0.31_{-0.05}^{+0.03}$\\
  & $\log_{10}(\Sigma_{\mathrm{bar}}) =\ \ [8.6, 9.1]$ & 90 & 1.52 & 10.23 & -8.88 & 0.46 & 8.84 & $0.13_{-0.03}^{+0.03}$\\
  & $\qquad\qquad\qquad\ \; [9.1, 13.0]$ & 24 & 1.52 & 10.31 & -8.79 & 0.20 & 9.22 & $-0.02_{-0.08}^{+0.09}$
\hglue -2pt \vspace{2pt} \\  \hline \vspace{2pt}
  & $\qquad\quad\;\,\,[8, 9.75]$ & 108 & 2.24 & 9.53 & -8.62 & 0.23 & 8.72 & $-0.07_{-0.02}^{+0.03}$\\
  & $\log_{10}(M_*)=[9.75, 10.25]$ & 152 & 2.29 & 9.94 & -8.76 & 0.34 & 8.90 & $-0.11_{-0.02}^{+0.02}$\\
  & $\qquad\qquad\;\,\,[10.25, 12]$ & 73 & 2.30 & 10.46 & -8.76 & 0.54 & 9.00 & $-0.03_{-0.05}^{+0.05}$
\hglue -2pt \vspace{2pt} \\  \cline{2-9} \vspace{2pt}
  & $\qquad\qquad\qquad\;[-11, -8.9]$ & 87 & 2.29 & 10.00 & -9.04 & 0.38 & 8.74 & $0.11_{-0.10}^{+0.04}$\\
  & $\log_{10}(\mathrm{sSFR})=[-8.9, -8.4]$ & 179 & 2.28 & 9.89 & -8.68 & 0.35 & 8.80 & $-0.08_{-0.03}^{+0.03}$\\
$2.0\leq z \leq 2.6$ & $\qquad\qquad\qquad\;[-8.4, -7.0]$ & 67 & 2.29 & 9.81 & -8.26 & 0.23 & 9.18 & $-0.21_{-0.04}^{+0.02}$
\hglue -2pt \vspace{2pt} \\  \cline{2-9} \vspace{2pt}
  & $\quad\qquad R_E =\ [0.0, 2.5]$ & 198 & 2.27 & 9.78 & -8.65 & 0.22 & 9.06 & $-0.12_{-0.02}^{+0.03}$\\
  & $\qquad\qquad\qquad\;[2.5, 15.0]$ & 135 & 2.29 & 10.12 & -8.79 & 0.57 & 8.63 & $-0.03_{-0.06}^{+0.03}$
\hglue -2pt \vspace{2pt} \\  \cline{2-9} \vspace{2pt}
  & $\qquad\qquad\qquad\,[6.0, 8.6]$ & 86 & 2.25 & 9.80 & -8.83 & 0.55 & 8.39 & $0.06_{-0.06}^{+0.06}$\\
  & $\log_{10}(\Sigma_{\mathrm{bar}}) =\ \ [8.6, 9.1]$ & 149 & 2.29 & 9.92 & -8.71 & 0.35 & 8.83 & $-0.09_{-0.02}^{+0.04}$\\
  & $\qquad\qquad\qquad\ \; [9.1, 13.0]$ & 98 & 2.29 & 9.94 & -8.46 & 0.14 & 9.31 & $-0.26_{-0.03}^{+0.02}$
\hglue -2pt \vspace{2pt} \\  \hline \vspace{2pt}
  & $\qquad\quad\;\,\,[8, 9.75]$ & 62 & 3.23 & 9.52 & -8.39 & 0.22 & 9.01 & $-0.21_{-0.08}^{+0.09}$\\
  & $\log_{10}(M_*)=[9.75, 10.25]$ & 63 & 3.18 & 9.94 & -8.58 & 0.29 & 9.02 & $-0.10_{-0.06}^{+0.02}$\\
  & $\qquad\qquad\;\,\,[10.25, 12]$ & 19 & 3.23 & 10.39 & -8.68 & 0.35 & 9.35 & $-0.25_{-0.13}^{+0.17}$
\hglue -2pt \vspace{2pt} \\  \cline{2-9} \vspace{2pt}
  & $\qquad\qquad\qquad\;[-11, -8.9]$ & 16 & 3.13 & 9.92 & -8.97 & 0.29 & 8.93 & $0.01_{-0.07}^{+0.06}$\\
  & $\log_{10}(\mathrm{sSFR})=[-8.9, -8.4]$ & 76 & 3.20 & 9.91 & -8.64 & 0.29 & 8.91 & $-0.11_{-0.05}^{+0.02}$\\
$2.9\leq z \leq 3.8$ & $\qquad\qquad\qquad\;[-8.4, -7.0]$ & 52 & 3.27 & 9.65 & -8.08 & 0.24 & 9.28 & $-0.28_{-0.10}^{+0.03}$
\hglue -2pt \vspace{2pt} \\  \cline{2-9} \vspace{2pt}
  & $\quad\qquad R_E =\ [0.0, 2.5]$ & 107 & 3.22 & 9.76 & -8.46 & 0.18 & 9.18 & $-0.23_{-0.04}^{+0.04}$\\
  & $\qquad\qquad\qquad\;[2.5, 15.0]$ & 37 & 3.21 & 9.96 & -8.62 & 0.49 & 8.69 & $-0.05_{-0.03}^{+0.12}$
\hglue -2pt \vspace{2pt} \\  \cline{2-9} \vspace{2pt}
  & $\qquad\qquad\qquad\,[6.0, 8.6]$ & 19 & 3.16 & 9.66 & -8.62 & 0.53 & 8.46 & $-0.04_{-0.05}^{+0.01}$\\
  & $\log_{10}(\Sigma_{\mathrm{bar}}) =\ \ [8.6, 9.1]$ & 66 & 3.20 & 9.88 & -8.59 & 0.31 & 8.87 & $-0.07_{-0.10}^{+0.04}$\\
  & $\qquad\qquad\qquad\ \; [9.1, 13.0]$ & 59 & 3.24 & 9.81 & -8.38 & 0.06 & 9.42 & $-0.29_{-0.08}^{+0.03}$
\hglue -2pt \vspace{2pt} \\   \enddata
\tablenotetext{a}{Including all galaxies, where $M_{\mathrm{dyn}}$ for 
unresolved/misaligned and velocity-limit galaxies are calculated using the median $V/\sigma$ for each respective bin.}
\end{deluxetable*}

\end{document}